# Generation, Annihilation and Flow of Structural Information in Ultrasonic Nondestructive Evaluation

Dr. Frank Schubert, Fraunhofer IKTS, Dresden, Germany

frank.schubert@ikts.fraunhofer.de

## Abstract

Nondestructive testing using ultrasound is based on the interaction of sound waves with the object being tested and any defects it may contain. The aim is to extract as much information as possible from the scattered wave field. In this paper, the concept of *information* in the context of ultrasonic testing is formalized and quantified physically for the first time. To this end, a balance equation for structural information is derived, analogous to Poynting's theorem for elastic energy. It contains an information density, an information flux vector as well as various information source terms. Various examples demonstrate how structural information is generated and annihilated within a component and along which pathways it travels from the defect to the sensor. Subsequently, the significance and potential impact of this new information concept for practical ultrasonic testing, structural health monitoring, numerical simulation, and machine learning are discussed. Finally, similarities and differences to mathematical Shannon information and statistical Fisher information are highlighted.

# 1. Introduction and Motivation

The American mathematician and electrical engineer Claude E. Shannon (1916–2001) laid the foundations for a systematic and quantifiable concept of information with his groundbreaking publication *A Mathematical Theory of Communication* [1] in 1948, thus paving the way for modern information and communication technology. Terms like *bits* and *bytes* are a direct consequence of this work and have entered both scientific and general usage. Against this backdrop, it is surprising that the term *information* in nondestructive evaluation (NDE) and structural health monitoring (SHM) has so far only been used colloquially, and that there is no physically formalized and quantifiable theory for it. NDE is fundamentally based on the interaction of physical fields with the object under investigation and any defects it may contain. The aim is to extract as much information as possible from the altered or scattered field. The concept of information in NDE is therefore so fundamental that a formalized theory could lead to groundbreaking new insights and systematic improvements in testing systems, inspection setups, and evaluation methods. This paper presents a first attempt at developing such a theory using ultrasonic testing as an example. However, the basic concept can, in principle, be applied to many other NDE and SHM methods, particularly elastodynamic and acoustic methods, electromagnetic and optical methods, eddy current testing, and thermography, to name just a few important examples.

Section 2 begins by discussing various reasonable requirements and criteria for an information measure suitable for ultrasonic NDE. Section 3 then examines in detail the significance of system changes and differential wave fields in a simplified, one-dimensional model using numerical wave field simulations. Section 4 formalizes the identified information measure, culminating in a one-dimensional balance equation for structural information. In this context, the concepts of information density, information flow, and information sources and sinks are introduced, and the similarities and differences to Poynting's theorem for elastic energy are discussed. Section 5 verifies the identified information quantities using selected simulations. Section 6 demonstrates the significance of information at the sensor location and in the time-domain signal detected, respectively. Here, the new concept of 'Change Bits', or shortly, *Cbits*, is introduced as a quantitative absolute unit of information. Finally, Section 7 discusses the relevance of the new information measure for ultrasonic NDE, SHM, machine learning, and numerical simulation. Section 8 addresses similarities and differences to Shannon information from mathematical information theory and Fisher information from statistics. The article concludes with a summary and outlook in Section 9. In the additional Appendix, a wave equation for the differential field is derived.

## 2. Requirements for a Suitable Information Measure

The better the interaction process between ultrasound waves excited by the probe and the defects present in the component is understood, the more optimally the testing system can be configured and the more effectively the time-domain signals detected by the sensor can be evaluated. For many years, numerical simulations have been used to elucidate this interaction, either in the form of commercially available software packages or in the form of individual in-house developments by various research groups.

Figure 1 shows an example of the interaction of a wave field emitted by an ultrasonic transducer with an elastic inclusion embedded in a homogeneous isotropic matrix medium, calculated using the author's EFIT simulation tool [2]. Due to the numerous diffraction and mode conversion effects, as well as the slight inclination of the elliptical inclusion, a very complex, asymmetric scattered field results, consisting of various longitudinal and transverse wave components. The scattered field depends on various parameters of the inclusion, such as its position, size, shape and orientation, its material properties, and - at a later stage - also on the position and shape of the component's back wall or boundaries (via secondary interactions).

In a sense, the wave field shown in Figure 1 therefore contains *structural information* about all these parameters, albeit in a cumulative, non-separable form. Based on the individual simulation, it is not possible to determine precisely where in the wave field the information, e.g., about the orientation of the scatterer or its material properties, is stored and how it propagates to a sensor.

From the considerations above, a first important requirement for the information measure to be found can be formulated:

1) The generation and flow of information should be describable for each structural parameter of the system individually.

Further meaningful requirements arise directly from the separation and characterization of the incident and scattered fields (see Figure 1):

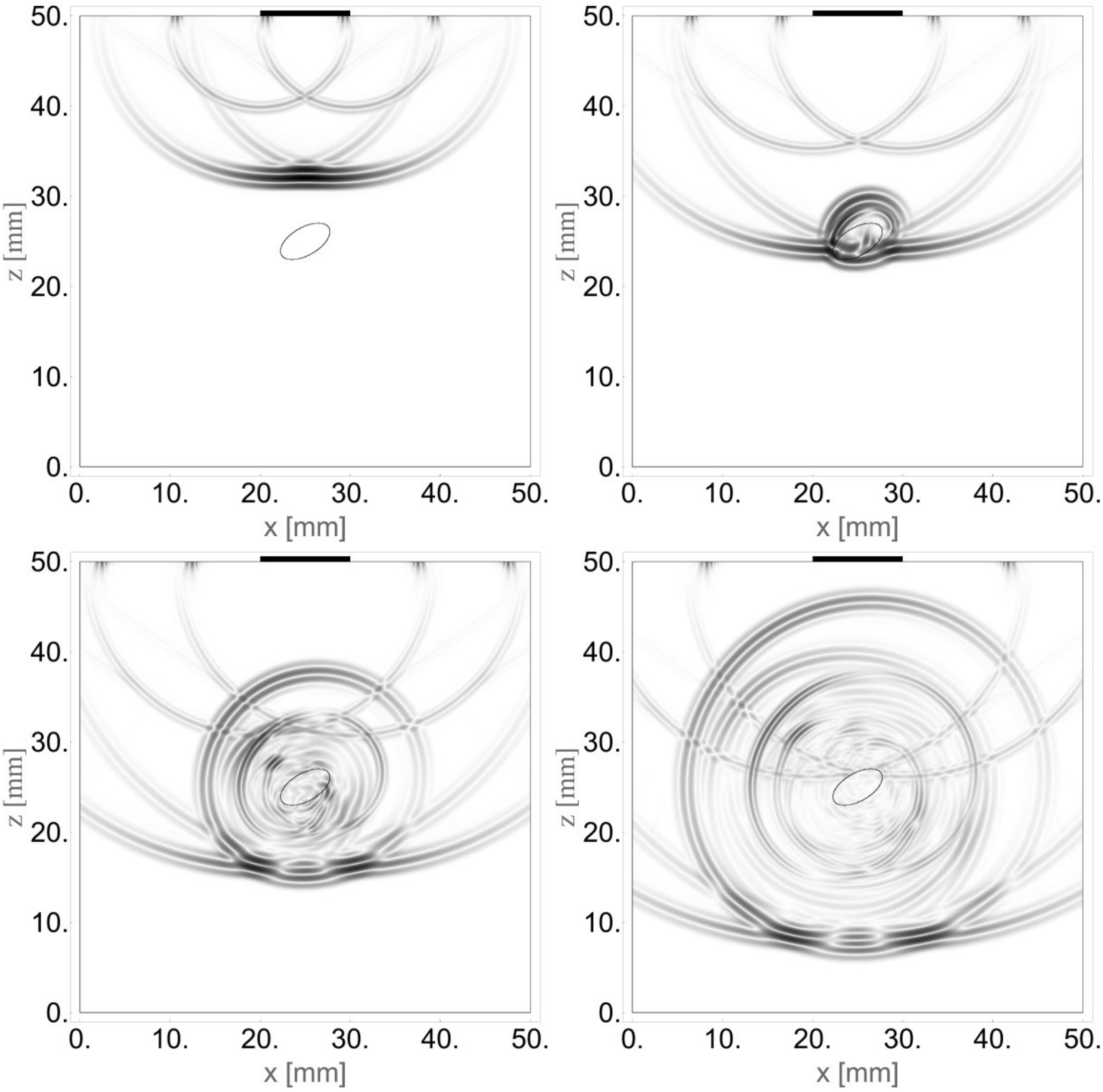

***Figure 1:*** *Two-dimensional plane strain simulation of the interaction of the wave field emitted by an ultrasonic probe with an elliptical elastic inclusion in a homogeneous, isotropic matrix medium, performed using elastodynamic finite integration technique (EFIT, [2]). Four time snapshots of the absolute magnitude of the 2D particle velocity vector are shown on a slightly logarithmic grayscale, approximately 5 µs (top left), 7 µs (top right), 9 µs (bottom left), and 11 µs (bottom right) after the start of excitation. The maximum magnitudes are shown in black, the minimum (zero) in white. The probe (black rectangle) has an active aperture of 10 mm and is designed as a piston oscillator. The center frequency of the excitation signal is 2 MHz, and the waveform is a raised cosine with two cycles (RC2). The material properties of the matrix are* $c_P =$ *4000 m/s (P-wave velocity),* $c_S =$ *2300 m/s (S-wave velocity), and* $\rho =$ *2400 kg/m³ (density), while those of the elastic inclusion are* $c_P =$ *5900 m/s,* $c_S =$ *3200 m/s, and* $\rho =$ *7800 kg/m³. These are typical values for a steel inclusion in a concrete matrix. The two principal axes of the elliptical scatterer, inclined at 30° to the horizontal, are 6 mm and 3 mm, respectively. The dimensions of the numerical grid are 870 × 870 grid cells and 1597 time steps (*$\Delta x = \Delta z =$ *57.47126 µm,* $\Delta t =$ *6.886826 ns). Perfectly Matched Layer (PML) boundary conditions were implemented at the two vertical model boundaries to suppress lateral reflections.*

2) Since the information about the individual structural parameters of a defect is only encoded in the scattered wave field, the incident wave field generated by the probe cannot carry any information as long as it has not interacted with any scatterer (see sub-figure 1 in Fig. 1).

3) Point 2 implies, in particular, that the source of the wave field (here, the probe aperture) cannot, in general, also be the source of information.
4) The information itself is apparently generated during the interaction of the incident wave field with the scatterer(s). This suggests that the scattering interfaces between two different materials, and possibly also the volume of the scatterer, can serve as sources of information (see sub-figure 2 in Figure 1).
5) It is plausible to assume that information is coupled to the propagation of the scattered waves, i.e., that the latter serve as carriers of information. Therefore, information should also propagate at the speed of the respective wave.

Further fundamental requirements for the information measure, which is yet to be defined, can be derived from the separate investigation of reflected and transmitted waves. For this purpose, we consider the simplified one-dimensional model with triple layering shown in Figure 2.

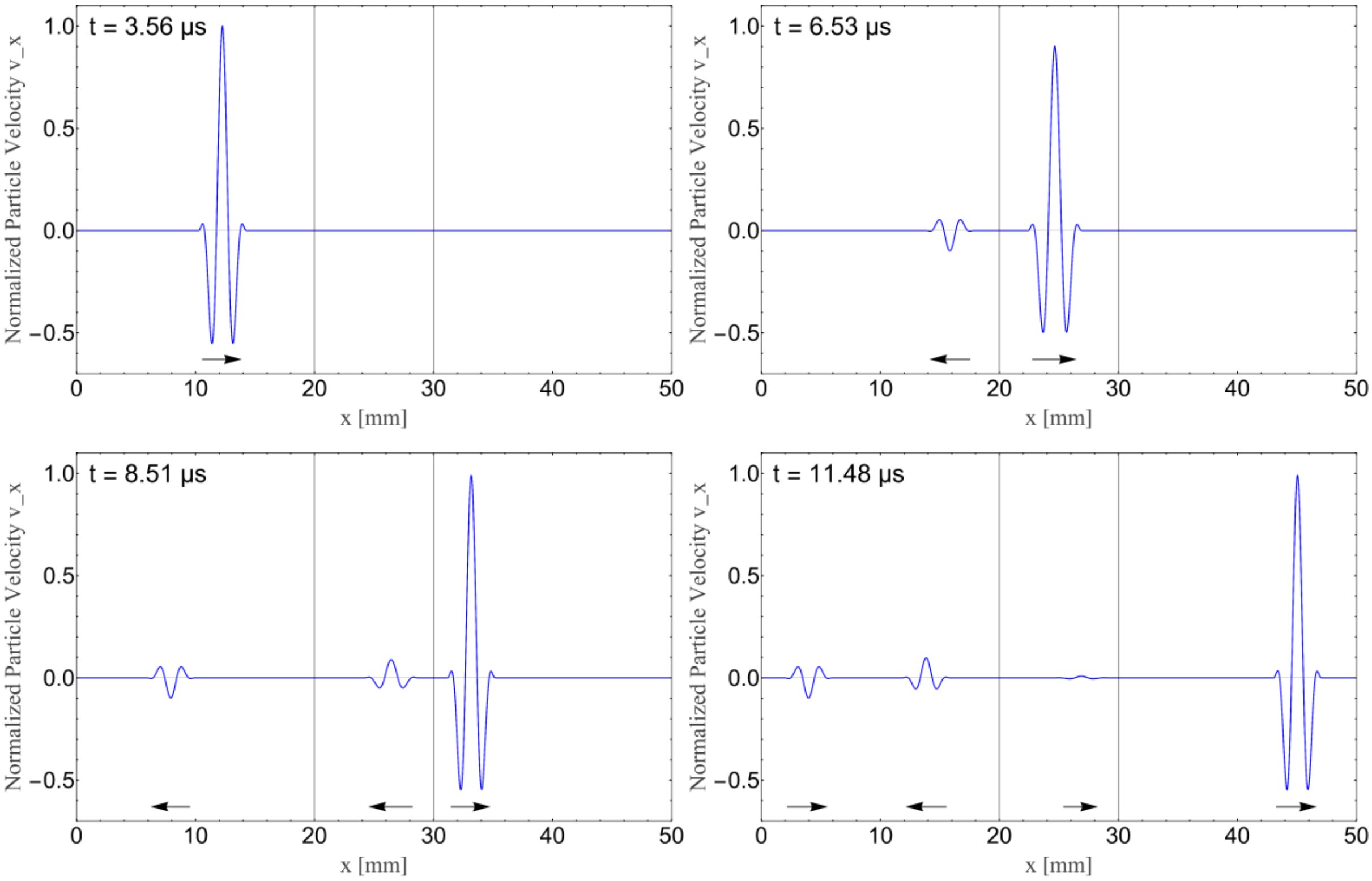

***Figure 2:*** *A one-dimensional simulation, performed using elastodynamic finite integration technique (EFIT), of the interaction between a plane longitudinal wave excited at the left boundary of the model (x = 0 mm) and an elastic layer, 10 mm thick, between x = 20 mm and x = 30 mm, embedded in a homogeneous, isotropic matrix medium (three-layer structure). Four time snapshots of the normalized particle velocity are shown: 3.56 μs (top left), 6.53 μs (top right), 8.51 μs (bottom left), and 11.48 μs (bottom right) after the start of excitation. The center frequency of the excitation signal is 2 MHz, and the waveform is a raised cosine with two cycles (RC2). The material properties of the matrix are* $c_P =$ *4000 m/s (P-wave velocity),* $c_S =$ *2300 m/s (S-wave velocity), and* $\rho =$ *2400 kg/m³ (density), while those of the layer are* $c_P =$ *4500 m/s,* $c_S =$ *2500 m/s, and* $\rho =$ *2600 kg/m³. These are typical values for a gravel aggregate in a concrete matrix. The dimensions of the numerical grid are 3480 cells and 3445 time steps (*$\Delta x = \Delta z =$ *14.367816 μm,* $\Delta t =$ *3.192848 ns). Stress-free boundary conditions were implemented at both lateral model boundaries. These are therefore ideally reflective. The left boundary of the model serves not only as an actuator but also as a sensor for the reflected waves, while the right boundary acts only as a sensor for the transmitted components. The small black arrows indicate the current direction of motion of the individual waveforms. In this model, the transmitted wave cannot carry any information about the position of the layer, as it is independent of it.*

The model describes the propagation and interaction of a plane longitudinal wave with a layer of finite thickness (the "scatterer") embedded in a homogeneous matrix medium. The material properties of the layer (e.g., density and speed of sound) differ only slightly from those of the matrix, so that in addition to weak primary reflections from both layer boundaries (in the rear direction), a strong transmitted wave also occurs in the forward direction. All further secondarily scattered waves that arise inside the layer are so small that they can be neglected here from a practical point of view. The reflection coefficient with respect to the particle velocity $v$ at both layer boundaries is $R_v = (Z_1 - Z_2)/(Z_1 + Z_2) \approx \pm 0.1$, where $Z_1 = \rho_1 c_1$ and $Z_2 = \rho_2 c_2$ represent the specific acoustic impedances of the matrix and layer medium, respectively. The corresponding transmission coefficient is calculated as $T_v = 2Z_1/(Z_1 + Z_2)$ and yields values of $T_v = 0.9$ and $T_v = 1.1$, respectively.

If, in this model, we are interested in the depth (or position) of the layer, i.e., the distance between the excitation probe (left boundary of the model) and the left or right layer boundary as a structural parameter, it becomes clear that only the two reflected waves from the left and right layer boundaries can carry information about the depth, not the transmitted wave. As long as the second receiving probe at the right boundary of the model is sufficiently far from the layer (to avoid reflections between the two), the detected transmitted wave field in this one-dimensional three-layer case remains independent of the layer's position and therefore cannot carry any information about it.

This changes, however, if we shift the second (right) layer boundary to the right edge of the model (Figure 3). In this two-layer case, the depth (or position) of the layer (more precisely: of the remaining left layer boundary) can be determined from the effective travel time of the transmitted wave, provided we assume the sound velocities of the matrix and the layer are known. In this case, the information about the layer position must also be encoded in the transmitted wave and must have already been generated at the (only existing) left layer boundary, since the influence of the sound velocity of the layer on the final travel time starts there. Since the left layer boundary is also present in the three-layer case, it follows that the position information generated there must be lost again in the forward-propagating wave in the three-layer system but must be preserved in the two-layer case.

The above considerations lead us to the final requirements for our information measure for now:

6) For individual waves, such as the forward wave in Figure 2 and Figure 3, there must be local sinks of information (e.g., the right layer boundary) in addition to information sources (here, the left layer boundary), where previously generated information is either annihilated or at least removed from this wave.
7) Point 6, together with requirement 4, suggests that, unlike (elastic) energy, information is not a globally conserved physical quantity of the system under consideration, but can apparently only be conserved between two interactions. As soon as another interaction occurs, the information can be lost and/or new information can be generated.

An information measure suitable for ultrasonic NDE should possess at least the seven properties formulated in this section in order to be practically meaningful and consistent in its application. In the next section, we will present a promising candidate and examine the extent to which it fulfills requirements 1-7.

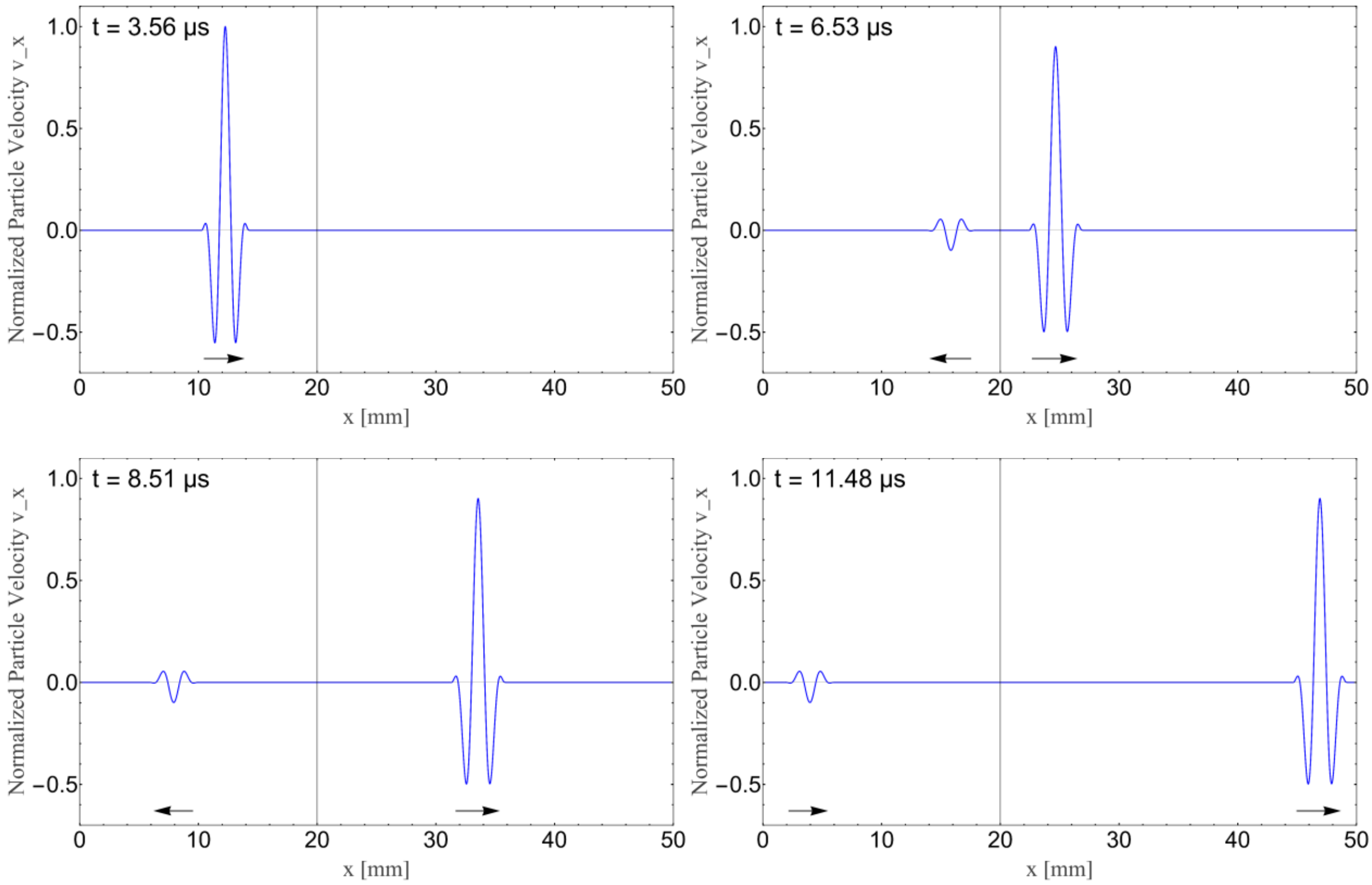

***Figure 3:*** *A one-dimensional simulation, performed using elastodynamic finite integration technique (EFIT), of the interaction between a plane longitudinal wave excited at the left model boundary ($x =$ 0 mm) and an elastic layer, 30 mm thick, between $x =$ 20 mm and $x =$ 50 mm, embedded in a homogeneous, isotropic matrix medium (two-layer structure). Four time snapshots of the normalized particle velocity are shown: 3.56 µs (top left), 6.53 µs (top right), 8.51 µs (bottom left), and 11.48 µs (bottom right) after the start of excitation. All other model parameters correspond to those of Figure 2. The transmitted wave carries time-of-flight-coded information about the position of the layer (more precisely: the position of the remaining layer boundary at $x =$ 20 mm).*

# 3. System Changes and Difference Wave Fields

We first address requirement 1 from Section 2 and consider how we can separately investigate the effects of the various structural system parameters on the scattered wave field. In practice, a single simulation is usually insufficient; instead, several calculations with varying parameters are performed to gain a better understanding of the effects of each quantity. We now formalize this approach and perform a small but finite absolute variation or change $\Delta S_P$ of the system $S$ with respect to the parameter $P$ of interest.

## 3.1 Change in Layer Position at Constant Layer Thickness

We return to the one-dimensional initial model with three layers from Figure 2 and are interested in the depth (or position) of the layer, which we will denote as $D$. A small change $\Delta S_D$ is now made to the setup shown in Figure 2. In the discretized model, it is convenient to simultaneously shift the left and right layer boundaries (and thus also the layer center) by one grid cell each (here by $\Delta x = 14.367816$ µs, which corresponds to approximately 1/40 of the smallest wavelength) to the right and left, respectively, without changing the layer thickness. This is shown schematically in Figure 4.

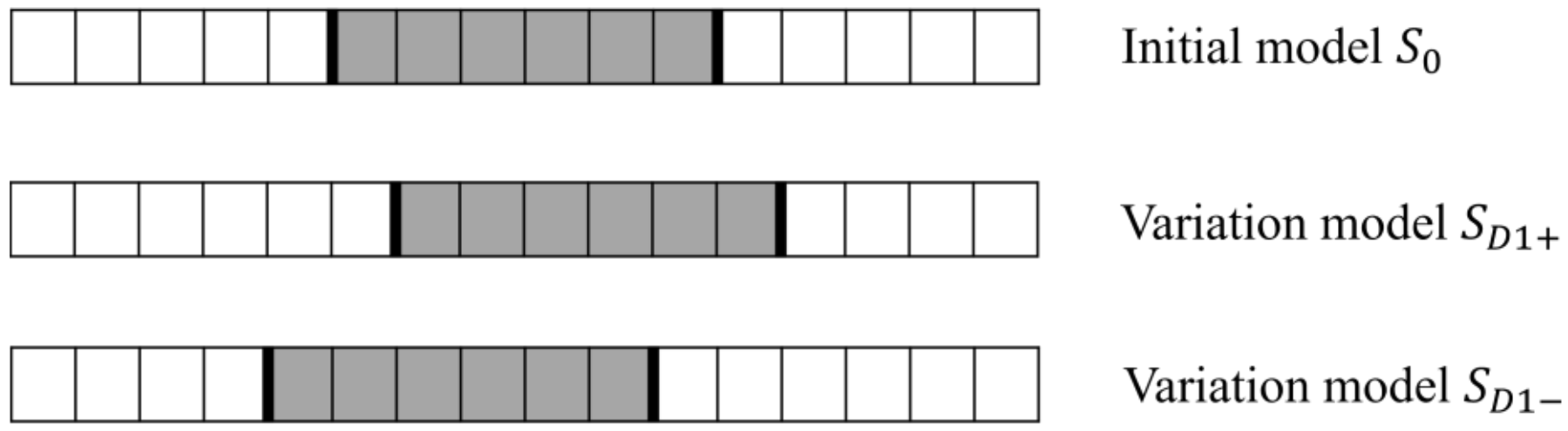

***Figure 4:*** *Discrete initial model of the layer (top) and the derived variation models regarding the layer position (middle and bottom) at constant layer thickness.*

For reasons of symmetry with the original model, we work with both variations and denote them as $\Delta S_{D1+}$ and $\Delta S_{D1-}$. We thus shift the two layer boundaries of the original model by one grid cell to the right and left, respectively, obtaining two new model systems, $S_{D1+}$ and $S_{D1-}$, whose results will differ only slightly from the original model $S_0$, since the EFIT grid cells are significantly smaller than the wavelengths predominant in the broadband signal. Nevertheless, there are differences, which can be most easily represented by the difference wave field, $W(S_{D1+} - S_{D1-})$. In Figure 5, this difference wave field with respect to the layer position (middle column) is compared with the wave field of the original model, $W(S_0)$ (left column). In the latter, the maximum amplitude of the excitation signal is normalized to one. In this example, the amplitudes of the difference waves are all less than 2%.

The first three images up to $t = 6.53$ µs show the moment shortly before, during, and after the interaction with the left layer boundary. The difference wave field with respect to position is initially zero ($t = 4.16$ µs) and is only generated at the moment of interaction at the left layer boundary ($t = 5.35$ µs). After the interaction, a difference wave propagates in both the backward and forward directions, with the forward wave exhibiting a significantly smaller amplitude and a different phase in the direction of propagation within the layer ($t = 6.53$ µs). The waveform of the difference field differs significantly from that of the particle velocity in the initial model and resembles its spatial derivative.

The following three images, from $t = 7.52$ µs to $t = 12.08$ µs, show the interaction with the right layer boundary. The newly generated difference wave field is radiated exclusively in the backward direction, not in the forward direction ($t = 7.52$ µs and $t = 8.91$ µs). The difference wave initially present, originating from the left layer boundary, therefore disappears in the forward direction. The new, backward-directed difference wave exhibits a higher amplitude within the layer than before the interaction with the right layer boundary (cf. $t = 6.53$ µs and $t = 8.91$ µs). After passing through the layer, both the difference wave from the left and the one from the right layer boundary again exhibit comparable amplitudes in the matrix medium ($t = 12.08$ µs). However, both waves have opposite phase relationships in the direction of propagation, apparently as a result of wave reflection at the acoustically soft (right layer boundary) and acoustically hard (left layer boundary) transition. It is also worth noting that the difference wave originating at the left layer boundary and propagating in the reverse direction is reflected at the left stress-free model boundary without a phase shift, analogous to the particle velocity in the original model.

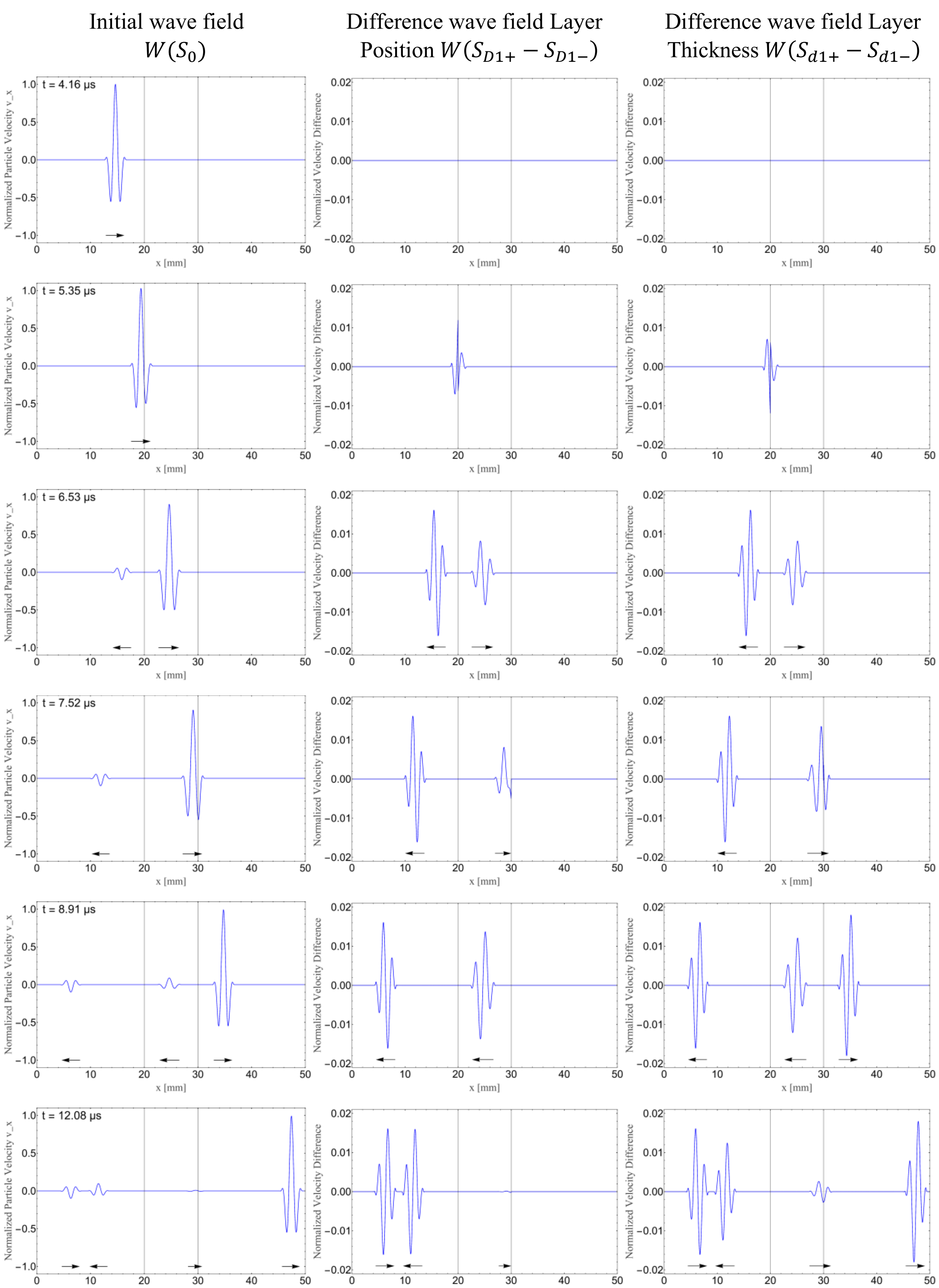

***Figure 5:*** *Representation of the wave field in the initial model,* $W(S_0)$ *(left column) as well as the difference wave field with respect to the layer position,* $W(S_{D1+} - S_{D1-})$ *(middle column) and with respect to the layer thickness,* $W(S_{d1+} - S_{d1-})$ *(right column) at the times* $t =$ *4.16 µs (1st row),* $t =$ *5.35 µs (2nd row),* $t =$ *6.53 µs (3rd row),* $t =$ *7.52 µs (4th row),* $t =$ *8.91 µs (5th row) and* $t =$ *12.08 µs (last row).*

In conclusion, it can be summarized that the difference wave field shown in Figure 5 (middle column), resulting from the change in layer position, fulfills the requirements 2-7 for an information measure (see Section 2) derived from general considerations remarkably well. To check whether this also applies to requirement 1, we will now consider the layer thickness as a further structural parameter of the model.

## 3.2 Change in Layer Thickness at Constant Layer Position

Analogous to the procedure in Section 3.1, we again modify the initial model from Figure 2, this time changing the thickness $d$ of the layer while leaving all other structural parameters unchanged. Among the latter is the depth $D$ (or position) of the layer, which, for practical reasons, we equate here with the distance of the layer's center from the left boundary of the model. Thus, the pure thickness change under consideration results in a shift of both layer boundaries outwards and inwards, respectively (Figure 6). We denote the new thickness change of the system by $\Delta S_d$.

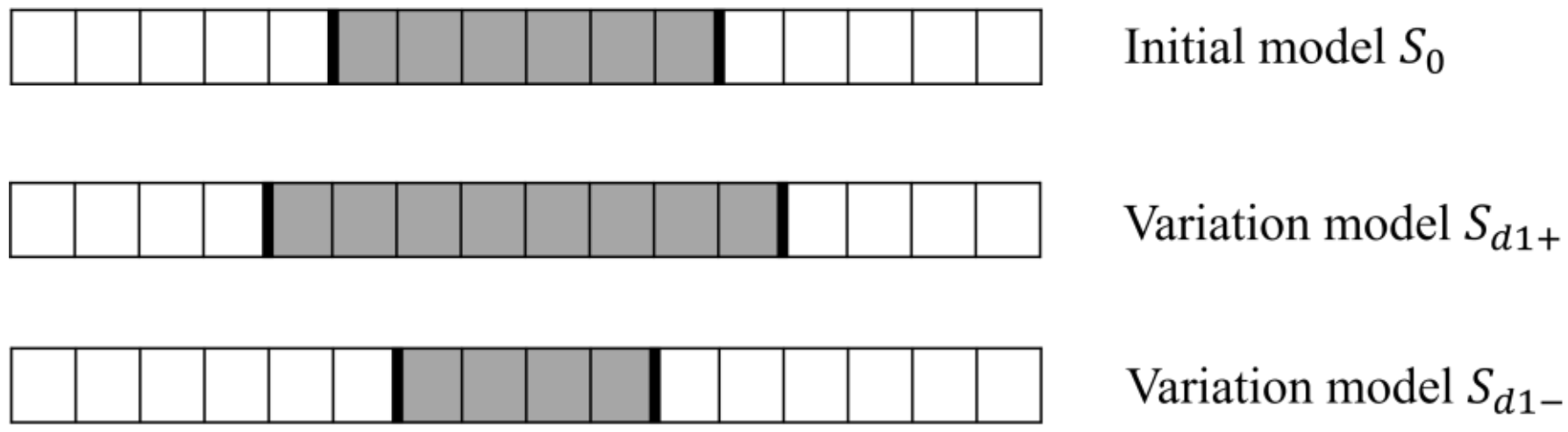

***Figure 6:*** *Discrete initial model of the layer (top) and the derived variation models regarding the layer thickness (middle and bottom) with constant layer position (≙ middle of the layer).*

If we increase the layer thickness by shifting the right layer boundary of the model by one grid cell to the right and the left layer boundary by one grid cell to the left, we obtain the change $\Delta S_{d1+}$. Conversely, if we decrease the layer thickness by shifting the right layer boundary to the left and the left layer boundary to the right, we obtain the change $\Delta S_{d1-}$. The resulting difference wave field, $W(S_{d1+} - S_{d1-})$, is shown in the right-hand column of Figure 5.

In the case of a layer thickness change, the difference wave field also arises first upon the interaction of the initial wave field with the left layer boundary (Figure 5, right column, $t = 5.35$ µs). Here, too, a difference wave is generated in both the reverse and forward directions ($t = 6.53$ µs), with the forward wave exhibiting a smaller amplitude within the layer. Compared to the layer position change (Figure 5, middle column), both difference waves here have a reversed phase relationship. This results from the fact that with the plus (+) variation of the layer position, the left layer boundary is shifted to the right, while with the layer thickness change, it is shifted to the left.

The greatest difference between the two difference wave fields in Figure 5 can be observed in the interaction with the right layer boundary ($t = 7.52$ µs and $t = 8.91$ µs). While only a backward-propagating difference wave is generated there when the layer position changes (middle column), a strong forward-propagating difference wave is also generated when the layer thickness changes. This corresponds to the considerations in Section 2 regarding Figure 3, namely that layer position information should only be evident in the reflected waves, while layer thickness information should also be evident in the transmitted wave. Thus, both difference wave fields from Figure 5 fulfill requirements 2-7 formulated in Section 2 for a suitable information measure and, considered together, also requirement 1, since they can be considered separately.

Assuming that the amplitudes of the difference waves in one and the same medium (here, the matrix medium) are directly correlated with the information content, initial qualitative assessments can be made for the two cases discussed so far. Accordingly, in the case of a change in layer position (Figure 5, middle column, $t = 12.08$ µs), the corresponding information is contained equally in the reflection from both the left and right layer boundaries (equal amplitudes). In the case of a change in layer thickness (Figure 5, right column, $t = 12.08$ µs), however, most of the information appears to be contained in the transmitted wave (largest amplitude), followed by the reflection at the left layer boundary, the reflection at the right layer boundary, and a first multiple reflection from within the layer, which can be observed at $x = 30$ mm and exhibits the smallest amplitude of all difference waves.

### 3.3 Change in the Speed of Sound in the Layer at Constant Density

In a further step, we now keep the geometric layer parameters position and thickness constant and instead vary the material properties of the layer, namely the longitudinal speed of sound $c_P$ and the density $\rho$. In the numerical EFIT model, this is achieved by first changing the speed of sound in the material cells of the layer by 1 m/s up and down, while density remains unchanged. The corresponding change $\Delta S_c$ is shown schematically in Figure 7.

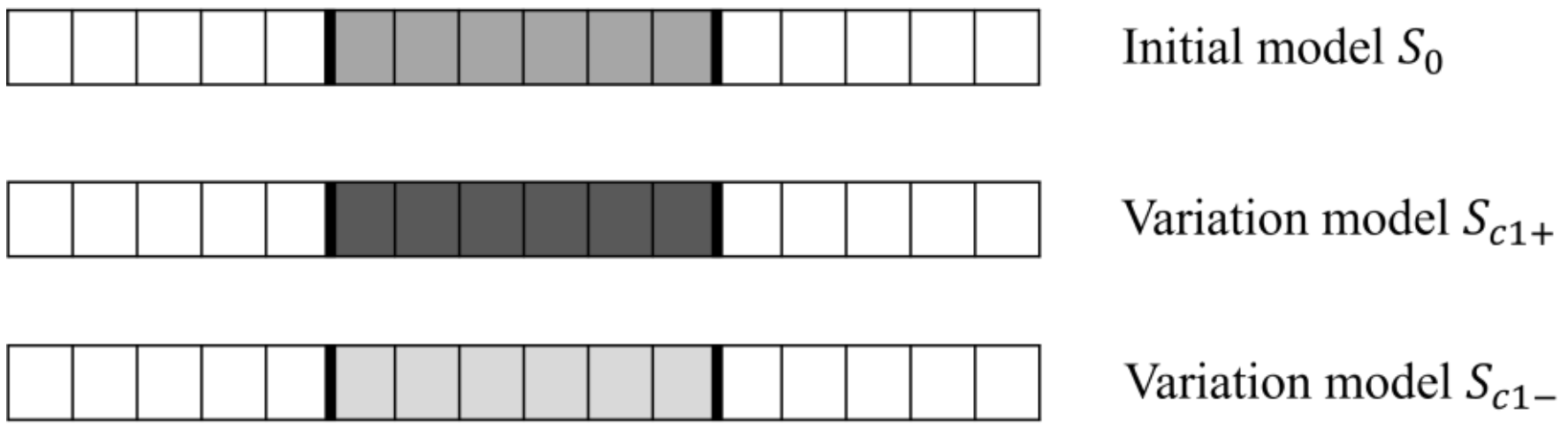

***Figure 7:*** *Discrete initial model of the layer (top) and the derived variation models with respect to the speed of sound in the layer (middle and bottom). The darker gray level in the middle indicates an increased speed of sound compared to the initial model, while the lighter gray level at the bottom indicates a reduced speed of sound. Changes in layer density can also be simulated analogously.*

Since the maximum speed of sound in the model influences the time step and thus also the stability of the numerical model, the initial speed of sound was slightly reduced to $c_P =$ 4499 m/s in all three sub-calculations to maintain a uniform grid, while the time step was set for a value of 4500 m/s. The change $\Delta S_c$ thus leads to speeds of sound of 4498 m/s and 4500 m/s, respectively, and therefore does not violate the stability criterion of the 1D-EFIT code, $\Delta t \leq \Delta x / c_P$, even in the plus (+) variation. This results in the difference wave field $W(S_{c1+} - S_{c1-})$, which is shown in the middle column of Figure 8. It exhibits maximum amplitudes of approximately 1% of the excited signal in the initial model.

At the first layer boundary, only a very small difference wave is generated in the backward direction ($t = 6.53$ µs). The difference wave in the forward direction is also initially rather small ($t = 5.35$ µs) but then becomes increasingly larger as it propagates through the layer ($t =$ 6.53 and $t = 7.52$ µs). This is due to the ever-increasing time delay between the two variation waves with increasing distance traveled through the layer. Only at the right boundary of the layer does this amplification process cease, and the difference wave continues to propagate with a constant amplitude in the matrix medium ($t = 8.91$ and $t = 12.02$ µs).

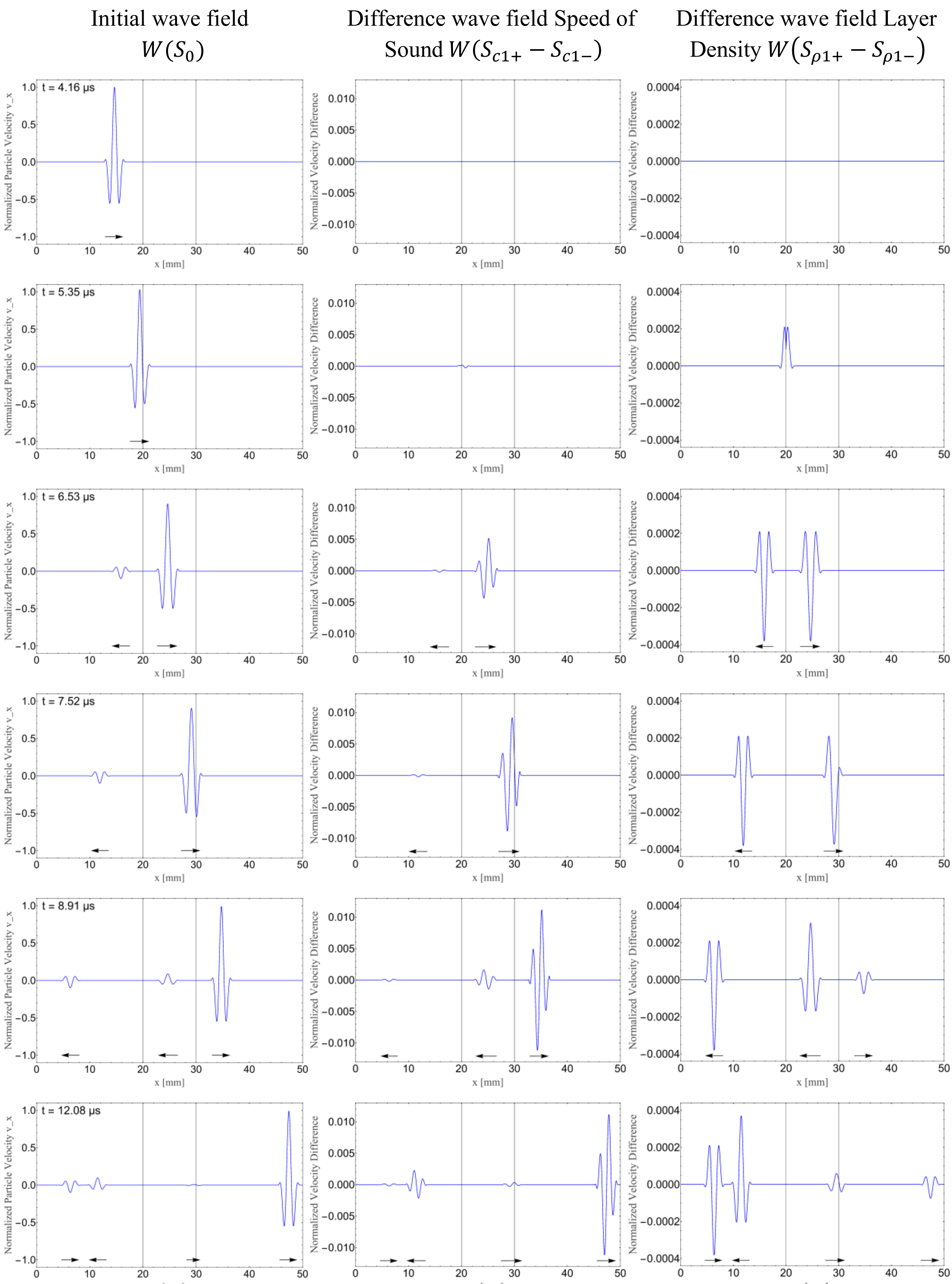

***Figure 8:*** *Representation of the wave field in the initial model,* $W(S_0)$ *(left column) as well as the difference wave field with respect to the speed of sound in the layer,* $W(S_{c1+} - S_{c1-})$ *(middle column) and with respect to the density of the layer,* $W(S_{\rho1+} - S_{\rho1-})$ *(right column) at the times* $t$ *= 4.16 µs (1st row),* $t$ *= 5.35 µs (2nd row),* $t$ *= 6.53 µs (3rd row),* $t$ *= 7.52 µs (4th row),* $t$ *= 8.91 µs (5th row) and* $t$ *= 12.08 µs (last row).*

From the right layer boundary, a strong difference wave is emitted in the forward direction and a smaller but significant difference wave in the reverse direction ($t = 8.91$ µs). These two waves also dominate the final difference wave field at $t = 12.08$ µs. The waveforms of the difference waves again resemble the spatial derivative of the excitation signal, analogous to the case of geometric parameter changes in Figure 5.

### 3.4 Change in the Density of the Layer at Constant Speed of Sound

Analogous to the procedure shown in Figure 7, the density of the layer is varied by 1 kg/m³ up and down while keeping the speed of sound constant, resulting in maximum changes of approximately 0.4% (right column of Figure 8).

During the density change, difference waves are generated at both layer boundaries. At the left layer boundary, two equally sized and in-phase difference waves are generated in the forward and reverse directions ($t = 6.53$ µs). At the right layer boundary, however, a large out-of-phase reverse wave and a smaller forward wave remain ($t = 8.91$ µs). It is striking in this example that inside the layer, the reflected difference wave from the right layer boundary ($t = 8.91$ µs) has a *smaller* amplitude than the originally incoming difference wave from the left layer boundary ($t = 6.53$ µs). This refers to a potential *sink* of information at the right layer boundary.

Ultimately, the two out-of-phase difference waves reflected at the layer boundaries exhibit the largest amplitudes, followed by the transmitted wave and a first multiple reflection from the layer at $x = 30$ mm ($t = 12.08$ µs). In contrast to the geometric layer changes from Sections 3.1 and 3.2, the difference waves here have the same waveform as the initial wave field.

### 3.5 Interim Conclusion

The investigations in this section have shown that the difference wave fields resulting from the parameter changes to the initial model fulfill all requirements 1-7 from Section 2 for the information measure to be determined.

Each structural parameter of the system can be considered separately (requirement 1) and each leads to entirely individual and clearly distinguishable difference wave fields, all of which result from one and the same initial wave field (see the last partial images of the difference wave fields in Figure 5 and Figure 8 for $t = 12.08$ µs). Difference wave fields are therefore obviously better suited for classification tasks than absolute wave fields, which differ only minimally from the initial wave field with each variation, a difference that is practically undetectable to the naked eye (and also to AI systems).

Since the considered difference formation of the variational models refers to the total wave field, which is composed of the incident and scattered fields, the two identical excitation fields are cancelled out in the difference calculation. This corresponds to requirements 2 and 3 from Section 2, namely that the excitation field carries no information before its interaction with the scatterer and that the source of the excitation wave (usually the transducer) cannot, in general, also be the source of information. The difference wave fields thus arise from the difference between the pure scattered waves of the two variations and are therefore only generated during the interaction of the incident wave field with the scatterer(s) (requirement 4). Not only can the scattering interfaces between two different materials act as sources of information, but also the volume of the scatterer itself, as in the considered case of the change in sound velocity from Figure 8 (right column). This, too, is plausible and obvious.

We further observed that all occurring difference waves are always coupled to a wave of the original model and therefore propagate at its speed (requirement 5). However, conversely, not every original wave is coupled to a difference wave. For instance, during the change in position shown in Figure 5, the difference wave originating at the left layer boundary and propagating forward is suppressed at the right layer boundary, so that no information is transferred into the transmitted wave. For such individual waves (here the forward wave), there can therefore apparently be effective sinks of information in addition to sources (requirement 6). However, since in this example the interaction with the right layer boundary results in *more* information overall than originally came from the left layer boundary (the reflected difference wave in the layer is larger than the incoming difference wave in the forward direction), it remains initially unclear whether there can also be *global* sinks of information where information is actually annihilated. The examples considered so far in the homogeneous and lossless model do not provide direct evidence for this. We will come back later to this point.

However, the examples clearly show that difference waves can, in principle, arise during any scattering process (e.g., during the thickness change in Figure 5 at both layer boundaries). This demonstrates that, unlike (elastic) energy, information is generally *not* a globally conserved quantity of the system under consideration but can apparently only be conserved between two interactions. Thus, difference wave fields also fulfill requirement 7 for the information measure to be found.

In the following section, we supplement and formalize our identified information measure and derive a specific balance equation for it.

# 4. Formalization of Information Concept and Balance Equations in 1-D

Despite the obvious suitability of difference wave fields as a measure of information, four formal problems still need to be resolved. These can be expressed as further requirements 8-11, thus supplementing the listing 1-7 from Section 2.

8) The information must be preserved even in the limiting case of vanishing changes $\Delta P$; otherwise, no information could be assigned to the initial wave field ($\Delta P \rightarrow 0$).
9) The information should not be described by a signed quantity (such as particle velocity) that can be positive or negative, but rather, like energy, should only have positive values.
10) In a linear, attenuation- and noise-free system, the information must not depend on the amplitude of the excitation signal but must be preserved under simple scaling of the excitation.
11) The information should be physically dimensionless and, as information density relative to the volume $V$, have the physical dimension $1/V$.

Our difference wave fields do not currently fulfill these new requirements. Let us first turn to requirement 8.

## 4.1 Information as Differential Wave Field with Respect to $P$

The difference wave fields considered so far are based on small, but finite, absolute parameter changes $\Delta P$. The smaller these are, the smaller the difference wave fields are as well. In the limiting case $\Delta P \rightarrow 0$, the difference wave field therefore also vanishes. However, this would mean that no information $I_P$ can be assigned to the initial wave field $W(S_0)$, the reference point of our variations. To avoid this, the difference wave field must evidently be related to the parameter change itself, i.e.,

$$I_P \sim \frac{W(S_{P+}) - W(S_{P-})}{\Delta P} \quad . \tag{1}$$

In this case, with a smaller change $\Delta P$, the difference wave field in the numerator also becomes smaller, so the entire quotient, which is obviously proportional to the information $I_P$, does not automatically vanish. We verify this by repeating the simulation for the layer position variation from Figure 5 (middle column) with the additional variations $\Delta S_{D2+}$ and $\Delta S_{D2-}$ as well as $\Delta S_{D3+}$ and $\Delta S_{D3-}$ . We thus shift the two layer boundaries not only by one, but also by two and three grid cells in both directions and generate the corresponding difference wave fields. These are shown in Figure 9 (top row) for $t = 6.53$ µs on a uniform scale. As expected, the amplitude of the difference field increases with increasing change.

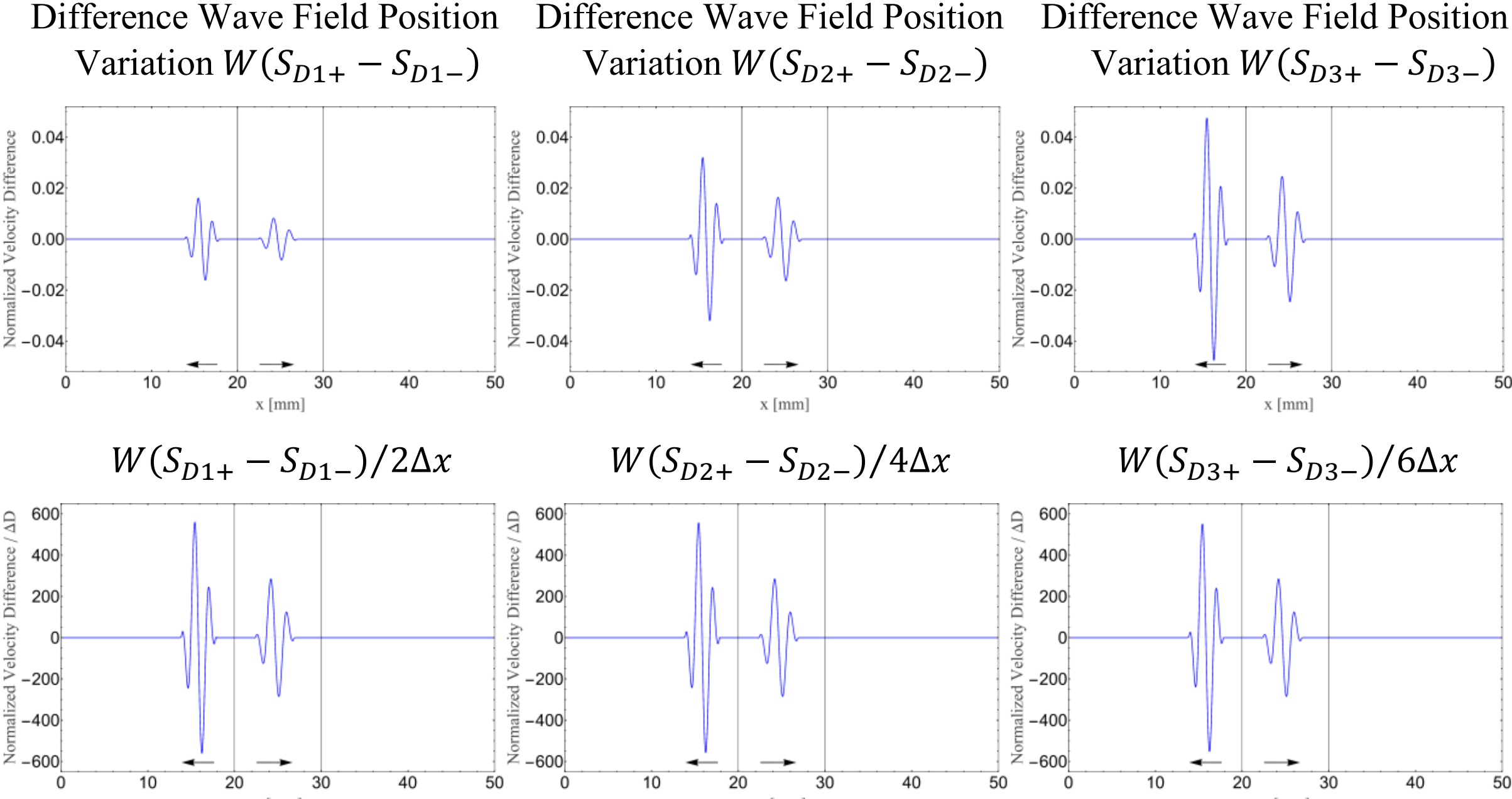

***Figure 9:*** *Representation of the difference wave fields with respect to the layer position, $W(S_{D1+} - S_{D1-})$ (left column), $W(S_{D2+} - S_{D2-})$ (middle column) and $W(S_{D3+} - S_{D3-})$ (right column) at time $t = 6.53$ µs, once uncorrected (top row) and once divided by the respective change $\Delta D$ (bottom row).*

In contrast, in the bottom row of Figure 9, all difference fields were divided by their respective changes, in this case by two ($\pm\Delta$x), four ($\pm 2\Delta$x),, and six ($\pm 3\Delta$x), times the grid spacing $\Delta$x. It is clearly evident that the quotient from Eq. (1) remains approximately the same in all three cases.

These findings suggest that the quotient also remains constant in the limiting case $\Delta P \rightarrow 0$, and that our information measure for the initial wave field $W(S_0)$ can be expressed by the change in the field quantity $W$ with parameter $P$, i.e., as a partial derivative with respect to $P$:

$$I_P \sim \frac{\partial W(S_0)}{\partial P} \tag{2}$$

Since we have used the particle velocity $v_x$ as the field quantity so far, we obtain the following expression as a preliminary measure for the information relevant in our previous examples:

$$I_P \sim \frac{\partial v_x}{\partial P} \tag{3}$$

With this extension, the new requirement 8 is fulfilled. However, the missing equals sign already indicates that expression (3) still cannot be identical to physical information. The change in particle velocity with parameter $P$ is generally signed, whereas one would expect only a single (positive) sign from physical information (requirement 9). This is reminiscent of the interplay between (signed) particle velocity and the associated (always positive) kinetic energy, which essentially results from squaring the particle velocity. As is well known, a balance equation can be formulated for kinetic energy, which, by including potential energy, can even be further developed into a continuity or conservation equation for total energy. This is expressed by the Poynting theorem (e.g., [3]). In the following two sections, we want to investigate whether an analogous balance equation, and perhaps even a continuity equation, can be derived for information. To do this, we first recap the procedure for kinetic energy for plane longitudinal waves in one dimension.

### 4.2 Balance Equation of Kinetic Energy and Poynting-Theorem in 1-D

In the following, we denote the particle velocity $v_x$ simply as $v$ and the stress $T_{xx}$ as $T$. Both are functions of position and time, i.e., $v = v(x,t)$ and $T = T(x,t)$. The density is considered here only as a time-invariant position function, i.e., $\rho = \rho(x)$. We begin our analysis with the linearized equation of motion in an absorption-free linear-elastic medium,

$$\rho \frac{\partial v}{\partial t} = \frac{\partial T}{\partial x} + f \quad . \tag{4}$$

Here, $f = f(x,t)$ represents an exciting volume force density acting on individual volume elements $\Delta V$, with the dimension "force per volume". The dependence of Eq. (4) on the elastic properties is also implicitly contained in the expression for the stress $T$, which, for example, in an isotropic medium takes the form $T = (\lambda + 2\mu)\, \partial u / \partial x = (\lambda + 2\mu)\, \varepsilon$. Here, $\lambda = \lambda(x)$ and $\mu = \mu(x)$ are the two Lamé constants, which are related to the speed of sound $c_P = c_P(x)$ of the longitudinal wave via the expression $c_P^2 = (\lambda + 2\mu)/\rho$. $u = u(x,t)$ denotes the displacement and $\varepsilon = \varepsilon(x,t)$ the deformation.

A rearrangement of Eq. (4) taking advantage of the temporal constancy of $\rho$ leads to

$$\frac{\partial}{\partial t}(\rho v) - \frac{\partial}{\partial x} T = f \quad . \tag{5}$$

This is a typical balance equation (here, momentum density balance) that links the time change of a spatial density function (here, the momentum $\rho v$ related to the volume element) with the spatial change of an associated flux or transport quantity (here, the stress $T$). The latter describes the forces acting on the surfaces of the volume element (more precisely: the flux of these forces through the surfaces).

The volume force density $f$ on the right-hand side of equation (5) represents an external source term, such as the external excitation by an ultrasonic probe on the boundary of the model or the excitation of acoustic emissions inside (which are generally also triggered by *external* forces). If the external excitation $f$ ends after a certain time, the balance equation (5) becomes a continuity equation:

$$\frac{\partial}{\partial t}(\rho v) - \frac{\partial}{\partial x}T = 0 \tag{6}$$

This is linked to a physical conserved quantity, which can be taken from the parentheses of the partial time derivative, here $\rho v$. Eq. (6) thus represents the differential form of the momentum conservation equation.

To arrive at the energy balance equation from Eq. (5) and Eq. (4), respectively, we multiply both sides of Eq. (6) by the particle velocity $v$ and obtain

$$\rho v \frac{\partial v}{\partial t} - v \frac{\partial T}{\partial x} = vf \tag{7}$$

and

$$\frac{\partial}{\partial t}\left(\frac{\rho}{2} v^2\right) - v \frac{\partial T}{\partial x} = vf \quad , \tag{8}$$

respectively. For the second term on the left side of the equation, we obtain using the product rule

$$v \frac{\partial T}{\partial x} = \frac{\partial}{\partial x}(v\, T) - T \frac{\partial v}{\partial x} \tag{9}$$

and thus, for the balance equation of kinetic energy density

$$\frac{\partial}{\partial t}\left(\frac{\rho}{2} v^2\right) + \frac{\partial}{\partial x}(-v\, T) = vf - T \frac{\partial v}{\partial x} \ . \tag{10}$$

The first expression in parentheses on the left-hand side denotes the quantity to be balanced, in this case the kinetic energy density $(\rho/2)\, v^2$ with the dimension "energy per volume". The second expression in parentheses, $-v\, T$, denotes the (here one-dimensional) Poynting vector, i.e., the energy flux through the surfaces of the volume element with the dimension "energy per area and time". The energy flux vector includes both kinetic ($v$) and potential ($T$) components.

The two terms on the right-hand side of Eq. (10) can be interpreted as source terms. The first term $vf$ denotes the power density of the external volume forces, the second term $-T(\partial v/\partial x)$ the power density of the internal surface forces associated with the stress tensor, both with the dimension "energy per volume and time". Even if the external volume force density $f$ vanishes, the second source term therefore remains, so that Eq. (10) does not become a continuity equation like Eq. (6). Thus, kinetic energy alone is not yet a conserved quantity. However, in elasticity theory it is shown (e.g. [3]) that the second source term can be expressed as the temporal change of the potential energy density $\Phi_{pot}$, i.e.,

$$\frac{\partial \Phi_{pot}}{\partial t} = T \frac{\partial v}{\partial x} = T\dot{\varepsilon} = (\lambda + 2\mu)\varepsilon\dot{\varepsilon} = \frac{1}{\lambda + 2\mu} T\dot{T} \ . \tag{11}$$

Thus, $\Phi_{pot}$ can be transformed into the left expression in parenthesis of Eq. (10) and we obtain

$$\frac{\partial}{\partial t}\left(\frac{\rho}{2} v^2 + \Phi_{pot}\right) + \frac{\partial}{\partial x}(-v\, T) = vf \tag{12}$$

as the balance equation for the total energy density, consisting of kinetic and potential components. When the external forces $f$ are removed, the continuity equation

$$\frac{\partial}{\partial t}\left(\frac{\rho}{2}v^2 + \Phi_{pot}\right) + \frac{\partial}{\partial x}(-v\,T) = 0 \tag{13}$$

results as a differential form of the energy conservation equation. Here too, the Poynting vector $-v\,T$ serves as the corresponding flux quantity.

With the displacement $u$ and the deformation $\varepsilon = \partial u/\partial x$ and $\dot{\varepsilon} = \partial v/\partial x$, respectively, an expression for the potential energy density $\Phi_{pot}$ can be directly derived from Eq. (11). For this, we consider the relationship $T = (\lambda + 2\mu)\,\varepsilon$ and obtain as the time antiderivative of Eq. (11)

$$\Phi_{pot} = \frac{1}{2}(\lambda + 2\mu)\varepsilon^2 = \frac{1}{2}(\lambda + 2\mu)\left(\frac{\partial u}{\partial x}\right)^2 = \frac{1}{2(\lambda + 2\mu)}T^2 \ . \tag{14}$$

The transition from the balance equation (10) of kinetic energy to the conservation equation (13) of total energy by eliminating the source term $-T(\partial v/\partial x)$ can be readily understood physically by considering that in an elastic wave field, there is a continuous conversion of kinetic energy into potential energy and vice versa, with the total energy remaining constant at all times. Accordingly, the time-dependent change of potential energy from Eq. (11) acts as a (positive or negative) source term for the kinetic energy, and conversely, the time-dependent change of kinetic energy acts as a source term for the potential energy. Indeed, in Eq. (12), the time-dependent change of kinetic energy can also be shifted to the right-hand side of the equation as the source term $\rho v\dot{v}$, thus obtaining a balance equation for potential energy analogous to Eq. (10):

$$\frac{\partial}{\partial t}\left(\frac{1}{2(\lambda + 2\mu)}T^2\right) + \frac{\partial}{\partial x}(-v\,T) = vf - \rho v\dot{v} \tag{15}$$

### 4.3 Balance Equation of Kinetic Information for the Case of Geometric System Parameters

In the examples in Section 3, we considered the difference or differential wave field $\partial v/\partial P$ as a potential candidate for the information measure $I_P$ to be found. In this section, we now want to clarify whether we can also derive a balance equation for this, analogous to the kinetic energy density from Eq. (10). We first define $\dot{v} = \partial v/\partial t$ and, after rearranging Eq. (4), obtain

$$\dot{v} - \frac{1}{\rho}\frac{\partial T}{\partial x} = \frac{f}{\rho} \ . \tag{16}$$

Let $P$ be a scalar, real, continuously and steadily variable system parameter that represents either purely geometric properties, such as position, thickness, or orientation, or material properties, such as density and speed of sound of certain system components. The partial derivative of both sides of Eq. (16) with respect to $P$ yields

$$\frac{\partial \dot{v}}{\partial P} - \frac{\partial}{\partial P}\left(\frac{1}{\rho}\frac{\partial T}{\partial x}\right) = \frac{\partial}{\partial P}\left(\frac{f}{\rho}\right) \ . \tag{17}$$

After multiplying both sides by $\rho\,\partial v/\partial P$, we obtain

$$\rho\frac{\partial v}{\partial P}\frac{\partial \dot{v}}{\partial P} - \rho\frac{\partial v}{\partial P}\frac{\partial}{\partial P}\left(\frac{1}{\rho}\frac{\partial T}{\partial x}\right) = \rho\frac{\partial v}{\partial P}\frac{\partial}{\partial P}\left(\frac{f}{\rho}\right) \tag{18}$$

and then

$$\frac{\partial}{\partial t}\left[\frac{\rho}{2}\left(\frac{\partial v}{\partial P}\right)^2\right] - \rho\frac{\partial v}{\partial P}\left[\frac{\partial}{\partial P}\left(\frac{1}{\rho}\right)\frac{\partial T}{\partial x} + \frac{1}{\rho}\frac{\partial}{\partial P}\left(\frac{\partial T}{\partial x}\right)\right] = \rho\frac{\partial v}{\partial P}\frac{\partial}{\partial P}\left(\frac{f}{\rho}\right) . \tag{19}$$

We denote the source term on the right-hand side of equation (19) by $Q_1$ and obtain with $\partial\rho^{-1}/\partial P = -\rho^{-2}\,\partial\rho/\partial P$:

$$\frac{\partial}{\partial t}\left[\frac{\rho}{2}\left(\frac{\partial v}{\partial P}\right)^2\right] + \frac{1}{\rho}\frac{\partial\rho}{\partial P}\frac{\partial v}{\partial P}\frac{\partial T}{\partial x} - \frac{\partial v}{\partial P}\frac{\partial}{\partial P}\left(\frac{\partial T}{\partial x}\right) = Q_1 . \tag{20}$$

Using the product rule, assuming the continuity of $T(x)$ and $T(P)$ (which allows the differentiation order to be interchanged[1]), and including another source term, we replace the last term on the left-hand side of Eq. (20) by

$$\frac{\partial v}{\partial P}\frac{\partial}{\partial P}\left(\frac{\partial T}{\partial x}\right) = \frac{\partial v}{\partial P}\frac{\partial}{\partial x}\left(\frac{\partial T}{\partial P}\right) = \frac{\partial}{\partial x}\left(\frac{\partial v}{\partial P}\frac{\partial T}{\partial P}\right) - Q_2 \text{ with } Q_2 = \frac{\partial}{\partial x}\left(\frac{\partial v}{\partial P}\right)\frac{\partial T}{\partial P} . \tag{21}$$

At first glance, the middle term on the left-hand side of Eq. (20) appears to be replaceable in a similar way by applying the product rule. However, this introduces expressions of the form $(\partial/\partial x)\,(\partial\rho(x)/\partial P)$, which must be treated with great mathematical care. For material-related system parameters (e.g., the density of a scatterer), $\partial\rho(x)/\partial P$ remains always differentiable with respect to $x$. For geometric system parameters, where the density change is caused by the displacement of an interface between two materials (e.g., between matrix and scatterer), $\partial\rho(x)/\partial P$ depends on the density of the two media involved and is only non-zero at the location of the interface itself. The expression $\partial\rho(x)/\partial P$ thus becomes discontinuous at the interface and is no longer differentiable with respect to $x$. Therefore, the substitution described above cannot be performed universally, and a distinction between geometric and material-related system parameters becomes necessary. For the former, the middle term in Eq. (20) must be moved to the right-hand side of the balance equation as a source term. Together with Eq. (21), this results in the balance equation for the kinetic energy density of the differential wave field, $(\rho/2)\,(\partial v/\partial P)^2$, for the case of geometric system parameters:

$$\frac{\partial}{\partial t}\left[\frac{\rho}{2}\left(\frac{\partial v}{\partial P}\right)^2\right] + \frac{\partial}{\partial x}\left(-\frac{\partial v}{\partial P}\frac{\partial T}{\partial P}\right) = Q_1 - Q_2 - \tilde{Q}_3 , \tag{22}$$

with

$$Q_2 = \frac{\partial}{\partial x}\left(\frac{\partial v}{\partial P}\right)\frac{\partial T}{\partial P} \text{ and } \tilde{Q}_3 = \frac{1}{\rho}\frac{\partial\rho}{\partial P}\frac{\partial v}{\partial P}\frac{\partial T}{\partial x} . \tag{23}$$

Compared to the expression for kinetic energy (10), the particle velocity $v$ in Eq. (22) has been replaced by its derivative with respect to $P$, ∂v/∂P. The flux quantity is the information Poynting vector, $-(\partial v/\partial P)\,(\partial T/\partial P)$. In this expression, $v$ has been replaced by $\partial v/\partial P$ and $T$ by $\partial T/\partial P$, compared to the energetic Poynting vector in Eq. (10).

[1] The continuity of the physical field quantity $T$ with respect to position $x$ is generally always given, as is the continuity with respect to $P$ in all the examples shown in Section 3. However, if, for example, the system parameter $P$ is defined as the presence or absence of a material boundary, $T(P)$ can also become discontinuous. Such special cases are excluded from this discussion.

## 4.4 Balance Equation of Total Information for the Case of Geometric System Parameters

The symmetry of the two balance equations (10) und (15) for the kinetic and potential energy densities, respectively, suggests that, analogous to balance equation (22) for the kinetic part of the information field, a balance for the potential part of information can also be derived. In this case, an expression of the form $(1/2)(\lambda+2\mu)^{-1}(\partial T/\partial P)^2$ would be expected on the left-hand side of the balance equation in the expression in parentheses, which is partially differentiated with respect to time. Such an expression can be realized if the time derivative of this expression, i.e., $(\lambda+2\mu)^{-1}(\partial T/\partial P)(\partial\dot{T}/\partial P)$, can be identified in the source terms on the right-hand side of Eq. (22). This term can indeed be found in the source term $Q_2$, because

$$Q_2=\frac{\partial}{\partial x}\left(\frac{\partial v}{\partial P}\right)\frac{\partial T}{\partial P}=\frac{\partial}{\partial P}\left(\frac{\partial v}{\partial x}\right)\frac{\partial T}{\partial P}=\frac{\partial}{\partial P}(\dot{\varepsilon})\frac{\partial T}{\partial P}=\frac{\partial}{\partial P}\left(\frac{\dot{T}}{\lambda+2\mu}\right)\frac{\partial T}{\partial P}$$

$$=\left[\frac{\partial}{\partial P}\left(\frac{1}{\lambda+2\mu}\right)\dot{T}+\left(\frac{1}{\lambda+2\mu}\right)\frac{\partial\dot{T}}{\partial P}\right]\frac{\partial T}{\partial P}=\tilde{Q}_2+\left(\frac{1}{\lambda+2\mu}\right)\frac{\partial T}{\partial P}\frac{\partial\dot{T}}{\partial P}$$

$$\text{with } \tilde{Q}_2=\frac{\partial}{\partial P}\left(\frac{1}{\lambda+2\mu}\right)\frac{\partial T}{\partial P}\dot{T}\quad . \tag{24}$$

We move the separated term to the left side of Eq. (22) as planned and thus obtain as an extended balance equation of the total information for geometric system parameters, consisting of kinetic and potential components:

$$\frac{\partial}{\partial t}\left[\frac{\rho}{2}\left(\frac{\partial v}{\partial P}\right)^2+\frac{1}{2}\left(\frac{1}{\lambda+2\mu}\right)\left(\frac{\partial T}{\partial P}\right)^2\right]+\frac{\partial}{\partial x}\left(-\frac{\partial v}{\partial P}\frac{\partial T}{\partial P}\right)=\ Q_1-\tilde{Q}_2-\tilde{Q}_3 \tag{25}$$

Unlike the energy balance equation (10), the right-hand side of this equation contains not just two, but three source terms. The first,

$$Q_1=\rho\frac{\partial v}{\partial P}\frac{\partial}{\partial P}\left(\frac{f}{\rho}\right)=\rho\frac{\partial v}{\partial P}\left[\frac{1}{\rho}\frac{\partial f}{\partial P}+f\frac{\partial}{\partial P}\left(\frac{1}{\rho}\right)\right]=\frac{\partial v}{\partial P}\left[\frac{\partial f}{\partial P}-\frac{f}{\rho}\left(\frac{\partial\rho}{\partial P}\right)\right], \tag{26}$$

practically always vanishes because the exciting external force density $f$ generally does not change when the parameter $P$ changes ($\partial f/\partial P=0$) and, moreover, usually only acts at the boundary of the model, where the density also usually does not change ($\partial\rho/\partial P=0$). The two other source terms on the right-hand side of Eq. (25), $\tilde{Q}_2$ and $\tilde{Q}_3$, do not vanish in general and therefore effectively represent source terms for the information. In contrast to Eq. (12) for energy, Eq. (25) therefore does not become a continuity equation even in the absence of external forces $f$. Thus, unlike the total energy, the structural information in the differential wave field generally does *not* represent a globally conserved physical quantity of the system.

## 4.5 Balance Equations of Total Information for the Case of Material-Related System Parameters

To derive the balance equations for material-related system parameters $P$, we follow the procedure in Section 4.3 up to Eq. (20). There, the middle term on the left-hand side can now also be replaced using the product rule, since in this case the differentiability of $\partial\rho(x)/\partial P$ with respect to $x$ is given. With

$$\frac{1}{\rho}\frac{\partial \rho}{\partial P}\frac{\partial v}{\partial P}\frac{\partial T}{\partial x}=\frac{\partial}{\partial x}\left(\frac{1}{\rho}\frac{\partial \rho}{\partial P}\frac{\partial v}{\partial P}T\right)-Q_3 \text{ and } Q_3=\frac{\partial}{\partial x}\left(\frac{1}{\rho}\frac{\partial \rho}{\partial P}\frac{\partial v}{\partial P}\right)T \tag{27}$$

we obtain a modified balance equation of the kinetic energy of the differential wave field, $(\rho/2)\,(\partial v/\partial P)^2$:

$$\frac{\partial}{\partial t}\left[\frac{\rho}{2}\left(\frac{\partial v}{\partial P}\right)^2\right]+\frac{\partial}{\partial x}\left(-\frac{\partial v}{\partial P}\frac{\partial T}{\partial P}+\frac{\partial \rho}{\partial P}\frac{\partial v}{\partial P}\frac{T}{\rho}\right)= Q_1-Q_2+Q_3\,. \tag{28}$$

This results in an additional term in the flux quantity on the left-hand side of the equation. To derive the total energy density, the source term $\tilde{Q}_2$ according to Eq. (24) can also be introduced in the case of material-related material parameters. From this it follows

$$\frac{\partial}{\partial t}\left[\frac{\rho}{2}\left(\frac{\partial v}{\partial P}\right)^2+\frac{1}{2}\left(\frac{1}{\lambda+2\mu}\right)\left(\frac{\partial T}{\partial P}\right)^2\right]+\frac{\partial}{\partial x}\left(-\frac{\partial v}{\partial P}\frac{\partial T}{\partial P}+\frac{\partial \rho}{\partial P}\frac{\partial v}{\partial P}\frac{T}{\rho}\right)= Q_1-\tilde{Q}_2+Q_3\,. \tag{29}$$

However, this does not yet complete the overall picture, because further time derivatives can be extracted from the source terms on the right-hand side. To do this, we first split the source term $Q_3$ from Eq. (27) into two terms according to $Q_3=Q_3^A+Q_3^B$, i.e.,

$$Q_3^A=\left[\frac{\partial}{\partial x}\left(\frac{1}{\rho}\right)\frac{\partial \rho}{\partial P}+\frac{1}{\rho}\frac{\partial}{\partial x}\left(\frac{\partial \rho}{\partial P}\right)\right]\frac{\partial v}{\partial P}T \tag{30}$$

and

$$Q_3^B=\frac{1}{\rho}\frac{\partial \rho}{\partial P}\frac{\partial}{\partial P}\left(\frac{\partial v}{\partial x}\right)T\,, \tag{31}$$

where in the last expression we have again reversed the order of differentiation with respect to $P$ and $x$. The term $Q_3^A$ in Eq. (30) no longer shows any obvious time derivatives with respect to the two field quantities $v$ or $T$. However, in Eq. (31) $\partial v/\partial x$ can be replaced by $\dot{T}/(\lambda+2\mu)$. Thus, after a short calculation, we obtain

$$Q_3^B=\frac{1}{\rho}\frac{\partial \rho}{\partial P}\frac{\partial}{\partial P}\left(\frac{1}{\lambda+2\mu}\right)T\dot{T}+\tilde{Q}_3^B \tag{32}$$

with

$$\tilde{Q}_3^B=\frac{1}{\rho}\frac{\partial \rho}{\partial P}\left(\frac{1}{\lambda+2\mu}\right)\frac{\partial \dot{T}}{\partial P}T\quad. \tag{33}$$

The first term on the right-hand side of Eq. (32) can be interpreted as a time derivative,

$$\frac{1}{\rho}\frac{\partial \rho}{\partial P}\frac{\partial}{\partial P}\left(\frac{1}{\lambda+2\mu}\right)T\dot{T}=\frac{\partial}{\partial t}\left[\frac{1}{2}\frac{1}{\rho}\frac{\partial \rho}{\partial P}\frac{\partial}{\partial P}\left(\frac{1}{\lambda+2\mu}\right)T^2\right], \tag{34}$$

which can therefore be integrated into the first term on the left-hand side of our current balance equation (29). Thus, together with expression (34), we obtain the modified balance equation for material-related parameters:

$$\frac{\partial}{\partial t}\left[\frac{\rho}{2}\left(\frac{\partial v}{\partial P}\right)^2+\frac{1}{2}\left(\frac{1}{\lambda+2\mu}\right)\left(\frac{\partial T}{\partial P}\right)^2-\frac{1}{2}\frac{1}{\rho}\frac{\partial \rho}{\partial P}\frac{\partial}{\partial P}\left(\frac{1}{\lambda+2\mu}\right)T^2\right]$$

$$+\frac{\partial}{\partial x}\left(-\frac{\partial v}{\partial P}\frac{\partial T}{\partial P}+\frac{\partial \rho}{\partial P}\frac{\partial v}{\partial P}\frac{T}{\rho}\right)= Q_1+Q_3^A-\tilde{Q}_2+\tilde{Q}_3^B \tag{35}$$

From the new source terms,

$$-\tilde{Q}_2+\tilde{Q}_3^B=-\frac{\partial}{\partial P}\left(\frac{1}{\lambda+2\mu}\right)\frac{\partial T}{\partial P}\dot{T}+\frac{1}{\rho}\frac{\partial \rho}{\partial P}\left(\frac{1}{\lambda+2\mu}\right)\frac{\partial \dot{T}}{\partial P}T\ , \tag{36}$$

another time derivative can be constructed for the left-hand side of the balance equation. For this, we use the relationship

$$\frac{1}{\rho}\frac{\partial}{\partial P}\left(\frac{\rho}{\lambda+2\mu}\right)=\frac{1}{\rho}\frac{\partial \rho}{\partial P}\left(\frac{1}{\lambda+2\mu}\right)+\frac{\partial}{\partial P}\left(\frac{1}{\lambda+2\mu}\right)\ , \tag{37}$$

in which both prefactors from the right-hand side of Eq. (36) appear. At this point, two cases must now be distinguished.

#### 4.5.1 Change of $\lambda+2\mu$

Assuming that $(\partial/\partial P)(\lambda+2\mu)^{-1}\neq 0$, we can rearrange Eq. (37) for exactly this factor:

$$-\frac{\partial}{\partial P}\left(\frac{1}{\lambda+2\mu}\right)=\frac{1}{\rho}\frac{\partial \rho}{\partial P}\left(\frac{1}{\lambda+2\mu}\right)-\frac{1}{\rho}\frac{\partial}{\partial P}\left(\frac{\rho}{\lambda+2\mu}\right) \tag{38}$$

Substituting into Eq. (36) yields

$$-\tilde{Q}_2+\tilde{Q}_3^B=\frac{1}{\rho}\frac{\partial \rho}{\partial P}\left(\frac{1}{\lambda+2\mu}\right)\left(\frac{\partial T}{\partial P}\dot{T}+\frac{\partial \dot{T}}{\partial P}T\right)-\ Q_4\ \ , \tag{39}$$

with

$$Q_4=\frac{1}{\rho}\frac{\partial}{\partial P}\left(\frac{\rho}{\lambda+2\mu}\right)\frac{\partial T}{\partial P}\dot{T}\ \ . \tag{40}$$

The first term on the right-hand side of Eq. (39) can again be interpreted as a time derivative according to

$$\frac{1}{\rho}\frac{\partial \rho}{\partial P}\left(\frac{1}{\lambda+2\mu}\right)\left(\frac{\partial T}{\partial P}\dot{T}+\frac{\partial \dot{T}}{\partial P}T\right)=\frac{\partial}{\partial t}\left[\frac{1}{\rho}\frac{\partial \rho}{\partial P}\left(\frac{1}{\lambda+2\mu}\right)\frac{\partial T}{\partial P}T\right] \tag{41}$$

and moved to the left-hand side of Eq. (35). Thus, in the case $(\partial/\partial P)(\lambda+2\mu)^{-1}\neq 0$, the following balance equation results for material-related system parameters:

$$\frac{\partial}{\partial t}\left[\frac{\rho}{2}\left(\frac{\partial v}{\partial P}\right)^2+\frac{1}{2}\left(\frac{1}{\lambda+2\mu}\right)\left(\frac{\partial T}{\partial P}\right)^2-\frac{1}{2}\frac{1}{\rho}\frac{\partial \rho}{\partial P}\frac{\partial}{\partial P}\left(\frac{1}{\lambda+2\mu}\right)T^2-\frac{1}{\rho}\frac{\partial \rho}{\partial P}\left(\frac{1}{\lambda+2\mu}\right)\frac{\partial T}{\partial P}T\right]$$

$$+\frac{\partial}{\partial x}\left(-\frac{\partial v}{\partial P}\frac{\partial T}{\partial P}+\frac{\partial \rho}{\partial P}\frac{\partial v}{\partial P}\frac{T}{\rho}\right)= Q_1+Q_3^A-Q_4\ \ , \tag{42}$$

with

$$Q_1=\rho\frac{\partial v}{\partial P}\frac{\partial}{\partial P}\left(\frac{f}{\rho}\right)=\frac{\partial v}{\partial P}\left(\frac{\partial f}{\partial P}-\frac{1}{\rho}\frac{\partial \rho}{\partial P}f\right)\ \ , \tag{43}$$

$$Q_3^A = \frac{\partial}{\partial x}\left(\frac{1}{\rho}\frac{\partial \rho}{\partial P}\right)\frac{\partial v}{\partial P}T = \left[\frac{\partial}{\partial x}\left(\frac{1}{\rho}\right)\frac{\partial \rho}{\partial P} + \frac{1}{\rho}\frac{\partial}{\partial x}\left(\frac{\partial \rho}{\partial P}\right)\right]\frac{\partial v}{\partial P}T \quad , \tag{44}$$

$$Q_4 = \frac{1}{\rho}\frac{\partial}{\partial P}\left(\frac{1}{c_P^2}\right)\frac{\partial T}{\partial P}\dot{T} = \frac{1}{\rho}\frac{\partial}{\partial P}\left(\frac{\rho}{\lambda + 2\mu}\right)\frac{\partial T}{\partial P}\dot{T} = \frac{1}{\rho}\left[\frac{\partial \rho}{\partial P}\left(\frac{1}{\lambda + 2\mu}\right) + \rho\frac{\partial}{\partial P}\left(\frac{1}{\lambda + 2\mu}\right)\right]\frac{\partial T}{\partial P}\dot{T} \tag{45}$$

While $Q_3^A$ is responsible for the generation of information through changes in density and particle velocity with parameter $P$, $Q_4$ describes the generation of information through changes in sound velocity and stress with $P$. As in the case of geometric system parameters, the balance equation (42) for material-related system parameters does not become a continuity equation even in the absence of external forces $f$. Thus, unlike the total energy, the structural information in the differential wave field does not generally represent a globally conserved quantity of the system.

For the case in Example 3.3 (change in the speed of sound of the layer at constant density), the equation can be simplified even further, as will be shown below. The case in Example 3.4 (change in the density of the layer at constant speed of sound) can alternatively be interpreted as a change in $\lambda + 2\mu$ at constant $c_P$ (due to the relationship $\varrho c_P^2 = \lambda + 2\mu$) which is also within the validity range of Eq. (42).

In the special case of a material change with constant density, which is realized, for example, in our example from Section 3.3, $\partial\rho/\partial P = 0$ holds, and in Eq. (42), the corresponding additional terms as well as the source term $Q_3^A$ vanish. Equations (42) - (45) thus simplify to

$$\frac{\partial}{\partial t}\left[\frac{\rho}{2}\left(\frac{\partial v}{\partial P}\right)^2 + \frac{1}{2}\left(\frac{1}{\lambda + 2\mu}\right)\left(\frac{\partial T}{\partial P}\right)^2\right] + \frac{\partial}{\partial x}\left(-\frac{\partial v}{\partial P}\frac{\partial T}{\partial P}\right) = Q_1 - Q_4 \tag{46}$$

with

$$Q_1 = \frac{\partial v}{\partial P}\frac{\partial f}{\partial P} \quad , \tag{47}$$

$$Q_4 = \frac{1}{\rho}\frac{\partial}{\partial P}\left(\frac{1}{c_P^2}\right)\frac{\partial T}{\partial P}\dot{T} = \frac{1}{\rho}\frac{\partial}{\partial P}\left(\frac{\rho}{\lambda + 2\mu}\right)\frac{\partial T}{\partial P}\dot{T} = \frac{\partial}{\partial P}\left(\frac{1}{\lambda + 2\mu}\right)\frac{\partial T}{\partial P}\dot{T} = \tilde{Q}_2 \quad . \tag{48}$$

Equations (46) und (47) are structurally similar to the total energy balance equation (12), except that all quantities appearing there have been replaced by their partial derivatives with respect to $P$. The important difference from equation (12) is that here the source term $Q_4 = \tilde{Q}_2$ additionally occurs on the right-hand side.

Another special case of Eq. (42) is that with constant speed of sound ($\partial c_P/\partial P = 0$), which is realized in the example from Section 3.4. Here, the source term $Q_4$ vanishes and equations (42) - (45) simplify to

$$\frac{\partial}{\partial t}\left[\frac{\rho}{2}\left(\frac{\partial v}{\partial P}\right)^2 + \frac{1}{2}\left(\frac{1}{\lambda + 2\mu}\right)\left(\frac{\partial T}{\partial P}\right)^2 + \frac{1}{2}\left(\frac{1}{\lambda + 2\mu}\right)\frac{1}{\rho^2}\left(\frac{\partial \rho}{\partial P}\right)^2 T^2 - \frac{1}{\rho}\frac{\partial \rho}{\partial P}\left(\frac{1}{\lambda + 2\mu}\right)\frac{\partial T}{\partial P}T\right]$$
$$+ \frac{\partial}{\partial x}\left(-\frac{\partial v}{\partial P}\frac{\partial T}{\partial P} + \frac{\partial \rho}{\partial P}\frac{\partial v}{\partial P}\frac{T}{\rho}\right) = Q_1 + Q_3^A \tag{49}$$

with

$$Q_1 = \rho \frac{\partial v}{\partial P} \frac{\partial}{\partial P} \left( \frac{f}{\rho} \right) = \frac{\partial v}{\partial P} \left( \frac{\partial f}{\partial P} - \frac{1}{\rho} \frac{\partial \rho}{\partial P} f \right) \quad , \tag{50}$$

$$Q_3^A = \frac{\partial}{\partial x} \left( \frac{1}{\rho} \frac{\partial \rho}{\partial P} \right) \frac{\partial v}{\partial P} T = \left[ \frac{\partial}{\partial x} \left( \frac{1}{\rho} \right) \frac{\partial \rho}{\partial P} + \frac{1}{\rho} \frac{\partial}{\partial x} \left( \frac{\partial \rho}{\partial P} \right) \right] \frac{\partial v}{\partial P} T \quad . \tag{51}$$

### 4.5.2 Change of $\rho$

We now return to our equations (36) und (37) and consider the other case where the expression $(1/\rho)(\partial\rho/\partial P)(\lambda + 2\mu)^{-1} \neq 0$ holds. Then we can rearrange equation (37) for this factor:

$$\frac{1}{\rho} \frac{\partial \rho}{\partial P} \left( \frac{1}{\lambda + 2\mu} \right) = \frac{1}{\rho} \frac{\partial}{\partial P} \left( \frac{\rho}{\lambda + 2\mu} \right) - \frac{\partial}{\partial P} \left( \frac{1}{\lambda + 2\mu} \right) \tag{52}$$

Substituting into Eq. (36) yields in this case

$$-\tilde{Q}_2 + \tilde{Q}_3^B = -\frac{\partial}{\partial P} \left( \frac{1}{\lambda + 2\mu} \right) \left( \frac{\partial T}{\partial P} \dot{T} + \frac{\partial \dot{T}}{\partial P} T \right) + \; Q_4 \tag{53}$$

with

$$Q_4 = \frac{1}{\rho} \frac{\partial}{\partial P} \left( \frac{\rho}{\lambda + 2\mu} \right) \frac{\partial \dot{T}}{\partial P} T \quad . \tag{54}$$

The first term on the right-hand side of Eq. (53) can again be interpreted as a time derivative according to

$$-\frac{\partial}{\partial P} \left( \frac{1}{\lambda + 2\mu} \right) \left( \frac{\partial T}{\partial P} \dot{T} + \frac{\partial \dot{T}}{\partial P} T \right) = -\frac{\partial}{\partial t} \left[ \frac{\partial}{\partial P} \left( \frac{1}{\lambda + 2\mu} \right) \frac{\partial T}{\partial P} T \right] \tag{55}$$

and moved to the left-hand side of Eq. (35). Thus, the following modified balance equation results as the general final outcome of our considerations regarding the total information contained in the wave field for the case $(1/\rho)(\partial\rho/\partial P)(\lambda + 2\mu)^{-1} \neq 0$:

$$\frac{\partial}{\partial t} \left[ \frac{\rho}{2} \left( \frac{\partial v}{\partial P} \right)^2 + \frac{1}{2} \left( \frac{1}{\lambda + 2\mu} \right) \left( \frac{\partial T}{\partial P} \right)^2 - \frac{1}{2} \frac{1}{\rho} \frac{\partial \rho}{\partial P} \frac{\partial}{\partial P} \left( \frac{1}{\lambda + 2\mu} \right) T^2 + \frac{\partial}{\partial P} \left( \frac{1}{\lambda + 2\mu} \right) \frac{\partial T}{\partial P} T \right] + \frac{\partial}{\partial x} \left( -\frac{\partial v}{\partial P} \frac{\partial T}{\partial P} + \frac{\partial \rho}{\partial P} \frac{\partial v}{\partial P} \frac{T}{\rho} \right) = \; Q_1 + Q_3^A + Q_4 \tag{56}$$

with

$$Q_1 = \rho \frac{\partial v}{\partial P} \frac{\partial}{\partial P} \left( \frac{f}{\rho} \right) = \frac{\partial v}{\partial P} \left( \frac{\partial f}{\partial P} - \frac{1}{\rho} \frac{\partial \rho}{\partial P} f \right) \quad , \tag{57}$$

$$Q_3^A = \frac{\partial}{\partial x} \left( \frac{1}{\rho} \frac{\partial \rho}{\partial P} \right) \frac{\partial v}{\partial P} T = \left[ \frac{\partial}{\partial x} \left( \frac{1}{\rho} \right) \frac{\partial \rho}{\partial P} + \frac{1}{\rho} \frac{\partial}{\partial x} \left( \frac{\partial \rho}{\partial P} \right) \right] \frac{\partial v}{\partial P} T \quad , \tag{58}$$

$$Q_4 = \frac{1}{\rho} \frac{\partial}{\partial P} \left( \frac{1}{c_P^2} \right) \frac{\partial \dot{T}}{\partial P} T = \frac{1}{\rho} \frac{\partial}{\partial P} \left( \frac{\rho}{\lambda + 2\mu} \right) \frac{\partial \dot{T}}{\partial P} T = \frac{1}{\rho} \left[ \frac{\partial \rho}{\partial P} \left( \frac{1}{\lambda + 2\mu} \right) + \rho \frac{\partial}{\partial P} \left( \frac{1}{\lambda + 2\mu} \right) \right] \frac{\partial \dot{T}}{\partial P} T \tag{59}$$

The changes compared to case 4.5.1 and equations (42) - (45) therefore lie in the modified fourth term in the time derivative and a modified source term $Q_4$. Similar to 4.5.1, two simplified special cases can again be considered.

In the case of a constant $\lambda + 2\mu$ it holds $(\partial/\partial P)(\lambda + 2\mu)^{-1} = 0$ and the above equations simplify to

$$\frac{\partial}{\partial t}\left[\frac{\rho}{2}\left(\frac{\partial v}{\partial P}\right)^2 + \frac{1}{2}\left(\frac{1}{\lambda + 2\mu}\right)\left(\frac{\partial T}{\partial P}\right)^2\right] + \frac{\partial}{\partial x}\left(-\frac{\partial v}{\partial P}\frac{\partial T}{\partial P} + \frac{\partial \rho}{\partial P}\frac{\partial v}{\partial P}\frac{T}{\rho}\right) = Q_1 + Q_3^A + Q_4 \quad (60)$$

with

$$Q_1 = \rho \frac{\partial v}{\partial P}\frac{\partial}{\partial P}\left(\frac{f}{\rho}\right) = \frac{\partial v}{\partial P}\left(\frac{\partial f}{\partial P} - \frac{1}{\rho}\frac{\partial \rho}{\partial P} f\right) \;, \quad (61)$$

$$Q_3^A = \frac{\partial}{\partial x}\left(\frac{1}{\rho}\frac{\partial \rho}{\partial P}\right)\frac{\partial v}{\partial P} T = \left[\frac{\partial}{\partial x}\left(\frac{1}{\rho}\right)\frac{\partial \rho}{\partial P} + \frac{1}{\rho}\frac{\partial}{\partial x}\left(\frac{\partial \rho}{\partial P}\right)\right]\frac{\partial v}{\partial P} T \;, \quad (62)$$

$$Q_4 = \frac{1}{\rho}\frac{\partial}{\partial P}\left(\frac{1}{c_P^2}\right)\frac{\partial \dot{T}}{\partial P} T = \frac{1}{\rho}\frac{\partial}{\partial P}\left(\frac{\rho}{\lambda + 2\mu}\right)\frac{\partial \dot{T}}{\partial P} T = \frac{1}{\rho}\frac{\partial \rho}{\partial P}\left(\frac{1}{\lambda + 2\mu}\right)\frac{\partial \dot{T}}{\partial P} T \;. \quad (63)$$

If, on the other hand, the speed of sound is constant ($\partial c_P/\partial P = 0$), then equations (56) - (59) yield:

$$\frac{\partial}{\partial t}\left[\frac{\rho}{2}\left(\frac{\partial v}{\partial P}\right)^2 + \frac{1}{2}\left(\frac{1}{\lambda + 2\mu}\right)\left(\frac{\partial T}{\partial P}\right)^2 + \frac{1}{2}\left(\frac{1}{\lambda + 2\mu}\right)\frac{1}{\rho^2}\left(\frac{\partial \rho}{\partial P}\right)^2 T^2 + \frac{\partial}{\partial P}\left(\frac{1}{\lambda + 2\mu}\right)\frac{\partial T}{\partial P} T\right]$$
$$+ \frac{\partial}{\partial x}\left(-\frac{\partial v}{\partial P}\frac{\partial T}{\partial P} + \frac{\partial \rho}{\partial P}\frac{\partial v}{\partial P}\frac{T}{\rho}\right) = Q_1 + Q_3^A \quad (64)$$

with

$$Q_1 = \rho \frac{\partial v}{\partial P}\frac{\partial}{\partial P}\left(\frac{f}{\rho}\right) = \frac{\partial v}{\partial P}\left(\frac{\partial f}{\partial P} - \frac{1}{\rho}\frac{\partial \rho}{\partial P} f\right) \;, \quad (65)$$

$$Q_3^A = \frac{\partial}{\partial x}\left(\frac{1}{\rho}\frac{\partial \rho}{\partial P}\right)\frac{\partial v}{\partial P} T = \left[\frac{\partial}{\partial x}\left(\frac{1}{\rho}\right)\frac{\partial \rho}{\partial P} + \frac{1}{\rho}\frac{\partial}{\partial x}\left(\frac{\partial \rho}{\partial P}\right)\right]\frac{\partial v}{\partial P} T \,. \quad (66)$$

This is identical to equations (49) - (51) in section 4.5.1, because for the prefactor in the last term of the time derivative, the following identity results in this case,

$$\frac{\partial}{\partial P}\left(\frac{1}{\lambda + 2\mu}\right) = \frac{\partial}{\partial P}\left(\frac{1}{\rho c_P^2}\right) = \frac{1}{c_P^2}\frac{\partial}{\partial P}\left(\frac{1}{\rho}\right) = -\frac{1}{\rho^2 c_P^2}\frac{\partial \rho}{\partial P} = -\frac{1}{\rho}\frac{\partial \rho}{\partial P}\left(\frac{1}{\lambda + 2\mu}\right) \;, \quad (67)$$

which corresponds to the prefactor from Eq. (49).

### 4.6 Validity of the Theory for Material-Related System Parameters

The cases considered in 4.5.1 and 4.5.2 with changing $\lambda + 2\mu$ and $\rho$, respectively, explicitly illustrate the limitations of the theory presented here. When formulating the new balance equation(s), we started with the partial derivative with respect to only *one* system parameter $P$ (see equations (16) und (17)). This means that the derived expressions can only describe cases in which one parameter changes independently of the others. If two or more parameters change simultaneously and independently, this can no longer be adequately captured by the equations. The two material quantities $\rho$ and $\lambda + 2\mu$ are elementary components of the model description and thus belong to the set of possible material-related system parameters. Therefore, the same applies to them as stated above: only one of the two quantities can change independently, but not both at the same time. A conditional change between these two material properties is allowed but must then be induced by a constant speed of sound, assumed to be known, according to $\rho c_P^2 = \lambda + 2\mu$.

The derivation of equation (34), described above, which includes a derivative with respect to both $\rho$ and $\lambda + 2\mu$, yields two competing balance equations. These equations become identical only if $c_P = const$ is claimed, i.e., if only one of the two quantities changes truly independently and the other undergoes only a conditional change induced by the constant speed of sound. If both $\rho$ and $\lambda + 2\mu$ change independently, i.e., if $\partial c_P/\partial P \neq 0$, two different solutions result, leading to a contradiction in the theory. From these considerations, the range of validity of the theory presented here for material-related system parameters $P$ is summarized in Table 1 below. Additionally, the case of pure geometric system parameters is listed as entry #0.

***Table 1:*** *Overview of the theory's domains of validity and the corresponding balance equations. It is* $c_P^2 = \lambda + 2\mu$. **Δ:** *independent change,* $\Delta_c$*: conditional change, const: constant value,* ×*: theory not valid.* (**42**)*: Overall balance equation for the case* $(\partial/\partial P)(\lambda + 2\mu)^{-1} \neq 0$ *(highlighted in orange).* (**56**)*: Overall balance equation for the case* $\partial\rho/\partial P \neq 0$ *(highlighted in green).*

| **Case #** | **Parameter *P*** | $\boldsymbol{\rho}$ | $\boldsymbol{c_P}$ | $\boldsymbol{\lambda + 2\mu}$ | **Valid Balance Equations** |
|---|---|---|---|---|---|
| 0 | geometric | $\Delta_c$ | $\Delta_c$ | $\Delta_c$ | (25) |
| 1a | $\boldsymbol{c_P}$ (Δ with constant $\rho$) | *const* | **Δ** | $\Delta_c$ | (**42**), (46) |
| 1b | $\boldsymbol{\lambda + 2\mu}$ (Δ with constant $\rho$) | *const* | $\Delta_c$ | **Δ** | (**42**), (46) |
| 2a | $\boldsymbol{\rho}$ (Δ with constant $c_P$) | **Δ** | *const* | $\Delta_c$ | (**42**), (**56**), (49) = (64) |
| 2b | $\boldsymbol{\lambda + 2\mu}$ (Δ with constant $c_P$) | $\Delta_c$ | *const* | **Δ** | (**42**), (**56**), (49) = (64) |
| 3a | $\boldsymbol{\rho}$ (Δ with constant $\lambda + 2\mu$) | **Δ** | $\Delta_c$ | *const* | (**56**), (60) |
| 3b | $\boldsymbol{c_P}$ (Δ with constant $\lambda + 2\mu$) | $\Delta_c$ | **Δ** | *const* | (**56**), (60) |
| 4 | $\boldsymbol{\rho}$ and $\boldsymbol{c_P}$ (independent Δ) | **Δ** | **Δ** | $\Delta_c$ | × |
| 5 | $\boldsymbol{\rho}$ and $\boldsymbol{\lambda + 2\mu}$ (independ. Δ) | **Δ** | $\Delta_c$ | **Δ** | × |
| 6 | $\boldsymbol{c_P}$ and $\boldsymbol{\lambda + 2\mu}$ (independ. Δ) | $\Delta_c$ | **Δ** | **Δ** | × |

Case 1a (change in the speed of sound in the layer at constant density), discussed in Section 3.3, falls within the scope of the overall balance equation (42), since the change in the speed of sound automatically entails a (conditional) change in $\lambda + 2\mu$, and the expression $(\partial/\partial P)(\lambda + 2\mu)^{-1}$ therefore does not vanish. Since $\partial\rho/\partial P = 0$, the simplified equation (46) applies. This is also valid for case 1b, which, within the framework of the theory presented here, is identical to case 1a, as it is irrelevant in the equations whether the change in $\lambda + 2\mu$ is independent or conditional.

Case 2a (density change of the layer at constant speed of sound), discussed in Section 3.4, is also described by the overall equation (42) because a (conditional) change of $\lambda + 2\mu$ occurs here as well. However, due to the constant speed of sound, the simplified equation (49) effectively applies. Case 2b is also to be considered identical to case 2a.

What is interesting about case 2 is that, in addition to $\lambda + 2\mu$, $\rho$ also changes, and thus the overall balance equation (56) can alternatively be used for the description. In the case of constant speed of sound, the simplified equation (64) results, which is identical to equation (49) due to the identity (67).

Cases 3a and 3b, in which $\lambda + 2\mu$ is kept constant, are rarely considered in practice because density and speed of sound are usually used to define the material parameters, and if one of these two quantities changes, a very specific adjustment of the other quantity is necessary to maintain the constancy of $\lambda + 2\mu$. Since cases 3a and 3b also belong to the category $\partial\rho/\partial P \neq 0$, they can likewise be described by the overall balance equation (56), but also, because of $\lambda + 2\mu = const$, more simply by Eq. (60).

Table 1 shows that the various valid cases for material-related system parameters can be described either by the two overall balance equations (42) and (56) or alternatively by the three simplified equations (46), (49) and (60). The choice depends on the practical implementation. For example, both examples from Sections 3.3 and 3.4 belong to the category $(\partial/\partial P)(\lambda + 2\mu)^{-1} \neq 0$, which suggests using balance equation (42) in both cases.

As described above, equations (42) and (56) contradict each other if independent changes in $\rho$ and $\lambda + 2\mu$ are allowed. Therefore, cases 4-6 in Table 1 can no longer be correctly addressed by the theory presented here. This would require an extended approach with two or more independent parameter changes, in which the deeper meaning of the two overarching equations (42) and (56) would then become clear, which in all valid cases 1-3 in Table 1 were ultimately replaced by simplified individual equations. However, such an extended theory would exceed the scope of the description presented here.

In the case of geometric system parameters, however, the restrictions described here to only one independently changing parameter $P$ are not always relevant. For example, the combination of a shift of the layer to the right (according to Example 3.1) and an increase in thickness (according to Example 3.2) leads to changes at the left layer boundary canceling each other out, leaving only a shift of the right layer boundary. This shift can be described by a new single parameter $P$ ("position of the right layer boundary") and is thus still describable within the framework of the theory presented here.

### 4.7 Sign of Balance Quantity

The different validity ranges of the theory presented in Section 4.6 now also allow us to prove that the time-derived terms on the left-hand side of the respective balance equation are always $\geq 0$, as we also required for a suitable measure of information in requirement 9 from Section 4.

For case 0 (geometric system parameters), equation (25) applies. There, the two squared expressions for the kinetic and potential information density are trivially $\geq 0$. For cases 1 and 2 in Table 1, equation (42) generally applies. It contains, firstly, the squared term of the kinetic energy density of the differential wave field. Terms 2-4 all depend quadratically on the stress $T$, which suggests that all three terms together represent the potential energy density of the differential wave field. The third term, like the fourth term, can only be effective if $\partial\rho/\partial P \neq 0$ .

We will show below that the assumed potential energy density, as the sum of terms 2-4 in the time derivative of Eq. (42), is always positive (or zero), and thus, together with the first term, always results in a balance quantity $\geq 0$. To do this, we must prove that

$$\frac{1}{2}\left(\frac{1}{\lambda+2\mu}\right)\left(\frac{\partial T}{\partial P}\right)^2-\frac{1}{2}\frac{1}{\rho}\frac{\partial\rho}{\partial P}\frac{\partial}{\partial P}\left(\frac{1}{\lambda+2\mu}\right)T^2-\frac{1}{\rho}\frac{\partial\rho}{\partial P}\left(\frac{1}{\lambda+2\mu}\right)\frac{\partial T}{\partial P}T\geq 0\quad . \tag{68}$$

In case 1 from Table 1, it is $\partial\rho/\partial P=0$, leaving only the first term in expression (68). This term is trivially $\geq 0$ due to the squaring of $\partial T/\partial P$. In case 2, we have $\partial\rho/\partial P\neq 0$, but the speed of sound is constant. Thus, we can again use identity (67) and rewrite the middle term on the left-hand side of inequality (68). We obtain

$$\frac{1}{2}\left(\frac{1}{\lambda+2\mu}\right)\left(\frac{\partial T}{\partial P}\right)^2+\frac{1}{2}\frac{1}{\rho^2}\left(\frac{\partial\rho}{\partial P}\right)^2\left(\frac{1}{\lambda+2\mu}\right)T^2-\frac{1}{\rho}\frac{\partial\rho}{\partial P}\left(\frac{1}{\lambda+2\mu}\right)\frac{\partial T}{\partial P}T\geq 0\,. \tag{69}$$

It is immediately apparent from this that the middle term, i.e., the third term in the time derivative of Eq. (42), is always positive within the validity of the theory. This is not the case for the remaining term on the far right, which even produces predominantly negative values. However, dividing expression (69) by $(1/2)(\lambda+2\mu)^{-1}$ yields

$$\left(\frac{\partial T}{\partial P}\right)^2+\frac{1}{\rho^2}\left(\frac{\partial\rho}{\partial P}\right)^2T^2-2\frac{1}{\rho}\frac{\partial\rho}{\partial P}\frac{\partial T}{\partial P}T=\left(\frac{\partial T}{\partial P}-\frac{1}{\rho}\frac{\partial\rho}{\partial P}T\right)^2\geq 0\,, \tag{70}$$

which was to be proven. Due to the identity of the right-hand term in Eq. (69) with that of the balance quantity in Eq. (56), the above proof also applies unchanged to case 3 in Table 1 and thus to all valid cases with material-related parameter changes.

With the considerations in this section, we have now also fulfilled requirement 9, formulated at the beginning of this chapter, for an unsigned information quantity. The following holds true for all cases 1-3 in Table 1 that can be described by the theory:

$$I_P\sim I_{p,kin}+I_{p,pot}\geq 0 \tag{71}$$

with the kinetic energy density of the differential wave field

$$I_{P,kin}\sim\frac{\rho}{2}\left(\frac{\partial v}{\partial P}\right)^2 \tag{72}$$

and its potential counterpart

$$I_{P,pot}\sim\frac{1}{2}\left(\frac{1}{\lambda+2\mu}\right)\left(\frac{\partial T}{\partial P}\right)^2 \tag{73}$$

for the cases 0, 1 und 3, and

$$I_{P,pot}\sim\frac{1}{2}\left(\frac{1}{\lambda+2\mu}\right)\left(\frac{\partial T}{\partial P}\right)^2-\frac{1}{\rho}\frac{\partial\rho}{\partial P}\left[\frac{1}{2}\frac{\partial}{\partial P}\left(\frac{1}{\lambda+2\mu}\right)T^2+\left(\frac{1}{\lambda+2\mu}\right)\frac{\partial T}{\partial P}T\right] \tag{74}$$

for case 2, respectively.

Thus, only requirements 10 and 11 remain: the independence of the information from the amplitude of the excitation signal and the requirement of a suitable physical dimension for the information. As it turns out, both requirements can be met by appropriate normalization and scaling of the existing expressions.

## 4.8 Appropriate Normalization and Scaling of Information

Based on our previous considerations, the information quantities found in expressions (72) - (74) depend on the partial derivatives of the two field quantities $v$ and $T$ with respect to the system parameter $P$, as well as on the field quantity $T$ itself. This leads to the surprising situation that the incident wave field generated by the excitation signal $f$ vanishes (leaving only the scattered field), but $f$ still has a direct influence on our information quantity. Since all terms in (72) - (74) are quadratic in both field quantities, a simple linear scaling of the excitation signal $f$ by a factor of $k$ results in a $k$-fold larger difference field and a $k^2$-fold increase in the information quantity, which is neither meaningful nor desirable. Furthermore, both information densities also exhibit the physical dimension of energy, which is additionally related to the square of the parameter $P$, since all terms show a quadratic dependence on $\partial/\partial P$.

To make the information dimensionless and thus assign the dimension $1/V$ to the information density, a first step is to normalize the energy density of the differential wave field to a suitable energy. To simultaneously fulfill requirements 10 and 11, it is therefore self-evident to perform the normalization to the energy $E_f$ of the excitation signal $f$. This quantity is independent of both time and position (since $f$ always acts only at the excitation point at $x = 0$) and can therefore, after division by both sides of the balance equation, appear in both partially derived expressions in parentheses on the left-hand side of the equation. $E_f$ can thus be used as a suitable normalizing factor for all relevant information quantities. This eliminates the physical dimension of energy and, at the same time, the dependence on the excitation signal.

As a final step, the dimension $1/P^2$ must now be eliminated from the equations. To do this, we recall that so far, in our equations, we have considered *absolute* changes $\Delta P$ of the system parameter $P$ via the $\partial/\partial P$ terms. Since $P$ can represent completely different parameters with different physical dimensions (such as length, length per unit time, or mass per unit volume), a meaningful comparison of the information quantities between the individual parameters is not possible. However, *relative* changes of the parameters of the form $\Delta P/P$ are indeed comparable. Thus, if we replace, for example, the difference quotient $\Delta v/\Delta P$ by $\Delta v/(\Delta P/P)$, we obtain $P(\Delta v/\Delta P)$, simply as a version of the original difference quotient scaled by $P$. In general, the transition from absolute to relative changes in $P$ is performed by changing $\partial/\partial P$ to $P(\partial/\partial P)$ and $(\partial/\partial P)^2$ to $P^2(\partial/\partial P)^2$, respectively. This means that the balance equations simply need to be multiplied by $P^2$, which effectively makes the information dimensionless.

With the normalization to the energy of the excitation signal $E_f$ described above and the scaling with $P^2$, our last two requirements 10 and 11 for a dimensionless information quantity independent of the excitation signal are also fulfilled, and we obtain as the final result the *information density* as a function of position and time.

$$
\begin{aligned}
I_P(x,t) = \frac{P^2}{E_f}\Bigg\{ & \frac{1}{2}\rho(x)\left(\frac{\partial v(x,t)}{\partial P}\right)^2 + \frac{1}{2}\left(\frac{1}{\lambda(x)+2\mu(x)}\right)\left(\frac{\partial T(x,t)}{\partial P}\right)^2 \\
& - \frac{1}{\rho(x)}\frac{\partial \rho(x)}{\partial P}\Bigg[\frac{1}{2}\frac{\partial}{\partial P}\left(\frac{1}{\lambda(x)+2\mu(x)}\right)T(x,t)^2 \\
& + \left(\frac{1}{\lambda(x)+2\mu(x)}\right)\frac{\partial T(x,t)}{\partial P}T(x,t)\Bigg]\Bigg\},
\end{aligned}
\tag{75}
$$

for case 2, and simplified

$$I_P(x,t) = \frac{P^2}{E_f}\left[\frac{1}{2}\rho(x)\left(\frac{\partial v(x,t)}{\partial P}\right)^2 + \frac{1}{2}\left(\frac{1}{\lambda(x)+2\mu(x)}\right)\left(\frac{\partial T(x,t)}{\partial P}\right)^2\right] \tag{76}$$

for cases 0, 1 and 3 in Table 1.

The physical dimension of the information density $I_P(x,t)$ is "information per volume." It can be obtained as a single value at a specific location $x$ at a specific time $t$, or as an integration over a defined volume of the overall system and/or a fixed time interval. If only the temporal maximum of the information is considered, the energy of the excitation signal's maximum is a suitable normalizing factor $E_f$. However, if $I_P(x,t)$ is integrated over the entire signal length, the integrated energy of the excitation signal should be used as the normalizing parameter instead.

The two additional terms in the potential part of Eq. (75) only become effective if $\partial\rho/\partial P \neq 0$. Where no material-related density changes occur in the system, the structure of the equation corresponds to that of the energy density from Section 4.2, where $v$ is replaced by $\partial v/\partial P$ and $T$ by $\partial T/\partial P$ (see Eq. (76)).

Another important quantity we obtain is the *information flux*, analogous to the (one-dimensional) energetic Poynting vector,

$$F_P(x,t) = \frac{P^2}{E_f}\frac{\partial v(x,t)}{\partial P}\left[-\frac{\partial T(x,t)}{\partial P} + \frac{1}{\rho(x)}\frac{\partial\rho(x)}{\partial P}T(x,t)\right], \tag{77}$$

for case 4, and simplified

$$F_P(x,t) = -\frac{P^2}{E_f}\frac{\partial v(x,t)}{\partial P}\frac{\partial T(x,t)}{\partial P}, \tag{78}$$

for cases 0–3, each with the physical dimension "information per unit area and time." The last expression again corresponds to the energetic Poynting vector with $\partial v/\partial P$ and $\partial T/\partial P$ instead of $v$ and $T$.

The following serve as source terms (and, if applicable, sinks) of the information:

$$Q_f(x,t) \coloneqq \frac{P^2}{E_f}Q_1(x,t) = \frac{P^2}{E_f}\rho(x)\frac{\partial v(x,t)}{\partial P}\frac{\partial}{\partial P}\left(\frac{f(x,t)}{\rho(x)}\right), \tag{79}$$

$$Q_{v0}(x,t) \coloneqq \frac{P^2}{E_f}\tilde{Q}_3(x,t) = \frac{P^2}{E_f}\frac{1}{\rho(x)}\frac{\partial\rho(x)}{\partial P}\frac{\partial v(x,t)}{\partial P}\frac{\partial T(x,t)}{\partial x}, \tag{80}$$

$$Q_v(x,t) \coloneqq \frac{P^2}{E_f}Q_3^A(x,t) = \frac{P^2}{E_f}\frac{\partial}{\partial x}\left(\frac{1}{\rho(x)}\frac{\partial\rho(x)}{\partial P}\right)\frac{\partial v(x,t)}{\partial P}T(x,t), \tag{81}$$

$$Q_{T0}(x,t) \coloneqq \frac{P^2}{E_f}\tilde{Q}_2(x,t) = \frac{P^2}{E_f}\frac{\partial}{\partial P}\left(\frac{1}{\lambda(x)+2\mu(x)}\right)\frac{\partial T(x,t)}{\partial P}\dot{T}(x,t), \tag{82}$$

$$Q_T(x,t) \coloneqq \frac{P^2}{E_f}Q_4(x,t) = \frac{P^2}{E_f}\frac{1}{\rho(x)}\frac{\partial}{\partial P}\left(\frac{1}{c_P(x)^2}\right)\frac{\partial T(x,t)}{\partial P}\dot{T}(x,t) \tag{83}$$

$$= \frac{P^2}{E_f} \frac{1}{\rho(x)} \frac{\partial}{\partial P} \left( \frac{\rho(x)}{\lambda(x) + 2\mu(x)} \right) \frac{\partial T(x,t)}{\partial P} \dot{T}(x,t) \ .$$

each with the dimension "information per volume and time".

$Q_f(x,t)$ describes all sources of information that arise from a change in the excitation signal $f$ or from a change in the density $\rho$, in each case at the location of the active transducer. A change in the excitation signal can be neglected in this case. This would only be meaningful in a non-linear medium, which is not the case here. Nevertheless, $Q_f(x,t) \neq 0$ can be valid if the density of the matrix medium is changed. Expression (79) is valid for all cases 0-4 from Table 1.

$Q_v(x,t)$ (cases 1-4) and $Q_{v0}(x,t)$ (case 0), respectively, describe the sources resulting from density changes and changes in particle velocity (with $P$). These terms become zero at constant density, and in the case of $Q_v(x,t)$, also if neither the density nor its change with respect to $P$ changes with position $x$. This is the case, for example, inside a scatterer with constant density or constant density change. Therefore, expressions (80) and (81), respectively, describe sources at the interface between two different materials.

$Q_T(x,t)$ (cases 1-4) and $Q_{T0}(x,t)$ (case 0), respectively, are responsible for the sources resulting from changes in the speed of sound (Eq. (83)) or $\lambda + 2\mu$ (Eq. (82) as well as changes in stress and disappear at a constant speed of sound or constant $\lambda + 2\mu$, respectively. $Q_T(x,t)$ describes volume sources resulting from changes in the speed of sound, e.g., inside a scatterer. $Q_{T0}(x,t)$, on the other hand, describes sources at the shifting interface between two media that are associated with changes in $\lambda + 2\mu$.

In the following section, we will use the simulation examples from Section 3 to check whether the quantities (75) - (83) derived here actually represent a physically meaningful and consistent description of the generation and propagation of information in the ultrasonic wave field.

# 5. Verification of the Derived Information Quantities

We revisit simulation examples 3.1–3.4 from Figure 5 and Figure 8, evaluating the information parameters determined in Section 4 for each case. For comparison purposes, we start with the energy balance in the initial model.

## 5.1 Energy Balance in the Initial Model

Figure 10 initially shows the analogous quantities from the balance equation (10) for the kinetic energy density in the initial model as time snapshots, in order to better assess similarities and differences with the subsequent information quantities. The individual energy quantities are each normalized to the maximum of the initial signal arriving from the left.

The kinetic energy density in the left column of Figure 10 shows only positive values due to the squaring of the particle velocity. Its maximum amplitude visibly decreases within the layer ($t = 6.53$ µs) and then increases again after passing through the layer ($t = 8.91$ µs). It should be noted that the signal waveform also broadens within the layer due to the longer wavelength present there. Therefore, the decrease in the maximum amplitude when passing through the layer does not mean that the total energy density in the signal decreases. As we will see below, the integral of the energy density over the total signal remains unchanged even within the layer!

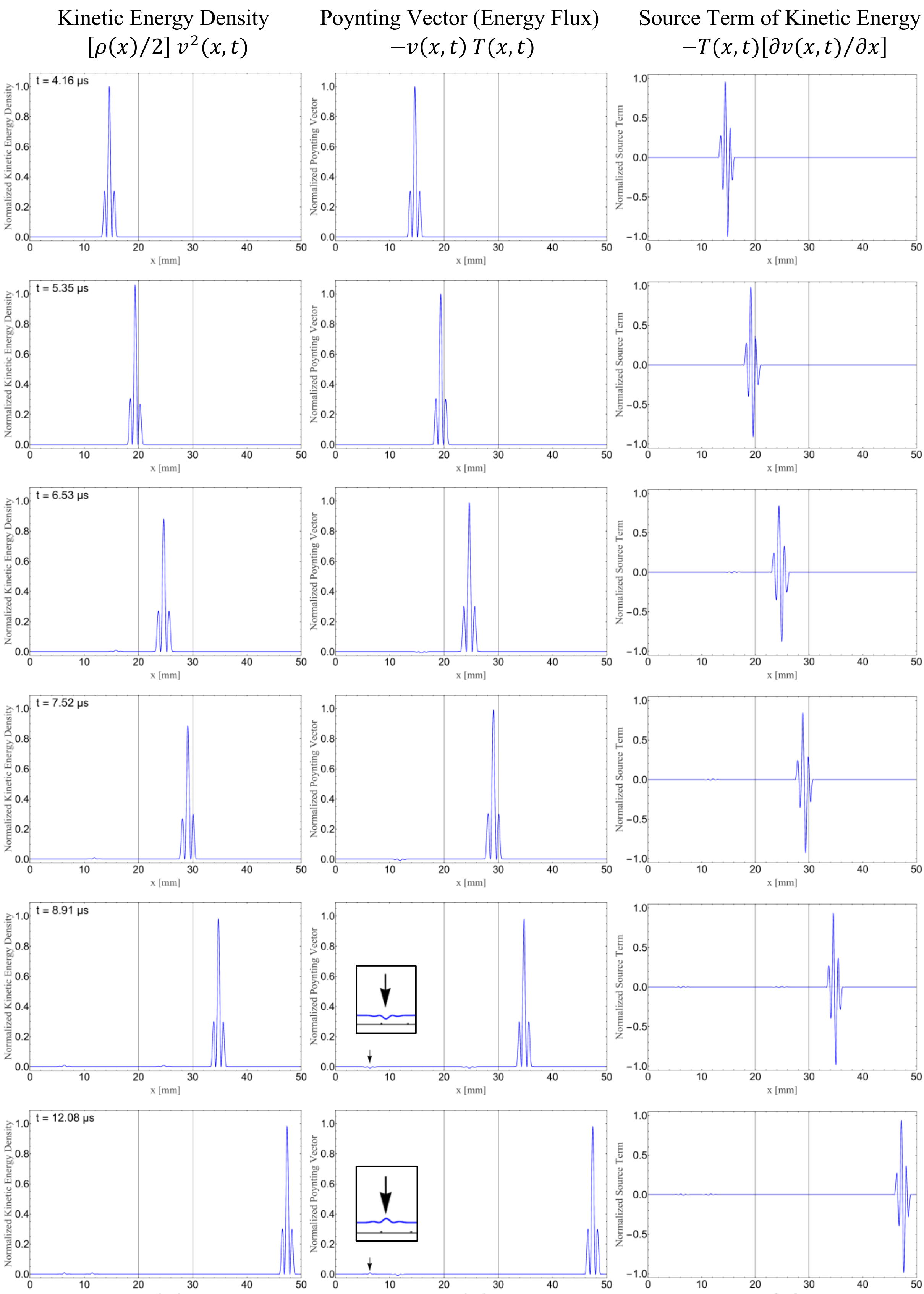

***Figure 10:*** *Normalized representation of the kinetic energy density (left column), the Poynting vector (middle column), and the source term of the kinetic energy (right column) in the initial model at times* $t$ = *4.16 μs (1st row),* $t$ = *5.35 μs (2nd row),* $t$ = *6.53 μs (3rd row),* $t$ = *7.52 μs (4th row),* $t$ = *8.91 μs (5th row), and* $t$ = *12.08 μs (last row).*

Slight fluctuations in the waveform are visible during the interaction with the layer boundaries ($t = 5.35$ µs and $t = 7.52$ µs). The final amplitude difference between the transmitted wave ($t = 12.08$ µs) and the incident wave ($t = 4.16$ µs) results from the energy loss when passing through both layer boundaries. In contrast to Figure 2, which depicted the particle velocity $v$, the reflection and transmission coefficients with respect to energy $E$ are decisive for the kinetic energy density given here. These coefficients are $R_E = R_v^2 = 0.01$ and $T_E = 1 - R_E = 0.99$. Accordingly, the two echoes from the left layer boundary appear very small in the images of Figure 10.

The energy flux, expressed by the Poynting vector in the middle column of Figure 10, has a similar shape to the energy density, but is only slightly affected by the interactions with the layer boundaries. Only a very small decrease is discernible within the layer. The small losses from the first ($t = 4.16$ µs) to the last image ($t = 12.08$ µs) result from the reflection losses at both layer boundaries.

Unlike energy density, the Poynting vector can take on both positive and negative values. In the one-dimensional case presented here, positive values indicate propagation from left to right (forward direction), while negative values indicate propagation in the opposite direction. In the images for $t \geq 6.53$ µs, the leftward-propagating reflections are just barely visible with small negative values in the left part of the model. The reflection from the left layer boundary exhibits a negative amplitude before interacting with the left model boundary ($t = 8.91$ µs), which is converted into a positive value ($t = 12.08$ µs) by the change in direction during the subsequent reflection. This is indicated in the two corresponding images by small vertical arrows and the corresponding insets.

The source term of the kinetic energy is shown in the right-hand column of Figure 10. It has a significantly different shape than the other two energy quantities, which is due to the spatial derivative of the particle velocity. The source term assumes both positive and negative values, reflecting the continuous conversion of kinetic energy to potential energy (negative source term) and from potential to kinetic energy (positive source term). It decreases slightly within the layer and also results in only small, barely visible reflections at both layer boundaries.

If the total energy density were displayed instead of the kinetic energy density in the left column of Figure 10, the source term in the right column would be omitted. It would then be zero at all times except for the excitation point at $x = 0$, where the ultrasonic energy input by the probe is implemented in the model. The snapshots for the total energy density according to Eqs. (12) and (14), which are not shown here, have essentially the same form as those of the kinetic energy density; however, there are slight deviations at the moment of interaction with the two layer boundaries, and the normalization factor is twice as large.

In addition to the image sequences from Figure 10, the energy content in the model under consideration can also be represented as a function of time. For this purpose, the kinetic energy, the potential energy according to Eq. (14), and the sum of both are added over all grid cells of the numerical model and displayed as a function of time for the period of interest up to $t = 12.08$ µs (Figure 11).

It can be seen that the energy content increases sharply during the duration of the excitation signal (up to $t = 1$ µs) and then remains essentially constant. In the case of kinetic and potential energy, temporary deviations from this plateau are observable, occurring at the time of the primary interaction of the incident wave with the left-hand layer boundary ($t \approx 5.5$ µs), with the right-hand layer boundary ($t \approx 7.7$ μs), and again with the interaction of the wave originally reflected at the right-hand layer boundary with the left-hand layer boundary ($t \approx 10.5$ μs). During these interactions, self-interference occurs between parts of the incident wave and parts of the already reflected wave, causing temporary deviations from the constant plateau. However, a comparison of the kinetic and potential energy curves shows that these deviations occur exactly opposite in phase in both cases, so that the sum of the two vanishes and the remaining plateau demonstrates the energy conservation according to Eq. (13) after removal of external forces (last image in Figure 11).

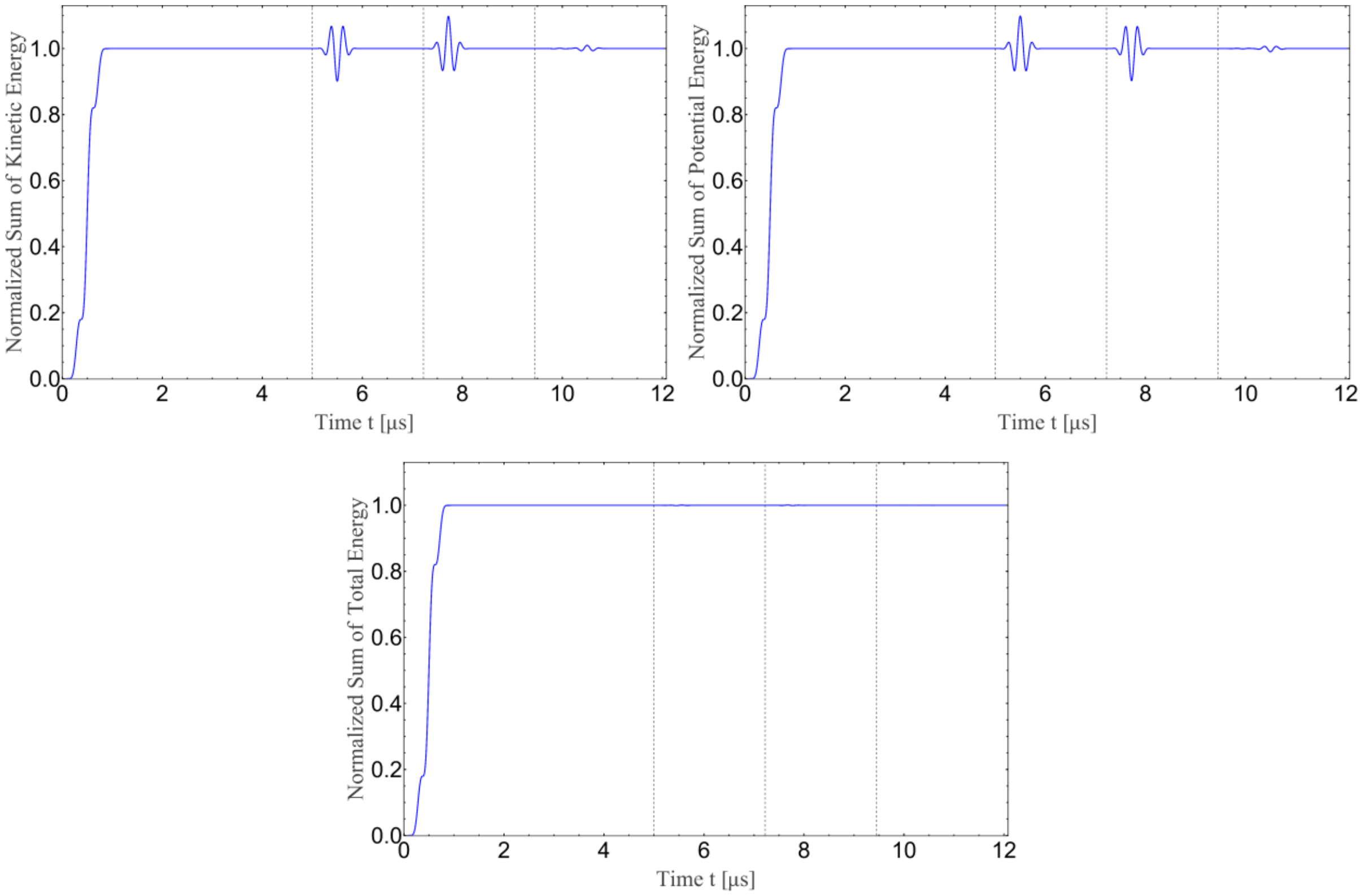

***Figure 11:*** *Normalized representation of kinetic energy (top left), potential energy (top right), and total energy (bottom) present in the model as a function of time. The first two curves are normalized to the final value of the potential energy at* $t$ *= 12.08 μs, and the overall curve to its own maximum. The vertical dashed lines serve as a guide and represent the theoretical arrival times of the signal at the left-hand layer boundary (*$t$ *= 5.0 μs), at the right-hand layer boundary (*$t$ *= 7.22 μs), and again at the left-hand layer boundary (*$t$ *= 9.45 μs). The self-interference effects in the individual balances are caused by the interaction of the finite-length transmitted pulse with the layer boundaries. They vanish in the total energy because they occur in the kinetic component in opposite phase to the potential component.*

## 5.2 Information Balance for the Parameter 'Layer position at constant layer thickness'

The left column of Figure 12 shows the time history of the total information density, consisting of kinetic and potential components, for the parameter 'layer position at constant layer thickness' according to Example 3.1. It can be seen how the information is generated at the left layer boundary ($t = 5.35$ μs), then propagates predominantly in the reverse direction, but also partly in the forward direction ($t = 6.53$ μs). New information is generated at the right layer boundary, but this only occurs in the reverse direction ($t = 7.52$ μs and $t = 8.91$ μs). On the second pass through the left layer boundary, the information in the second reflected wave increases again, but does not quite reach the level of the first reflection from the left layer boundary ($t = 12.08$ μs).

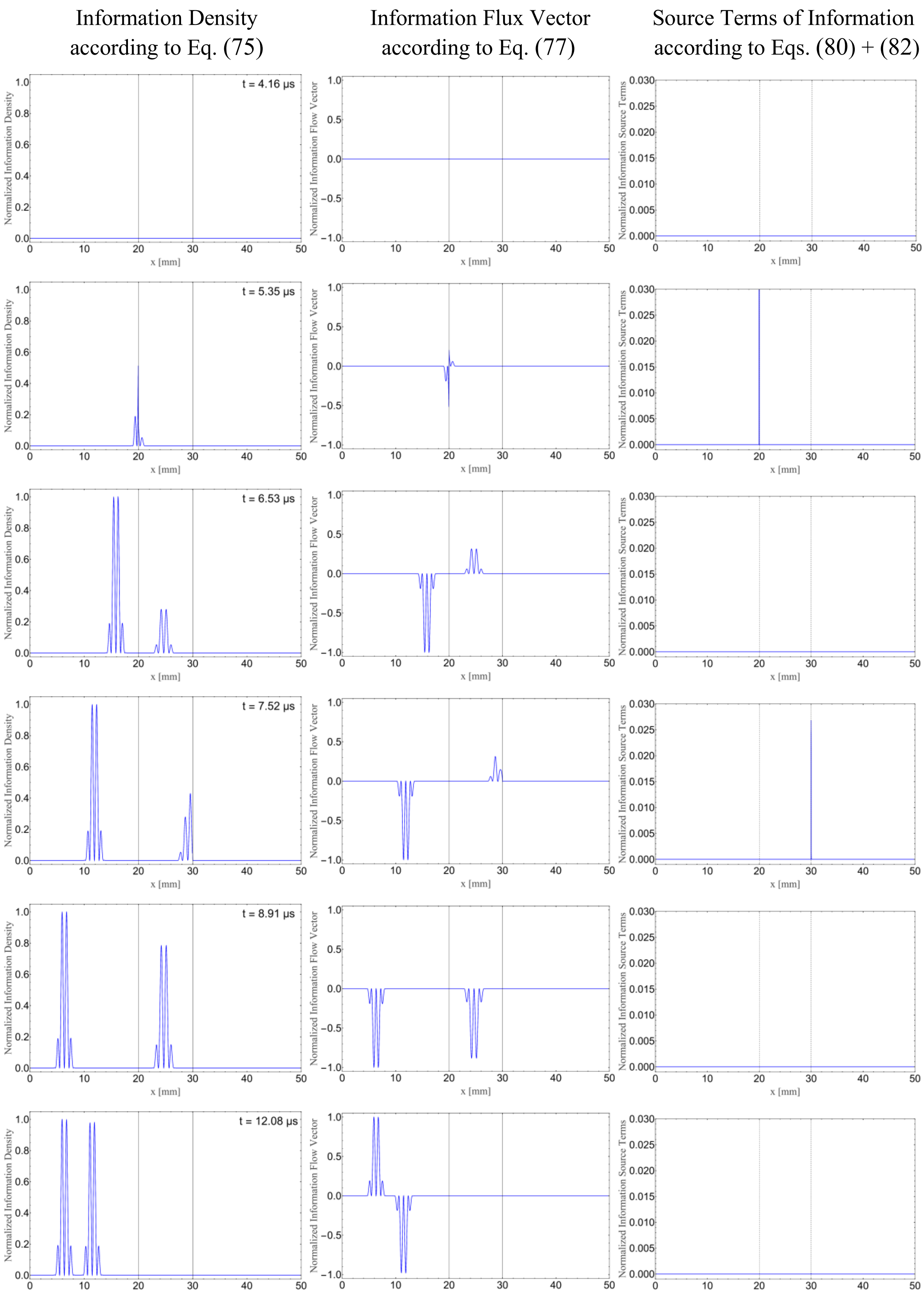

***Figure 12:*** *Normalized representation of the information density (left column), the information flux (middle column) and the cumulative source terms of the information (right column) for the parameter 'layer position at constant layer thickness' at the times* $t$ *= 4.16 µs (1st row),* $t$ *= 5.35 µs (2nd row),* $t$ *= 6.53 µs (3rd row),* $t$ *= 7.52 µs (4th row),* $t$ *= 8.91 µs (5th row) and* $t$ *= 12.08 µs (last row).*

The signed information flux in the middle column of Figure 12 also shows the generation of information, first at the left boundary ($t = 5.35$ µs) and then at the right boundary ($t = 7.52$ µs and $t = 8.91$ µs). In both cases, the respective direction of the information flux (positive sign = flow to the right, negative sign = flow to the left) is correctly represented. This also applies to the reflection of the first layer boundary echo at the left model boundary, where a sign change occurs. Finally, the magnitude of the information flux in both boundary echoes is almost the same ($t = 12.08$ µs). The described behavior confirms the qualitative and quantitative suitability of expression (77) as an information flux quantity, analogous to the Poynting vector of energy flux.

The source terms of information according to equations (80) and (82) in the right-hand column of Figure 12 are only temporarily effective at the two layer boundaries ($t = 5.35$ µs and $t = 7.52$ µs) in this case and are otherwise equal to zero. The amplitude shown at the right layer boundary at $t = 7.52$ µs is in this case only slightly less than 3% of the deflection at the left layer boundary at $t = 5.35$ µs (here normalized to one and overdriven), which, however, is not representative but is merely due to the specific selection of the points of time for the wave front snapshots. A better impression of the dynamic behavior of the source terms is obtained if they are shown as a function of time at the location of the left and right layer boundaries (Figure 13).

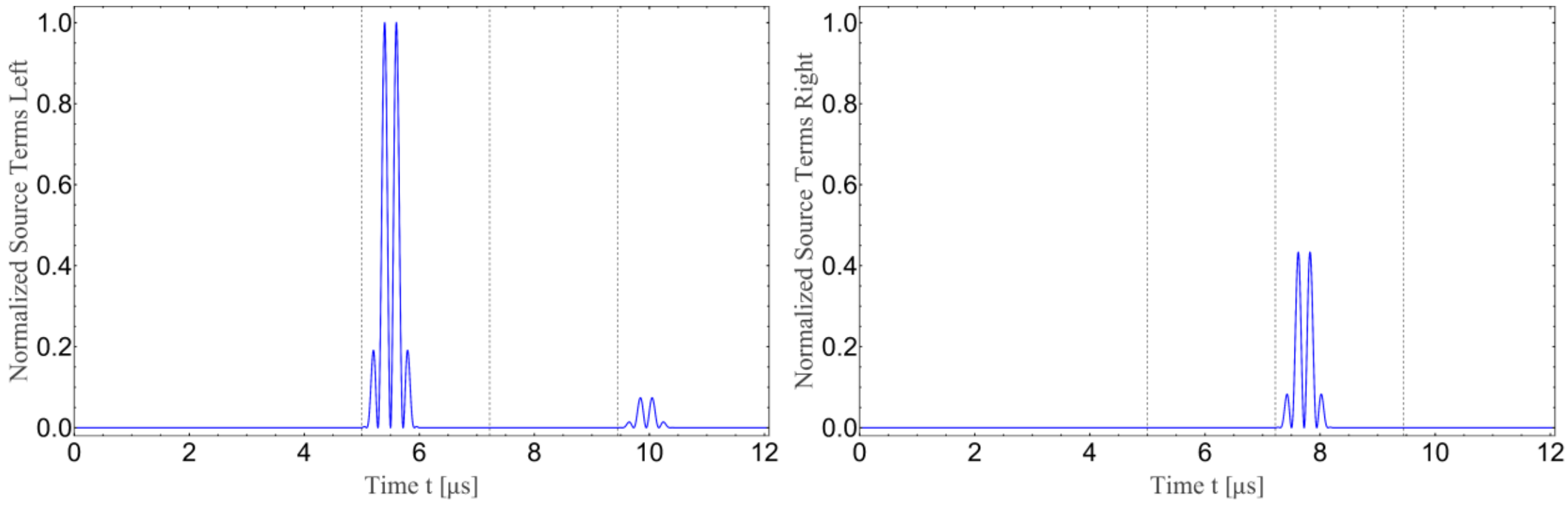

***Figure 13:*** *Normalized time-dependent representation of the source terms appearing in the model at the left-hand (left) and right-hand layer boundary (right). Both curves are normalized to the maximum of the left source term. The vertical dashed lines serve as a guide and represent the theoretical arrival times of the signal at the left-hand layer boundary ($t = 5.00$ µs), at the right-hand layer boundary ($t = 7.22$ µs), and again at the left-hand layer boundary ($t = 9.45$ µs). All source contributions are positive and therefore represent effective sources of information.*

At the left layer boundary, information is generated in two different time intervals: during the primary interaction between $t = 5.0$ µs and $t = 6.0$ µs, and to a significantly lesser extent during the secondary interaction between $t = 9.45$ µs and $t = 10.45$ µs (Figure 13, left). At the right layer boundary, within the time window shown here, there is only one contribution from the primary interaction between $t = 7.22$ µs and $t = 8.22$ µs (image on the right). It can be seen that, cumulatively (primary + secondary contribution), almost 70% of the information is generated at the left layer boundary, while only about 30% is generated at the second layer boundary. All source contributions are consistently positive and therefore represent effective sources of information.

The signal shape of the information quantities in Figure 12 und Figure 13 is noteworthy in comparison to that of the energy quantities in Figure 10. While the latter exhibit a distinct maximum in the center of the signal (where the particle velocity is also at its maximum), the information becomes zero at this point because the difference signals cross zero at this location. As a result, the information quantities exhibit two equally sized, closely spaced maxima on either side of the signal center.

Figure 14 (top left) shows the time history of the kinetic part of information present in the model according to the first term of Eq. (75). Three plateaus above the zero line are clearly visible. The first arises immediately after the interaction with the left layer boundary at $t \approx 5.5$ µs, the second after the interaction with the right layer boundary at $t \approx 7.7$ µs, and the third after the secondary interaction with the left layer boundary at $t \approx 10.5$ µs. The transitions from the first to the second and from the second to the third plateau are generated by the source terms and, similar to the energy analysis in Figure 11, are masked by self-interference effects.

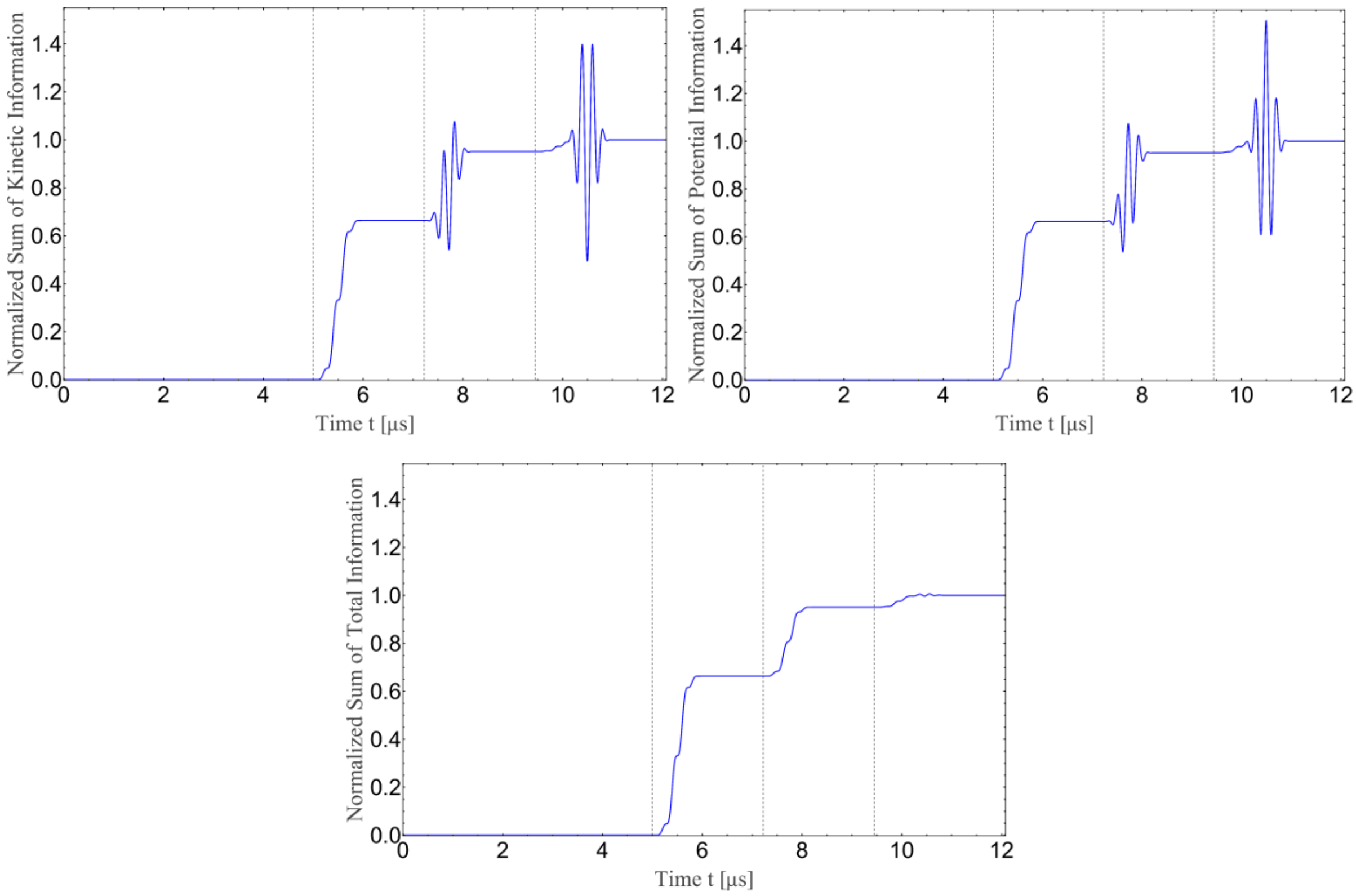

***Figure 14:*** *Representation of the kinetic information (top left), potential information (top right), and total information (bottom) present in the model for the parameter 'layer position at constant thickness' as a function of time. All curves are normalized to their respective final values at* $t$ *= 12.08 µs. The vertical dashed lines serve as orientation and represent the theoretical arrival times of the signal at the left-hand layer boundary (*$t$ *= 5.00 µs), at the right-hand layer boundary (*$t$ *= 7.22 µs), and again at the left-hand layer boundary (*$t$ *= 9.45 µs). The self-interference effects in the individual balances are caused by the interaction of the finitely long transmitted pulse with the layer boundaries. They vanish in the total information because they occur in the kinetic component in opposite phase to the potential component.*

Figure 14 (top right) shows the behavior of the potential part of information according to the second term of Eq. (75). Analogous to the behavior of the energy in Figure 11, the potential component generally shows the same behavior as the kinetic component. Here, too, the self-interference effects are in opposite phase to the kinetic part. As a consequence, they cancel each other out in the total sum of information, so that only the three plateaus and the three risings are recognizable in the overall balance (Figure 14, bottom).

The representation shows that most of the information (almost 70% of the total information) is generated at the first rise, i.e., at the primary interaction with the left layer boundary. The rise from the first to the second plateau (interaction with the second layer boundary) is smaller, and the smallest increase in information is shown at the transition from the second to the third plateau (secondary interaction with the left layer boundary), in accordance with the behavior of the source terms in Figure 13.

Figure 14 also confirms that the total information according to Eq. (75), unlike total energy, is not a conserved quantity, but can change with each new interaction. In the time window shown here up to $t = 12.08$ µs, the information content increases stepwise with each interaction.

### 5.3 Information Balance for the Parameter 'Layer thickness at constant layer position'

The left column of Figure 15 shows the time history of the information density for the parameter 'layer thickness at constant layer position' according to Example 3.2. Similar to the previous case, the information is initially generated at the left layer boundary ($t = 5.35$ µs) and is then transmitted predominantly in the reverse direction, but also partly in the forward direction ($t = 6.53$ µs). At the right layer boundary, new information is generated, which this time occurs not only in the reverse direction but also, and especially, in the forward direction ($t = 7.52$ µs and $t = 8.91$ µs). The transmitted portion of the information exhibits the largest amplitude of all generated contributions. On the second pass through the left layer boundary, the information in the second reflected wave appears to decrease slightly ($t = 12.08$ µs).

The signed information flux in the middle column of Figure 15 also shows the generation of information, first at the left ($t = 5.35$ µs) and then at the right boundary ($t = 7.52$ µs and $t = 8.91$ µs). In both cases, the respective direction of the information flux (positive sign = flow to the right, negative sign = flow to the left) is correctly represented. This also applies to the reflection of the first boundary layer echo at the left model boundary, where a sign change occurs. Finally, the magnitude of the information flux in both boundary layer echoes is different ($t = 12.08$ µs).

The source terms of information according to Eqs. (80) + (82) in the right column of Figure 15 are again only effective at the two layer boundaries and are otherwise zero. Their time-dependent behavior is shown in Figure 16. Again, the two contributions from the primary interaction with the left and right layer boundaries are visible, but this time the source term from the right layer boundary dominates (right-hand picture). Interestingly, the secondary contribution from the left layer boundary exhibits *negative* amplitudes in the time interval between $t = 9.45$ µs and $t = 10.45$ µs (left-hand picture). It thus represents an effective *sink* of information in the overall system!

Figure 17 (top left) shows the time history of the kinetic part of information present in the model according to the first term of Eq. (75). Three plateaus above the zero line are again visible. The first arises immediately after the interaction with the left layer boundary at $t \approx 5.5$ µs, the second after the interaction with the right layer boundary at $t \approx 7.7$ µs, and the third after the secondary interaction with the left layer boundary at $t \approx 10.5$ µs. The transitions from the first to the second and from the second to the third plateau are, as in the previous example, masked by self-interference effects.

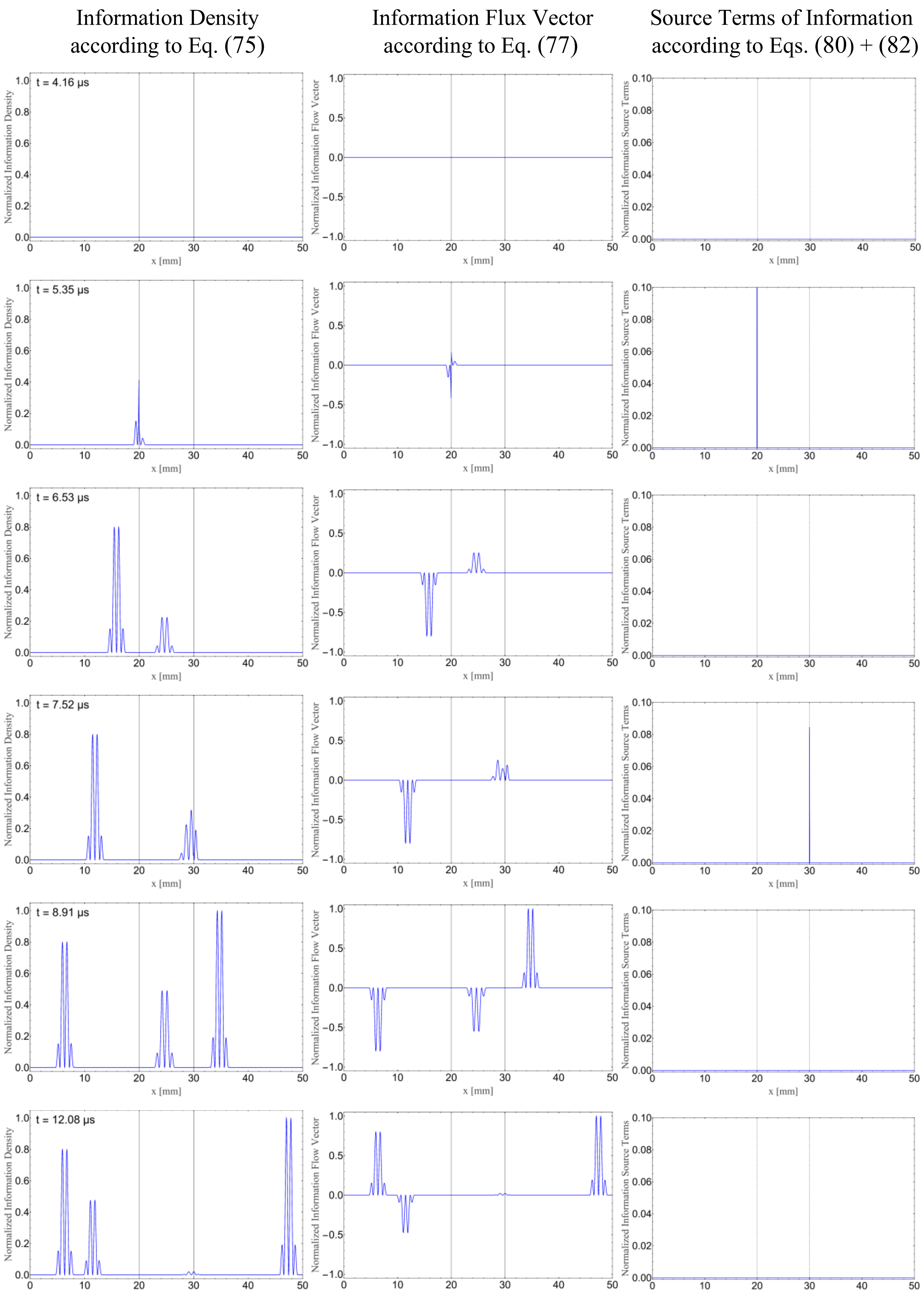

***Figure 15:*** *Representation of the information density (left column), the information flux (middle column) and the cumulative source terms of the information (right column) for the parameter 'layer thickness at constant layer position' at the times* $t$ *= 4.16 µs (1st row),* $t$ *= 5.35 µs (2nd row),* $t$ *= 6.53 µs (3rd row),* $t$ *= 7.52 µs (4th row),* $t$ *= 8.91 µs (5th row) and* $t$ *= 12.08 µs (last row).*

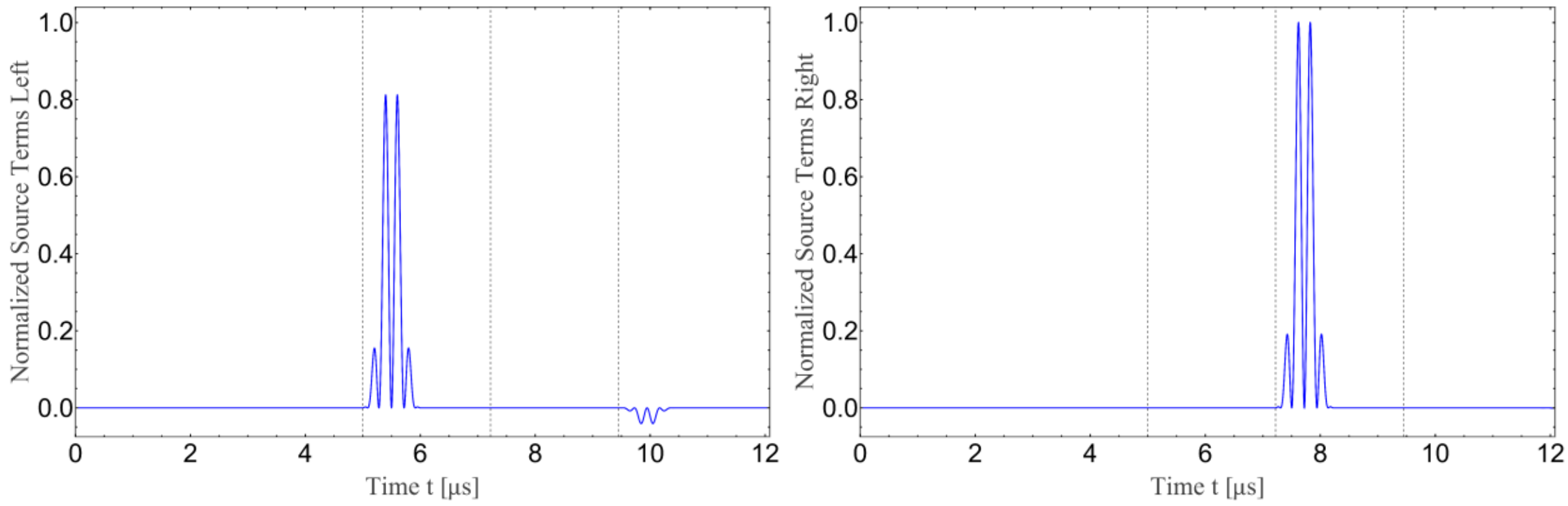

***Figure 16:*** *Normalized time-dependent representation of the source terms appearing in the model at the left-hand (left) and right-hand layer boundary (right). Both curves are normalized to the maximum of the right source term. The vertical dashed lines serve as a guide and represent the theoretical arrival times of the signal at the left-hand layer boundary ($t$ = 5.00 µs), at the right-hand layer boundary ($t$ = 7.22 µs), and again at the left-hand layer boundary ($t$ = 9.45 µs). The second source contribution at the left layer boundary between $t$ = 9.45 µs and $t$ = 10.45 µs is negative and therefore represents an effective sink of information.*

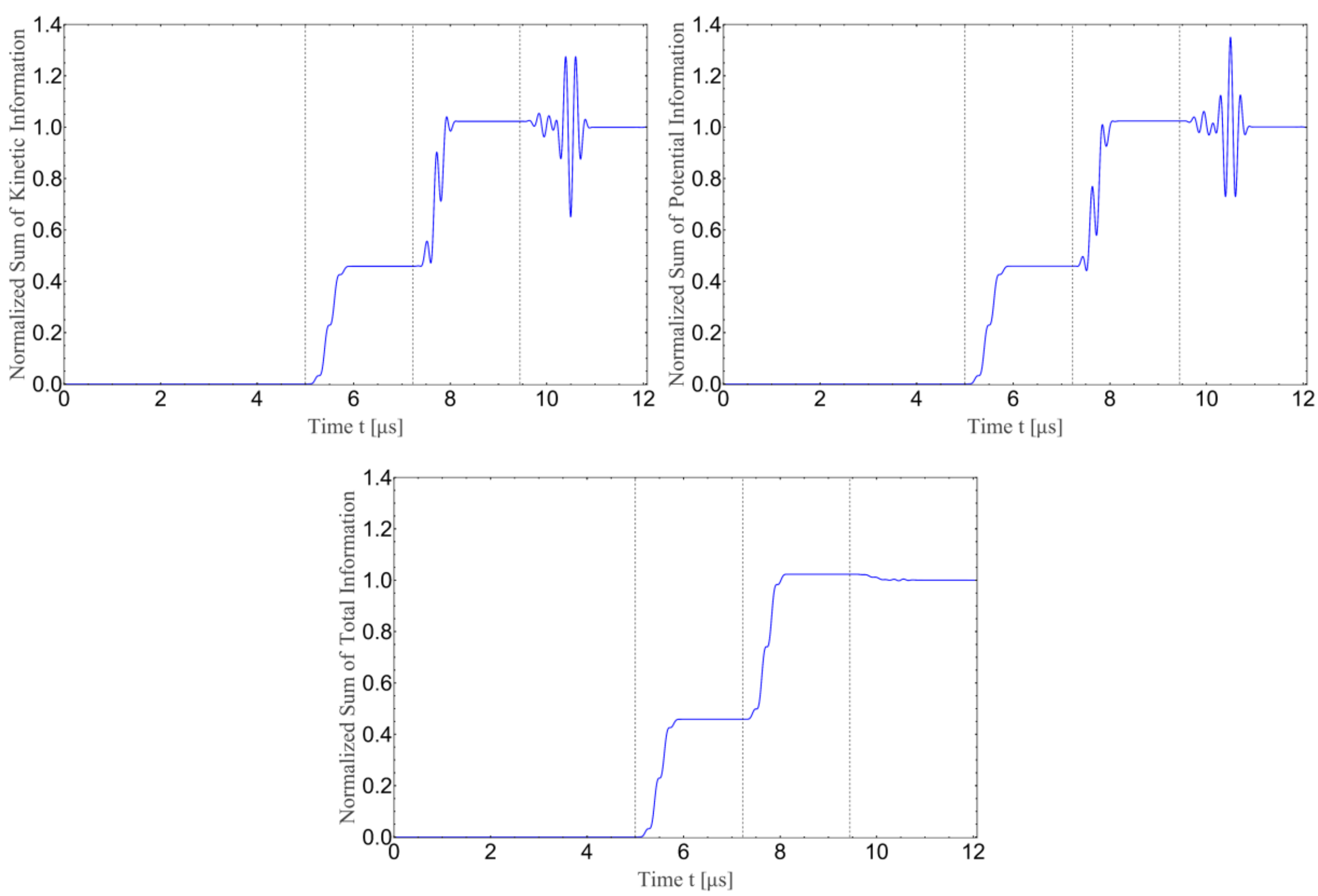

***Figure 17:*** *Representation of the kinetic information (top left), potential information (top right), and total information (bottom) present in the model for the parameter 'layer thickness at constant layer position' as a function of time. All curves are normalized to their respective final values at $t$ = 12.08 µs. The vertical dashed lines serve as orientation and represent the theoretical arrival times of the signal at the left-hand layer boundary ($t$ = 5.00 µs), at the right-hand layer boundary ($t$ = 7.22 µs), and again at the left-hand layer boundary ($t$ = 9.45 µs). The self-interference effects in the individual balances are caused by the interaction of the finite-length transmitted pulse with the layer boundaries. They vanish in the total information because they occur in the kinetic component in opposite phase to the potential component.*

Figure 17 (top right) shows the time history of the potential part of information according to the second term of Eq. (75). Analogous to the previous example, the potential part basically shows the same course as the kinetic part, but here too the self-interference effects occur in opposite phase. As a consequence, the latter cancel each other out in the overall summation, so that only the three plateaus and the three transitions are visible in the overall balance (Figure *17*, bottom). In contrast to the first example, most of the information here is generated only at the second rise, i.e., at the primary interaction with the right-hand layer boundary. The rise to the first plateau (interaction with the first layer boundary) is somewhat smaller.

Particularly noteworthy is that the transition from the second to the third plateau (secondary interaction with the left layer boundary) leads to a *lower* information content. Here, information is evidently annihilated again in the model, which confirms the assumption made above in connection with Figure 16 that, in addition to sources, there can also be global *sinks* of information, even though no dissipative physical effects are inherent to the model. Since there is initially no information in the system at $t = 0$, this also means that there is generally always a maximum of information within a chosen time window, i.e., a time range in which the information content of the system with respect to the parameter $P$ is maximized. In the present example, this is the case for the plateau between approximately $t = 8$ µs and $t = 10$ µs.

## 5.4 Information Balance for the Parameter 'Speed of sound in the layer at constant layer density'

The left column of Figure 18 shows the time history of information density for the parameter 'speed of sound in the layer at constant layer density' according to Example 3.3. Unlike the previous cases, the information here is not only generated at the layer boundaries, but gradually as the sound propagates through the layer ($t = 6.53$ µs and $t = 7.52$ µs). It therefore represents a volume source (here in 1-D). The information content increases continuously from the left to the right layer boundary and reaches its maximum only when passing the right layer boundary ($t = 8.91$ µs and $t = 12.08$ µs). The majority of information is thus carried by the forward wave in this case. In the direction of reflection, only a small secondary contribution is visible in this representation, which initially arises at the right layer boundary and then propagates back through the layer. Since the speed of sound in the layer, and thus its acoustic impedance, is changed in this example, the question arises as to why information from the left layer boundary is not also reflected backwards. Indeed, this contribution is present in the signals (see inset at $t = 6.53$ µs). However, as a source term from the individual interface, it is much smaller than the information contributions from the volume of the layer, which dominate here.

The middle column of Figure 18 describes the temporal evolution of the information flux. Here, too, continuously increasing contributions appear only as the wave traverses the layer, reaching their maximum after passing the right-hand layer boundary. The signs of the flux vector for the transmitted (positive sign) and reflected waves (negative sign) are correctly represented and correspond to the directions of propagation of the associated information waves in the left-hand column.

In contrast to the previous examples, the single source term in the right-hand column of Figure 18 does not show abrupt jumps at the two layer boundaries, but instead continuously increasing contributions with a positive sign, entirely consistent with the behavior of the information density and the information flux. The occurrence of the source term across the entire layer thickness clarifies that it represents indeed a volume source within the layer.

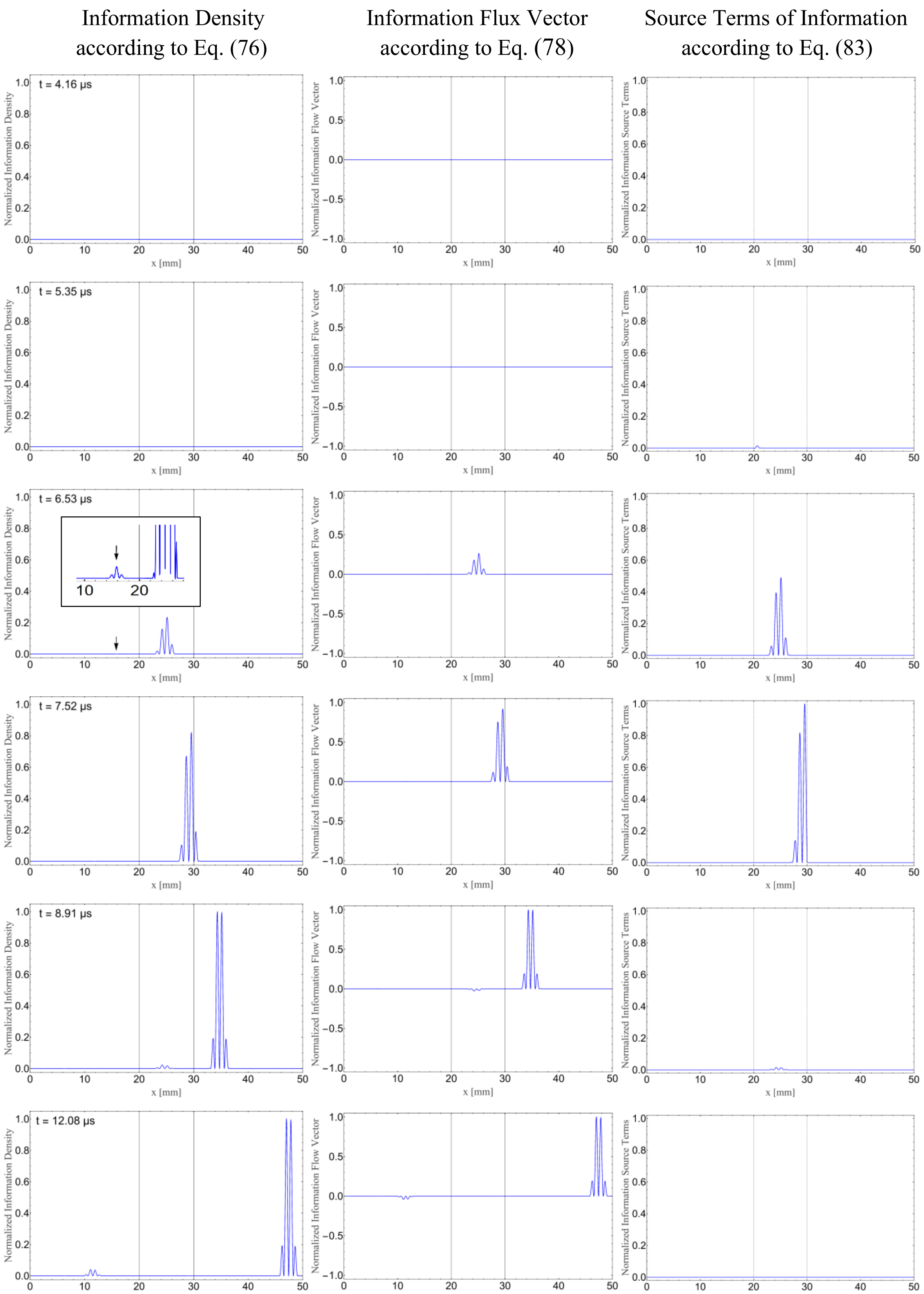

***Figure 18:*** *Representation of the information density (left column), the information flux (middle column) and the cumulative source terms of the information (right column) for the parameter 'layer sound velocity at constant layer density' at the times* $t$ *= 4.16 μs (1st row),* $t$ *= 5.35 μs (2nd row),* $t$ *= 6.53 μs (3rd row),* $t$ *= 7.52 μs (4th row),* $t$ *= 8.91 μs (5th row) and* $t$ *= 12.08 μs (last row).*

Figure 19 shows the time history of the source term at both layer boundaries. The maximum is reached at the right-hand layer boundary after the information within the layer has accumulated to its maximum (right-hand picture). The second source contribution at the left-hand layer boundary between $t = 9.45$ µs and $t = 10.45$ µs, generated by the wave reflected at the right-hand layer boundary and traveling back through the layer, also exhibits consistently positive amplitudes. However, these are significantly smaller than those of the primary transmitted signal at the right-hand layer boundary. In contrast to all previously observed contributions, the first source contribution at the left-hand layer boundary exhibits signed behavior (see left-hand picture). The positive amplitudes are slightly larger in magnitude than the negative values over the entire signal length, resulting in a small net source of information. This also includes the reflection discussed in Figure 18 at $t = 6.53$ µs due to the small impedance step at the left-hand layer boundary.

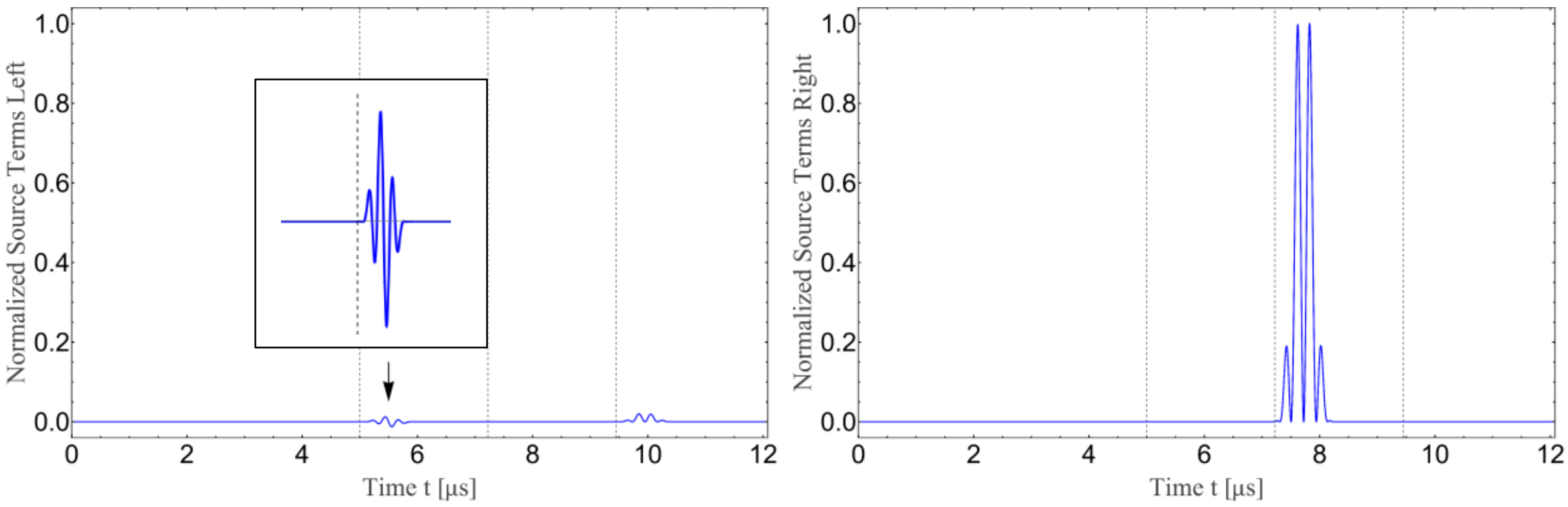

***Figure 19:*** *Normalized time-dependent representation of the source term appearing in the model at the left-hand (left) and right-hand (right) layer boundary. Both curves are normalized to the maximum of the right-hand source term. The vertical dashed lines serve as a guide and represent the theoretical arrival times of the signal at the left-hand layer boundary ($t$ = 5.00 µs), at the right-hand layer boundary ($t$ = 7.22 µs), and again at the left-hand layer boundary ($t$ = 9.45 µs). The initial source contribution at the left-hand layer boundary is signed, with positive amplitudes slightly exceeding the negative ones over the entire signal length, leaving a small net source of information.*

Figure 20 shows the temporal evolution of the information present in the model, again divided into the kinetic component (top left), the potential component (top right), and the total information (bottom). Instead of the steep jumps from the previous examples, a longer, continuous increase can be seen during the primary passage through the layer up to approximately $t = 8$ µs. After reaching the right-hand layer boundary, there is a second, significantly smaller increase between about $t = 8$ µs and $t = 10$ µs, which results from the secondary passage of the reflected wave through the layer. From about $t = 10$ µs onward, the information in the model reaches its maximum in this time window. As in the previous examples, a self-interference effect also occurs in this case, this time shortly after the interaction with the right-hand layer boundary. Here, too, the kinetic and potential components exhibit opposite phase behavior, so that after summation, a smooth, monotonic curve is obtained for the total information (bottom of the figure).

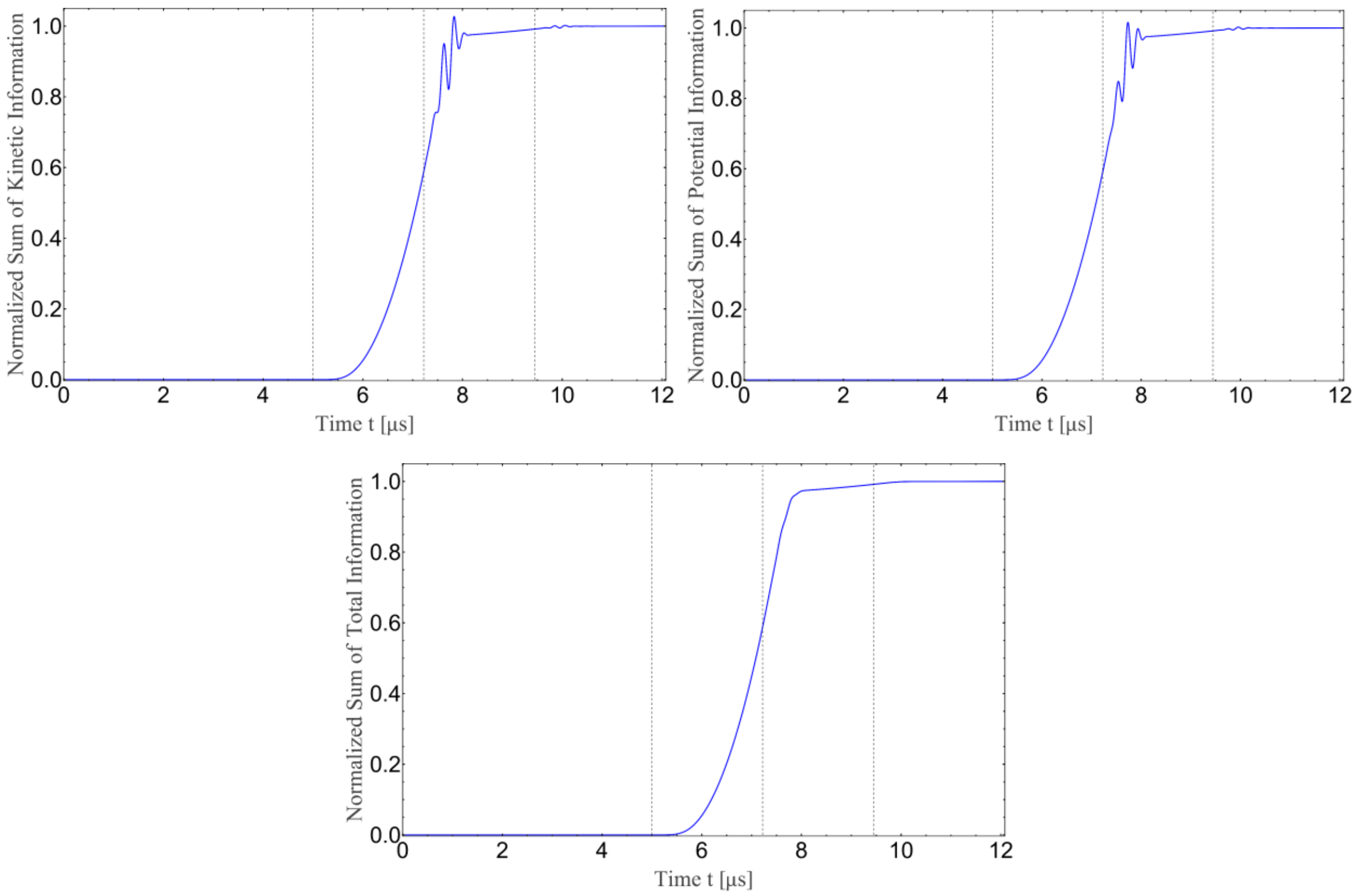

***Figure 20:*** *Representation of the kinetic information (top left), potential information (top right), and total information (bottom) present in the model for the parameter 'layer sound velocity at constant layer density' as a function of time. All curves are normalized to their respective final values at $t$ = 12.08 µs. The vertical dashed lines serve as orientation and represent the theoretical arrival times of the signal at the left-hand layer boundary ($t$ = 5.00 µs), at the right-hand layer boundary ($t$ = 7.22 µs), and again at the left-hand layer boundary ($t$ = 9.45 µs). The self-interference effects in the individual balances are caused by the interaction of the finite-length transmitted pulse with the right-hand layer boundary. They vanish in the total information because they occur in the kinetic component in opposite phase to the potential component.*

## 5.5 Information Balance for the Parameter 'Layer Density at constant layer sound velocity'

Figure 21 shows the time history of the three information quantities for the parameter 'layer density at constant speed of sound' according to Example 3.4. The first column illustrates that the information is generated at the first layer boundary and then propagates both forwards and backwards ($t$ = 5.35 µs and $t$ = 6.53 µs). In contrast to the speed of sound variation from Section 5.4, however, there is no continuous information generation as the wave traverses the layer. The information wave in the layer maintains its amplitude until it interacts with the right-hand layer boundary. There, the largest portion of the incoming wave is reflected, while only a small part remains in the forward direction ($t$ = 7.52 µs and $t$ = 8.91 µs).

As the wave reflected at the right-hand layer boundary passes through the left-hand layer boundary, its maximum amplitude increases. However, the width of the transmitted pulse also decreases. Additionally, by reflection another small information density wave in the forward direction is generated ($t$ = 12.08 µs).

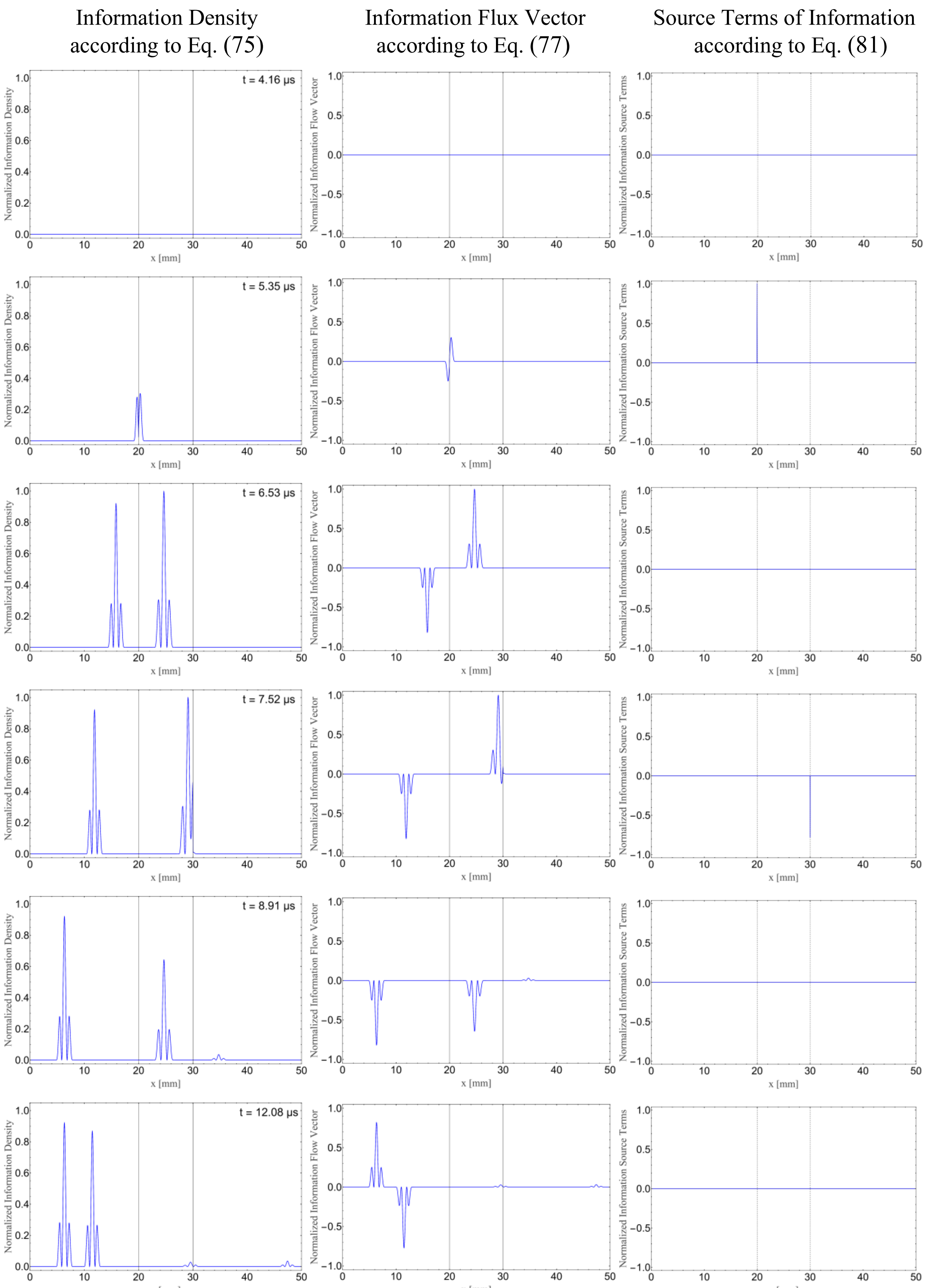

***Figure 21:*** *Representation of the information density (left column), the information flux (middle column) and the cumulative source terms of the information (right column) for the parameter 'layer density at constant speed of sound' at the times* $t$ *= 4.16 µs (1st row),* $t$ *= 5.35 µs (2nd row),* $t$ *= 6.53 µs (3rd row),* $t$ *= 7.52 µs (4th row),* $t$ *= 8.91 µs (5th row) and* $t$ *= 12.08 µs (last row).*

The information flux vector in the middle column of Figure 21 exhibits the same fundamental behavior as the information density, with the signs of the waves also correctly indicating their respective directions of propagation (positive sign = direction of propagation to the right, negative sign = direction of propagation to the left).

The cumulative source terms of information in the right-hand column of Figure 21 show only contributions at the two layer boundaries, very similar to the case of the geometric parameters from Sections 5.2 and 5.3. Their time history, shown in Figure 22, reveals two temporally separated, positive contributions from the left-hand layer boundary and one negative contribution from the right-hand layer boundary.

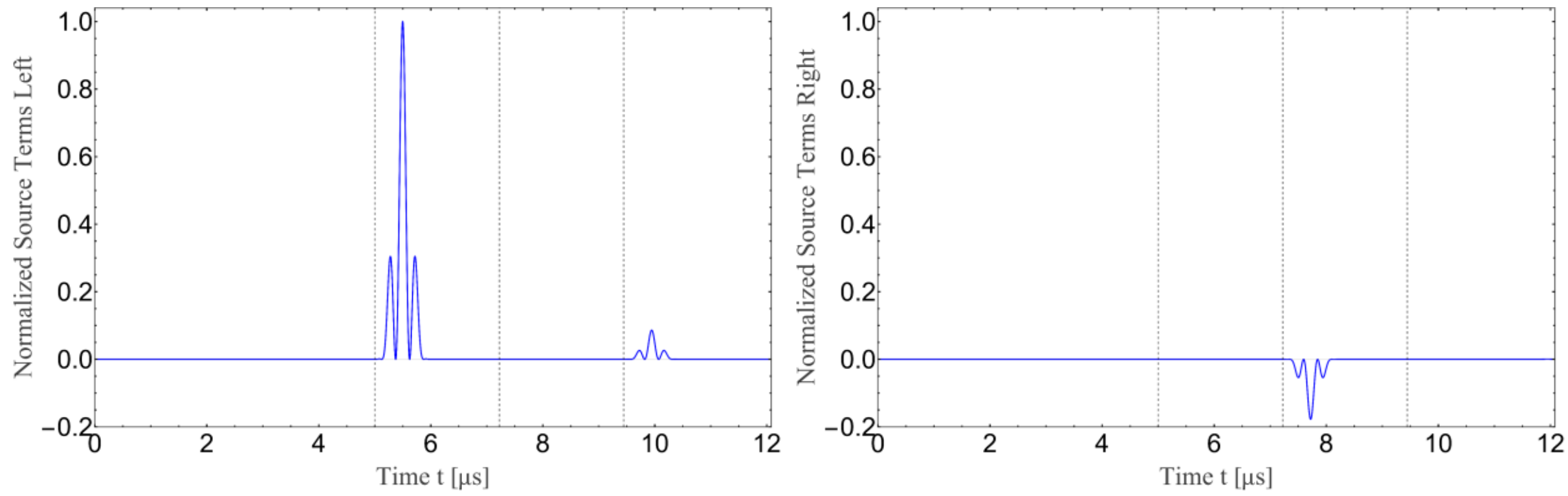

***Figure 22:*** *Normalized time-dependent representation of the source term appearing in the model at the left-hand (left) and right-hand layer boundary (right). Both curves are normalized to the maximum of the left-hand source term. The vertical dashed lines serve as a guide and represent the theoretical arrival times of the signal at the left-hand layer boundary ($t$ = 5.00 µs), at the right-hand layer boundary ($t$ = 7.22 µs), and again at the left-hand layer boundary ($t$ = 9.45 µs). The source contribution at the right-hand layer boundary is negative, thus this layer boundary acts as a sink of information.*

Figure 23 again shows the temporal evolution of the information present in the model, divided into the kinetic (top left), potential (top right), and total components (bottom). The phenomenon of self-interference effects can again be observed; these occur in opposite phase and disappear after summation. The drop in information from the first to the second plateau is clearly visible, mediated by the negative source term at the right-hand boundary of the layer.

What is remarkable about this last example is that all information quantities exhibit a maximum in the center of the signal, in contrast to the previous examples, where a minimum was present at the same location. These observations are in complete agreement with the difference wave fields in Figures 5 and 8, where a similar behavior can be observed.

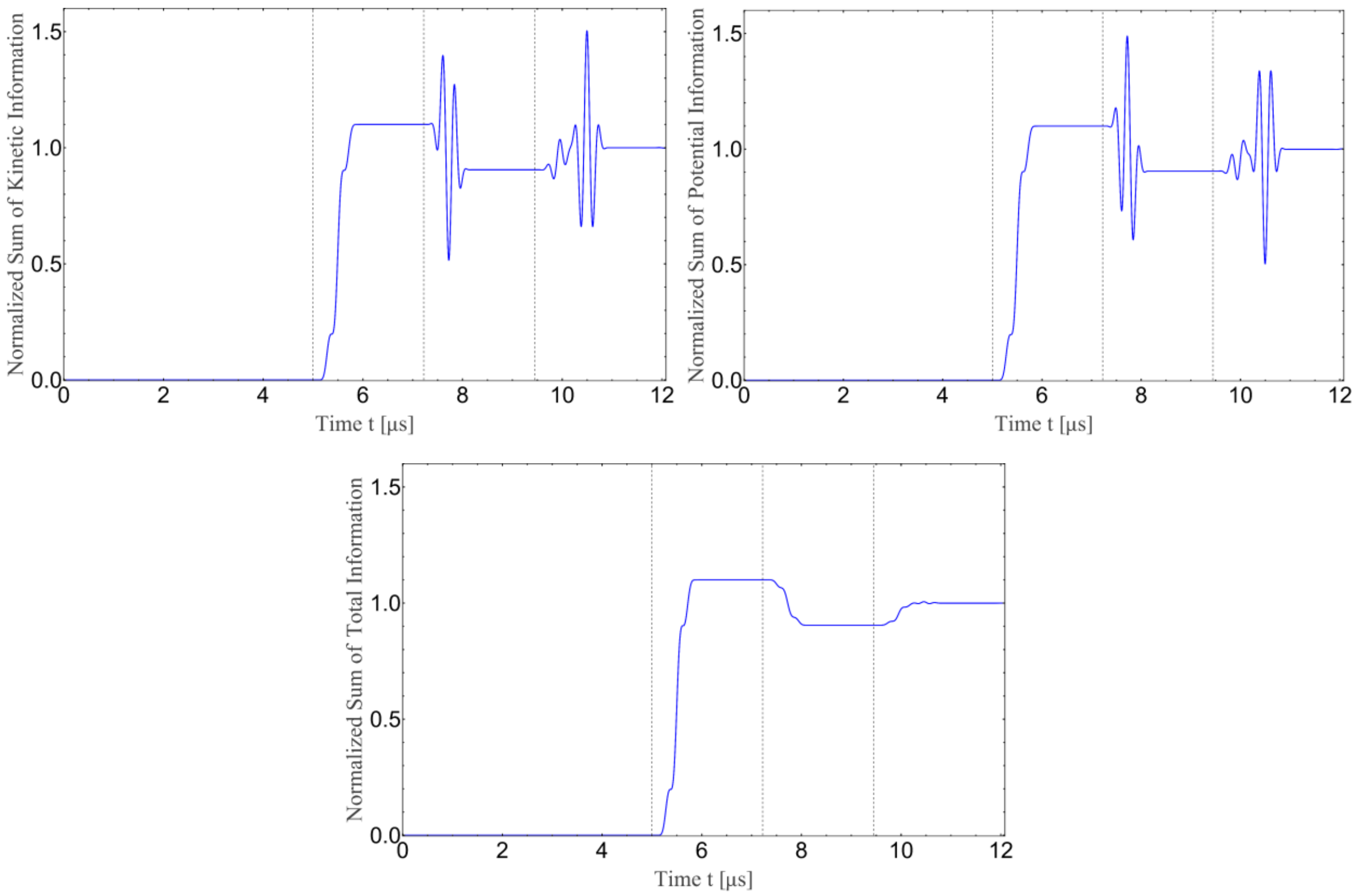

***Figure 23:*** *Representation of the kinetic information (top left), potential information (top right), and total information (bottom) present in the model for the parameter 'density of the layer at constant speed of sound' as a function of time. All curves are normalized to their respective final values at $t$ = 12.08 µs. The vertical dashed lines serve as orientation and represent the theoretical arrival times of the signal at the left-hand layer boundary ($t$ = 5.00 µs), at the right-hand layer boundary ($t$ = 7.22 µs), and again at the left-hand layer boundary ($t$ = 9.45 µs). The self-interference effects in the individual curves are caused by the interaction of the finite-length transmitted pulse with the two layer boundaries. They vanish in the total information because they occur in the kinetic component in opposite phase to the potential component.*

# 6. Quantitative Information at the Sensor and the Information Unit ‚Change Bit' (Cbit)

Having focused so far on the generation, annihilation, and propagation of information within the model, we now turn to the part of the information that is most readily accessible through measurement, i.e. the information signals at the two sensor positions on the left-hand and right-hand model boundaries. Since we assume stress-free boundary conditions in our simulation and abstain from an explicit transducer model with direct coupling, we detect the pure particle velocity signal, as it would be recorded, for example, by a contactless laser vibrometer. This has the advantage that all stress components at the boundary vanish, and the information can be described solely by the kinetic term in Eq. (76).

Figure 24 shows the time history of the information at the left-hand (reflection) and right-hand model boundaries (transmission) for all four presented examples. Here, we have increased the simulation time window from $t$ = 12.08 µs to $t$ = 20.0 µs to capture all relevant signal contributions. The information signals are compared to the time signals of the initial model (first row). The relevant time windows for the excitation signal ($t$ = 0 – 1 µs), the reflections at both layer boundaries ($t$ = 10 – 11 µs and $t$ = 14.44 – 15.44 µs, respectively), the transmitted signal ($t$ = 12.22 – 13.22 μs), and the first multiple echo from the layer ($t$ = 18.9 – 19.9 μs and $t$ = 16.67 – 17.67 μs, respectively) are each highlighted by vertical dotted lines.

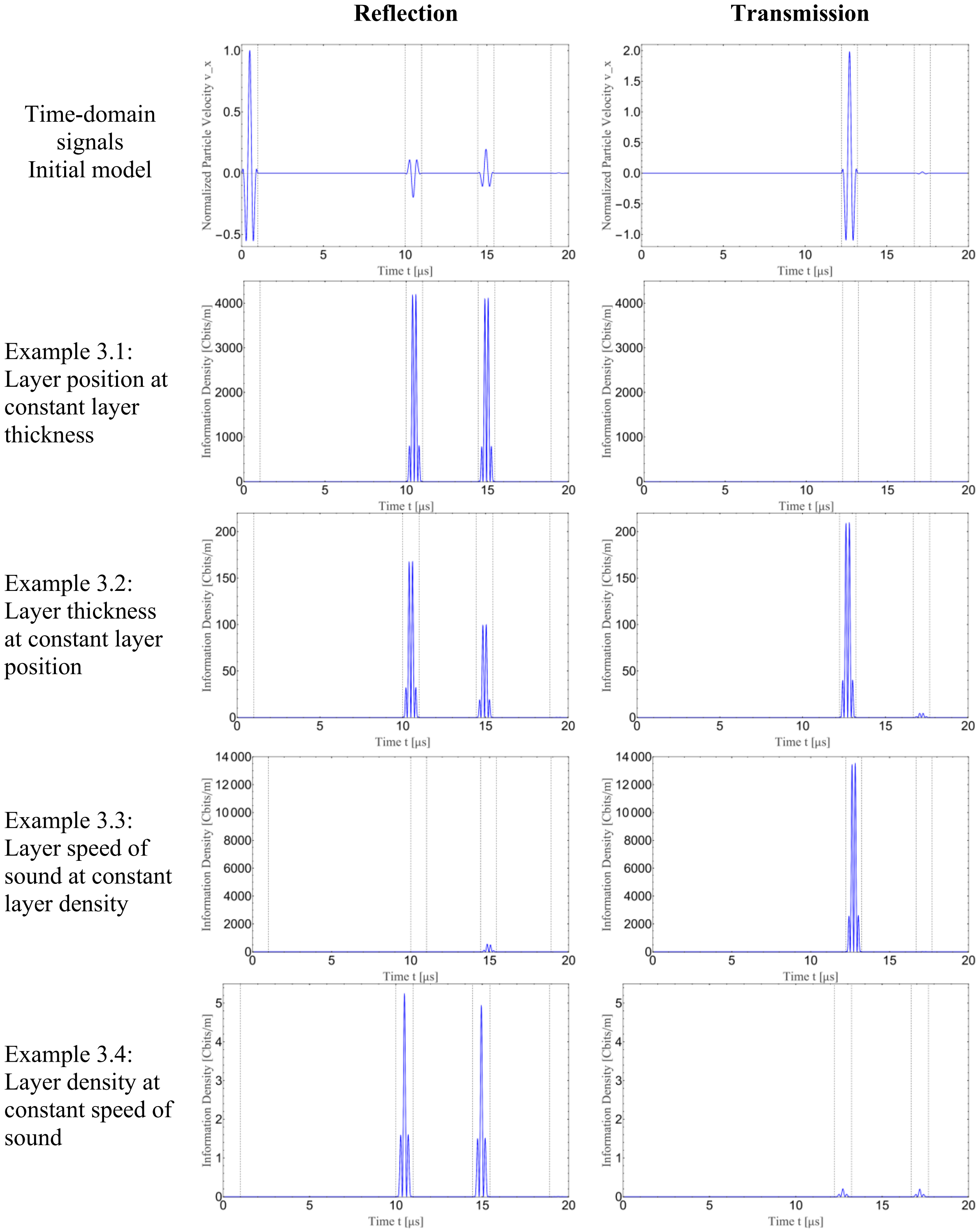

***Figure 24:*** *Time histories of information densities calculated at both sensor positions for the four examples described in sections 3.1–3.4. For reference, the two time-domain signals in the initial model are also shown (top row). All information densities are given in absolute information units in Cbits/m.*

In contrast to all previous representations, in which we depicted information on a normalized scale, Figure 24 makes use of one of the greatest advantages of the information theory presented here: the existence of an absolute information scale. This allows for a direct quantitative comparison between physically different parameters and different measurement setups, such as reflection and transmission. As we explained above, information is a physically dimensionless quantity. Nevertheless, it makes sense to introduce a base unit for information in analogy to the *bits* and *bytes* of Shannon's information theory. We refer to this fundamental unit of structural information as the 'Change Bit', or *Cbit* for short, because it represents a measure of the (absolute) change of the wave field for a given (relative) parameter change. Other suitable terms would be 'Structural Bit' or 'Parameter Bit'.

All information densities in Figure 24 are given in Cbits/m in the present 1D model (usually Cbits/m³ in 3-D), allowing for quantitative comparisons. The decisive factor is not so much the maximum of the signals, but rather the integral over all information contributions occurring within the time window of interest, here up to $t = 20$ µs.

For Example 3.1 (layer position at constant layer thickness), 100% of the information occurs on the reflection side, since no information about the layer position is encoded in the transmitted signal. A maximum peak of 4184.2 Cbits/m is recorded in the reflected signal; the integral over both reflections comprises 102.41 Cbits/m, based on (or multiplied by) the 20 µs time window.

For Example 3.2 (layer thickness at constant layer position), the maximum amplitude is indeed on the transmission side (208.94 Cbits/m compared to 167.37 Cbits/m in reflection), but the integral of the information across both reflections comprises a total of 3.2991 Cbits/m compared to only 2.6356 Cbits/m on the transmission side. Thus, 55.6% of the information is provided in the reflected signal and 44.4% in the transmitted signal. Overall, the integrated value from both signals is 5.9347 Cbits/m, which is less than 6% of the total information from Example 3.1. Therefore, with the same relative parameter change, the initial model examined here generates approximately 17 times more information in the case of a change in position than in the case of a change in thickness. This difference becomes clear when we look again at the corresponding difference wave fields in Figure 5. For example, the maximum amplitudes of the first layer echo were practically identical in both examples (approximately 1.6% of the initial signal). In Example 3.1, however, this was caused by a relative parameter change of $2\Delta x/25$ mm, while in Example 3.2, the relative change was five times larger, at $4\Delta x/10$ mm and $10\Delta x/25$ mm, respectively. Squaring the factor of 5 yields a value 25 times larger in terms of information, which corresponds almost exactly to the observed ratio of the maximum amplitudes of 4184.2 Cbits/m (Example 3.1) to 167.37 Cbits/m (Example 3.2) for the first reflection.

In Example 3.3 (speed of sound in the layer at constant density), 96.2% of the information is provided in the transmitted signal and only 3.8% in the reflected signal. This results in integrated information values of 166.48 Cbits/m in transmission and 6.5552 Cbits/m in reflection, cumulatively 173.04 Cbits/m. The maximum peaks are 13491.0 Cbits/m for the transmitted signal and 550.85 Cbits/m for the reflected signal. Example 3.3 thus generates the most information of all four examples, which is primarily due to the fact that the source present here is effective not only at the layer boundaries but also throughout its volume.

For the last example 3.4 (density of the layer at constant speed of sound), the lowest information values of all the examples are obtained. The integrated values are 0.0953 Cbits/m (96.3%) for the reflection side and 0.0037 Cbits/m (3.7%) for the transmission side, cumulatively 0.099

Cbits/m. The maximum values amount to 5.2304 Cbits/m for reflection and 0.203 Cbits/m for transmission.

Across all four examples, a total of 112.36 Cbits/m are generated on the reflection side, compared to 169.12 Cbits/m on the transmission side. This result underscores the well-known fact that both measurement setups have their place. While transmission setups are well-suited for thickness and sound velocity measurements, reflection methods are suitable for density and position changes, as well as thickness changes. The latter are apparently more difficult to measure than position changes. Density changes are much harder to extract than sound velocity changes and exhibit the lowest information output of all four examples. If the integrated information for the density change is normalized to one, the resulting multipliers are 60 for thickness changes, 1000 for position changes, and 1750 for sound velocity changes. The Cbit values thus span a range of more than three orders of magnitude. It should be emphasized that these values are only valid for the specific configuration of the initial model presented here and cannot be generalized. Since the square of the parameter $P$ under consideration appears as a scaling factor in all information-related quantities, the specific choice of $P$ has a significant influence on the absolute Cbit values.

# 7. Significance of Information Measure for NDE, SHM, ML and Numerical Simulation

The information concept presented in this work is so universal and fundamental that it could have a significant impact on inspection techniques, evaluation methods, and numerical simulations. The most important aspects are briefly discussed below.

## 7.1 Nondestructive Evaluation

Naturally, not all parameter changes implemented in this work using numerical simulations can be reproduced in practical testing. Nevertheless, there are several opportunities to modify individual parameters experimentally. For example, artificially introduced defects in test specimens, such as notches or cylindrical boreholes, could be gradually enlarged, and the resulting information extracted from the difference wave fields. Changes in the position of defects relative to the probes can also be achieved through small adjustments to the probe positions. Local temperature changes of inclusions could modify their elastic properties, thus enabling material-related parameter changes as well.

Furthermore, it is expected that new evaluation methods and imaging algorithms will be developed in the future that directly utilize difference wave fields instead of absolute wave fields. Since the difference signals encode pure information regarding individual parameters, more customized and accurate evaluation and imaging methods are conceivable.

## 7.2 Structural Health Monitoring

SHM has long used difference wave fields and difference signals to evaluate system changes relative to an existing baseline. The information concept presented in this work changes the perspective on differential fields and enables a more targeted approach to extracting individual parameters. Extending the theory to several parameters that change simultaneously and independently may offer the possibility of targeted separation of various superimposed influencing factors, such as crack length and temperature-dependent changes in sound velocity within the matrix medium.

## 7.3 Numerical Simulation

Numerical simulation is the method of choice for implementing a wide variety of parameter changes in a model, even if these are difficult or impossible to achieve experimentally. It can therefore be assumed that the role of numerical simulations in NDE and SHM will change significantly. Instead of calculating a series of absolute wave fields as before, we will likely see more applications that focus on difference wave fields and the information they carry. This will allow measurement setups to be developed and specifically optimized to extract individual system parameters with maximum information-theoretic efficiency.

## 7.4 Machine Learning

As we have already shown elsewhere in the text, difference wave fields carry individual information about the influence of each individual system parameter. While absolute wave fields are almost identical for small parameter changes, each difference field with respect to a specific parameter differs significantly from the difference fields of the other parameters. Therefore, difference wave fields appear to be much better suited for classification tasks than absolute wave fields.

An even more crucial aspect of the new information measure for machine learning concerns the interplay between parameter space and feature space (Figure 25). Every existing system configuration can be viewed as a single point $\underline{P}_1$ in a generally high-dimensional parameter or configuration space, which completely determines the system by means of geometric and material-related parameters. Each individual ultrasound measurement on the system links this single point in parameter space to a time signal or A-scan with $N$ samples. The time signal can therefore be viewed as an $N$-dimensional vector or as a point $\underline{v}_1$ in an $N$-dimensional raw data feature space. If we now perform a small but finite variation in the parameter space to a second point $\underline{P}_2 = \underline{P}_1 + \underline{\Delta P}$, the subsequent measurement on the varied system yields a second time signal and thus a second vector $\underline{v}_2$ in feature space. The Euclidean distance between the two vectors is $| \underline{v}_2 - \underline{v}_1 |$, i.e. the magnitude of the difference signal.

The greater the distance between the two time signals in the feature space for a given parameter change $\underline{\Delta P}$, the more information about the effect of parameter $P$ (here $P = P_j$) is stored in the time signal for the configuration $\underline{P}_1 + ½ \, \underline{\Delta P}$ (the midpoint between P1 and P2). The information concept presented in this work thus allows the purely intuitive notion that a larger distance between signals in feature space enables better classification or a better separation of classes to be replaced by a quantitative, information-theoretic argument. With this insight, it becomes possible to assign individual information for each parameter to each point in parameter space, and thus a weighting factor, thereby distinguishing between important and less important configurations. This should enable the future development of targeted search strategies in parameter space for the efficient generation of new training data, generating as much information as possible with as little new data as possible.

It is also conceivable to use the information to quantitatively assess the effectiveness of dimensionality reduction methods that are important for practical application, or to develop new methods that minimize information loss during dimensionality reduction. Last but not least, it should be possible to integrate the causal-deterministic information into the cost function of an ML model and thus to significantly increase its performance.

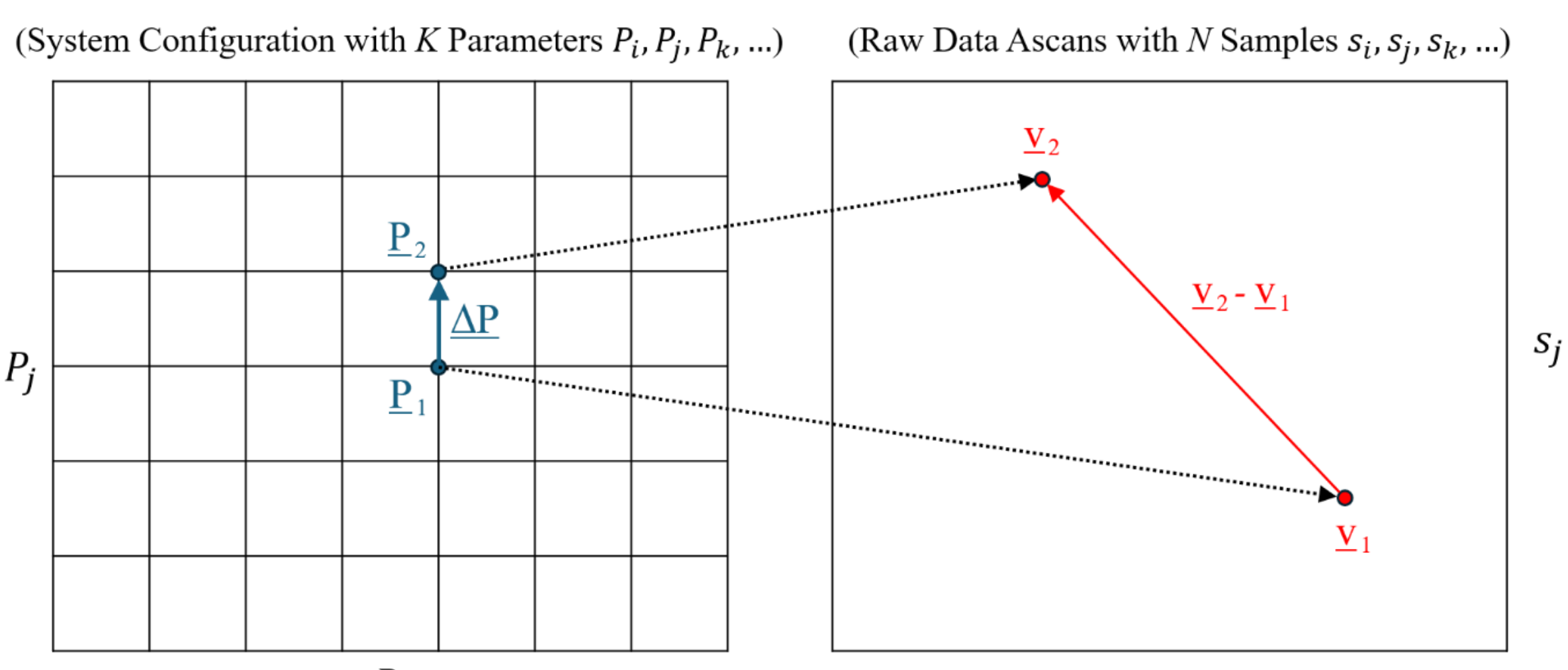

***Figure 25:*** *Relationship between a parameter change* $\underline{\Delta P}$ *in parameter/configuration space (on the left) and the associated information* $I_P \sim (\underline{v}_2 - \underline{v}_1)^2$ *in the raw data feature space (on the right).*

## 8. Relationship with Shannon and Fisher Information

The unique aspect of the “change information” introduced in this paper lies in the fact that, unlike all other known information measures such as Shannon or Fisher information, it does not require a probabilistic or statistical basis but can instead be motivated and described purely deterministically. All simulations used are based on physical-deterministic fundamental equations and yield the same result with each repetition (apart from minimal and insignificant numerical deviations). Neither expected values and variances nor other ensemble-averaged thermodynamic quantities need to be used. This suggests that the change information represents fundamental physical theory, which only takes on a statistical character in conjunction with real, noisy measurement data. In the following, we briefly discuss the mathematically motivated Shannon information and the statistical Fisher information in order to explore the differences and potential connecting factors with the new change information.

### 8.1 Shannon Information: Information as a Surprise

The mathematical information theory established by Claude Shannon (see, e.g., [1, 4]) deals with concepts such as information and entropy, transmission of information, as well as data compression and coding. It is of paramount importance, particularly for telecommunications, but also for all types of communication systems. Shannon's theory uses the thermodynamic concept of entropy to characterize the information content of messages transmitted over a noisy communication channel. For example, if the message to be transmitted is given by a set of characters $X = \{x_1, x_2, \dots, x_n\}$ with $i = 1, \dots, n$, and by the probabilities $p_i(x_i)$, with which the individual characters appear in the message, then, according to Shannon, the information content of a single character is

$$I(p_i) = -\log_2(p_i)\,. \tag{84}$$

This term is often referred to as "self-information," "surprise value," or "information gain." The less probable a character is in the entire string, the greater its surprise value and the more information is conveyed when that character appears. Conversely, very frequent characters encode little surprise or information.

The Shannon entropy $H$ of the entire string $X$ is defined as the expected value of the total information gain, or as the weighted average of all surprise values across all possible symbols in the string:

$$H(X) = \sum_{i=1}^{n} p_i I(p_i) = -\sum_{i=1}^{n} p_i \log_2(p_i) \leq \log_2(n) \tag{85}$$

$H$ has the unit 'bit' or 'Shannon (Sh)' and describes the information contained in the random variable $X$ or its distribution $p$. If $X$ is uniformly distributed, i.e., if $p_i = 1/n$ for all $i$, then $H$ is maximized and equals $H = \log_2(n)$. Thus, the more uniform the probabilities of the individual characters in a message, the higher the entropy of the entire message.

Shannon himself described his work as a "mathematical theory" of communication. It explicitly excludes semantic, meaning- and knowledge-related aspects - that is, statements about the usable content of transmitted messages and their meaning for the recipient. This means that a meaningful message for the recipient can be transmitted just as easily as a completely meaningless, random sequence of characters. Although Shannon's theory is usually referred to as an information theory, it therefore makes no direct statements about the information content of transmitted messages.

In contrast, the “change information” presented in this work is different. Here, the occurrence of large absolute field changes in response to certain relative parameter changes does indeed represent usable and meaningful information for the receiver or inspector, because it allows the sensitivity of the system, or more precisely, the measurement performed on it, to specific physical parameter changes to be quantified. Against this background, the change information can be considered not only a mathematical but also a physical information theory.

The principles of Shannon theory listed in equations (84) und (85) also have no significant relevance for the change information or the underlying difference wave fields, since the information content there does not correlate with the relative frequency of the amplitude values occurring in the difference signal, but rather with the squared amplitude values or their time integral. For these reasons, a closer connection between Shannon information and change information cannot be established at this point.

## 8.2 Fisher Information: Information as a Precise Statistical Estimate

Fisher information, named after the British statistician Ronald Fisher, is a measure in mathematical statistics that provides information about the best possible quality of parameter estimates [5, 6]. It plays an important role in the theory of maximum likelihood estimation and is also used in Bayesian statistics for calculating prior distributions. Fisher information is a measure of how much information an observable random variable $X$ contains about a (generally unknown) parameter $\theta$ on which the probability distribution of $X$ depends.

Let $f(x;\theta)$ be the probability density function of $X$ assuming a fixed value of $\theta$. It describes the probability of $X = x$ occurring when the value of $\theta$ is known. From this, the so-called Likelihood function can be derived according to

$$L(\theta; X = x) = c(x)\, f(x;\, \theta)\ . \tag{86}$$

It describes the *plausibility* of the parameter $\theta$ when the event $X = x$ has been observed. Here, $c(x)$ is an arbitrary constant function that does not depend on $\theta$ [7]. Since the theory of maximum likelihood estimation usually only considers likelihood ratios with respect to different $\theta$ values, $c(x)$ generally cancels out.

The natural logarithm of the likelihood function, $l(\theta;x) = \ln L(\theta;x)$, is called the Log-Likelihood and is generally the more convenient and important tool. With it, the so-called score function $S(\theta;x)$ can be calculated:

$$S(\theta;x) = \frac{\partial}{\partial\theta} \ln L(\theta;x)\,, \tag{87}$$

i.e., the partial derivative of the log-likelihood function with respect to $\theta$. Assuming weak regularity conditions, it can be shown that the expected value $\mathbb{E}$ of the score function vanishes:

$$\mathbb{E}[S(\theta;x)] = \mathbb{E}\left[\frac{\partial}{\partial\theta} \ln L(\theta;x)\right] = 0 \tag{88}$$

Fisher information $I(\theta)$ is defined as the variance of the score function with respect to $\theta$, which, due to the vanishing mean, is identical to the expected value of the squared score function. Therefore,

$$I(\theta) = \mathrm{Var}_\theta[S(\theta;x)] = \mathbb{E}\left[\left(\frac{\partial}{\partial\theta} \ln L(\theta;x)\right)^2\right]\,. \tag{89}$$

If the log-likelihood function is twice differentiable with respect to $\theta$, it can be shown, assuming additional regularity conditions, that the Fisher information can also be expressed as a twofold derivative of the log-likelihood [7]:

$$I(\theta) = -\mathbb{E}\left[\frac{\partial^2}{\partial\theta^2} \ln L(\theta;x)\right] \tag{90}$$

Fisher information can thus be interpreted as the curvature of the graph of the log-likelihood function at the location $\theta$. Near the maximum, low Fisher information indicates that the maximum is broad, meaning there are many points in the vicinity that yield a similar log-likelihood. Therefore, the true value of $\theta$ can only be determined imprecisely with a given number of measurements; or in other words, many measurements are needed to determine the true value with a given accuracy. Conversely, high Fisher information indicates that the maximum of the log-likelihood function is sharply defined. The true value of $\theta$ can then be determined very accurately with a given number of measurements, or in other words, fewer measurements are needed to achieve a given accuracy.

If we now identify the statistical parameters $\theta$ of Fisher's theory with our system parameters $P$ from the change information and consider each measured, noisy ultrasound time signal as an observable random variable $X(t)$, an interesting connection between the two concepts can be established using Eq. (89), which also features a squared partial derivative with respect to the parameter $\theta$. However, instead of the deterministic field quantities of the change information, namely the particle velocity $v$ or the stress $T$, a statistical log-likelihood function appears here. In the limiting case of an infinite number of measurements, the expected value of the squared score function from Eq. (89) can thus be considered as the true value of the Fisher information or with the deterministic value of the physical change information.

While we considered the parameter $P$ (or $\theta$) to be known and constant when verifying the change information using numerical simulations, Fisher's theory, according to equation (89), assumes the observation or measurement of the time signal $X = x$ is given, while the (unknown) parameter $\theta$ (or $P$) is to be estimated as precisely as possible. Eq. (89) can thus be seen as an inverse, statistical representation of the deterministic change information, which could open up new and interesting possibilities for the evaluation of real measurement data.

# 9. Summary and Outlook

## 9.1 Summary

In this work, a new information theory for non-destructive ultrasonic testing based on difference and differential wave fields, respectively, was developed and numerically verified using a simplified one-dimensional numerical model. Analogous to Poynting's theorem for energy transported in an ultrasonic field, a balance equation for structural information was derived and successfully tested for physical plausibility. Key components of this balance equation are the information density as a balance quantity and an information flux vector as the counterpart to the Poynting vector of energy flow. Unlike the energy balance equation, the information balance includes additional source terms, so that it does not become a continuity equation even after the excitation signal ceases. This contrasts with the statement in [8], where a continuity equation is assumed for the electromagnetic field. The present work clearly demonstrates that, unlike energy, information in an elastodynamic field is *not* a conserved physical quantity. Information can only be temporarily conserved between two interactions. During the interaction itself, information can be generated as well as annihilated, even if no energy dissipation is inherent in the system. Generation of information occurs primarily at the interfaces between two materials, but in the case of changes in sound velocity, also within the volume of the changing material. The information propagates from the sources to the sensors, with the ultrasound waves themselves acting as information carriers. Information therefore propagates at the speed of sound.

The investigations further demonstrated that information, like energy, can be split into kinetic and potential components. Except for temporary self-interference effects occurring within the model volume, kinetic and potential information exhibit qualitatively and quantitatively similar behavior, such that the total information essentially results from twice the kinetic or potential component. While the kinetic component of the information assumes the same mathematical expression in all the examples considered, the potential component is described by an extended expression in the case of material-related system parameters that are associated with density changes. This also applies to the information flux vector. At the stress-free model boundaries, the information is completely given by the kinetic term and therefore take a particularly simple form there, which is proportional to the kinetic energy of the differential wave field.

By appropriately scaling with the system parameter $P$ under consideration (or its square), which corresponds to introducing a relative parameter change, and normalizing to the energy of the excitation signal, it is ensured that the final information quantities in the linear-elastic model under consideration do not depend on the excitation amplitude and that the information itself becomes physically dimensionless. Nevertheless, the derived quantities provide an absolute information scale, the base units of which are called 'Change Bits', or *Cbits* for short. This allows for the quantitative comparison of different system parameters as well as different measurement setups. As it turns out, both changes in sound velocity and changes in position generate a particularly large amount of information. While reflection setups are particularly well-suited from an information-theoretical perspective for density and position changes, transmission setups are the preferred method for sound velocity changes. Thickness changes can be readily detected in both transmission and reflection.

The change information presented in this work has no connection with the information-technical Shannon information, but it does show interesting analogies to statistical Fisher information. Unlike Fisher information, however, the change information has a purely deterministic basis, suggesting that it represents the more fundamental theory from a physical perspective. Applying this theory to real, noisy measurement signals will likely lead to forms of Fisher information, but it cannot be equated with it, as it is claimed for the electromagnetic field in [8]. The possible relationship between these two information theories should be the subject of future investigations.

## 9.2 Outlook

The information theory presented in this paper was derived using the simplified case of plane longitudinal waves and is therefore only valid in idealized one-dimensional systems. An extension to more realistic 2-D and 3-D systems is the subject of ongoing work.

As already described, certain finite system variations can also be implemented experimentally. Thus, the generation, annihilation, and flow of information can also be visualized and investigated experimentally. This, too, is the subject of ongoing work.

The examples shown referred to a single initial model, i.e., a specific model configuration in parameter space. However, each point in this parameter space generally exhibits different derivatives of the field quantities $v$ and $T$ with respect to the parameters under consideration. This results in multidimensional information landscapes and thus information-theoretic weighting factors in the parameter space, which can serve as a starting point for targeted search and coverage strategies for the most efficient generation of new experimental or simulated training data with high quality.

Further interesting applications were already discussed in Section 7. These include, in particular, extending the theory to several simultaneously changing parameters, developing new evaluation methods and imaging techniques based on difference wave fields, and developing new information-based machine learning methods.

The concept of change information presented in this work is the first information theory to be based on a deterministic, i.e., non-statistical, foundation and therefore represents a fundamental theory from a physical perspective. Its scope is by no means limited to ultrasonic testing and can be readily applied to all situations in NDE and physics in which physical fields (e.g., acoustic, electromagnetic, thermal, gravitational, etc.) interact with objects. The specific balance equations differ in all cases, but the fundamental concept that the information results from the squared differential field should remain valid in all the aforementioned cases.

The principle of change information is so universal that extending it to measurements not describable by physical fields, as well as to non-technical systems such as health, social, financial, and ecosystems, etc., seems both possible and worthwhile. The essential components and hypotheses of such a higher-level systemic or physical information theory are illustrated in Figure 26.

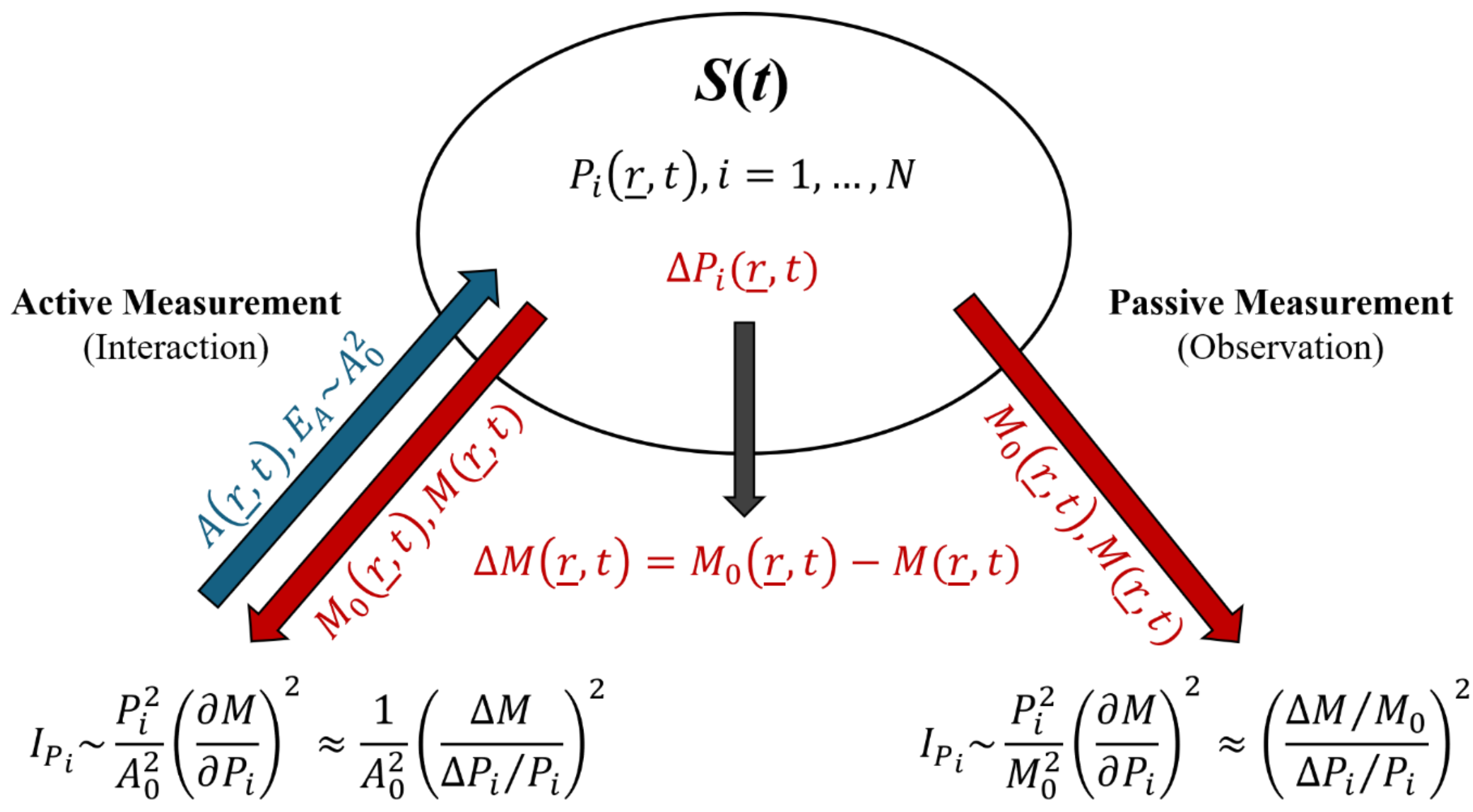

***Figure 26:*** *Basic principles and hypotheses of a systemic deterministic information theory.*

Consider an arbitrary technical or non-technical, generally time-dependent system $S(t)$, whose state is determined by a set of $N$, generally space- and time-dependent system parameters $P_i(\underline{r},t)$ and on which active or passive measurements or observations $M(\underline{r},t)$ are performed. Every natural or induced parameter change $\Delta P_i(\underline{r},t)$ within the system leads to a change in the measured signal, $\Delta M(\underline{r},t) = M_0(\underline{r},t) - M(\underline{r},t)$, where $M_0(\underline{r},t)$ represents the measurement before and $M(\underline{r},t)$ the measurement after the parameter change. Our hypothesis is that all such measurements provide information $I_{P_i}$ about the influence of the parameter $P_i$ on the system $S$, which is proportional to $P_i^2(\partial M/\partial P_i)^2$ or, in real systems, to $P_i^2(\Delta M/\Delta P_i)^2$. In addition, a normalization or scaling factor is included, which can be chosen differently depending on the type of measurement.

The examples discussed in this work were exclusively active ultrasound measurements, in which the system interacts with an excitation field or excitation signal $A(\underline{r},t)$ and provides the information as a system response that can be measured by the sensor(s) (left-hand side in Figure 26). The information balance equation derived in this work was normalized by the energy of the excitation signal $E_A \sim A_0^2$ to eliminate the physical dimension of the energy and the dependence of the information on the strength of the excitation signal. Since the energy of the excitation signal represents a constant for the system that is independent of position and time, it can be applied as a fixed value to all (position- and time-dependent) terms of the balance equation. It therefore takes on the character of a scaling factor for the information in the balance equation. Despite normalization to $A_0^2$, the information itself can then be seen as an *absolute* change in the measured quantity $\Delta M$, which is related to a *relative* parameter change $\Delta P_i/P_i$:

$$I_{P_i} \sim \frac{P_i^2}{A_0^2}\left(\frac{\partial M}{\partial P_i}\right)^2 \approx \frac{1}{A_0^2}\left(\frac{\Delta M}{\Delta P_i/P_i}\right)^2 \qquad (91)$$

In passive measurements (right-hand side of Figure 26), the energy of the excitation signal is either unknown (as in acoustic emission testing) or no excitation is present at all. In both cases, the energy of the excitation signal is therefore unsuitable as a normalizing quantity. Instead, the only plausible alternative is the square of the measured quantity before the parameter change, i.e., $M_0^2 = M_0^2(\underline{r}, t)$ in general. Unlike the constant $A_0^2$, this expression represents a spatially and temporally dependent normalizing quantity that also changes the nature of the information itself. The latter can then be interpreted as a *relative* change in the measured quantity $\Delta M/M$, which is related again to a relative parameter change $\Delta P_i/P_i$:

$$I_{P_i} \sim \frac{P_i^2}{M_0^2}\left(\frac{\partial M}{\partial P_i}\right)^2 \approx \left(\frac{\Delta M/M_0}{\Delta P_i/P_i}\right)^2 \tag{92}$$

This also provides a plausible basis for an information measure, as the following simple example shows. If, for instance, the number of acoustic emission events per minute in the system increases by $\Delta M = 2$ events per minute due to a parameter change, then stating this absolute value is not particularly meaningful as long as it is unclear whether, e.g., $M_0 = 10$ events per minute or $M_0$ =1000 events per minute were measured before the parameter change. Only after dividing by $M_0$, informative statements emerge, such as a 20 percent change or a 0.2 percent change.

Since the information concept presented here is based on changes in measured field quantities, and a change can in principle be defined as either an absolute or a relative quantity, it is not surprising that the derived information measure also exists in both absolute and relative formulations. While the relative formulation appears to be the only possible one for passive measurements, both formulations are conceivable for active measurements. In this work, the absolute formalism with normalization to the energy of the excitation signal was preferred, but a relative formulation would also be possible in principle. The differences between the two formulations are best illustrated by a simple example.

For instance, if we obtain two echoes of different strengths with amplitudes of 1 and 1/2, respectively, as a result of an active pulse-echo measurement on the initial model, and then perform a parameter change that alters both echoes by the same factor (i.e., halves them), the absolute formulation of the information presented in this work according to Eq. (91) yields four times the information content for the first echo as for the second. From a measurement perspective, this makes perfect sense because large absolute changes are generally easier to measure than small ones, especially when the latter are almost hidden in noise. In contrast, the relative formulation of the information with normalization to $M_0^2$ according to Eq. (92) leads to the same information content for both echoes. This can also be useful if both echoes are well above the noise threshold. The specific choice of absolute or relative formulation of information therefore depends on the respective application, the applicable framework conditions, and the research questions being pursued. The advantages and disadvantages of both approaches can only be properly assessed once experience with further applications is available in the future.